\documentclass[twocolumn]{aa} 

\usepackage{mathtools, amsmath}
\usepackage{nccmath}
\usepackage{float}
\usepackage{graphicx}
\usepackage{hyperref}
\usepackage{placeins}

\usepackage{svg}
\newcommand{\orcid}[1]{\textsuperscript{\href{https://orcid.org/#1}{\includegraphics[width=8pt]{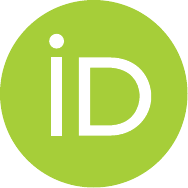}}}}

\usepackage{xcolor}
\hypersetup{
    colorlinks,
    linkcolor={red!50!black},
    citecolor={blue!50!black},
    urlcolor={blue!80!black}
}
\usepackage{natbib}

\newcommand{\fsig}{f{\sigma_8}}
\newcommand{\hmpc }{$h^{-1}$Mpc}
\newcommand{\ihmpc}{$h\, \text{Mpc}^{-1}$}
\newcommand{\snsim}{\textsc{snsim}}
\newcommand{\sn}{SN Ia}
\newcommand{\sns}{SNe Ia}

\usepackage[switch]{lineno} 

\begin{document} 

%\linenumbers

   \title{Growth-rate measurement with type-Ia supernovae using ZTF survey simulations}

   %\subtitle{...}

   \author{Bastien Carreres\orcid{0000-0002-7234-844X}\fnmsep
      	\inst{1}\fnmsep\thanks{Corresponding author: \email{carreres@cppm.in2p3.fr}}
      	\and
      	Julian E. Bautista\orcid{0000-0002-9885-3989}\fnmsep\inst{1}
        \and
        Fabrice Feinstein\inst{1}
        \and 
        Dominique Fouchez\inst{1}
        \and
        Benjamin Racine\orcid{0000-0001-8861-3052}\fnmsep\inst{1}
        \and
        Mathew Smith\inst{2}
        \and
        Melissa Amenouche\inst{3}
        \and 
        Marie Aubert\inst{3}
        \and
        Suhail Dhawan\inst{4}
        \and
        Madeleine Ginolin\inst{2}
        \and
        Ariel Goobar\inst{5}
        \and
        Philippe Gris\inst{3}
        \and
        Leander Lacroix\inst{5}\fnmsep\inst{6}
        \and 
        Eric Nuss\inst{7}
        \and
        Nicolas Regnault\inst{6}
        \and
        Mickael Rigault\inst{2}
        \and
        Estelle Robert\inst{2}
        \and
        Philippe Rosnet\inst{3}
      	\and
        Kelian Sommer\inst{7}
        \and
        Richard Dekany\inst{8}
        \and
        Steven L. Groom\orcid{0000-0001-5668-3507}\fnmsep\inst{9}
        \and
        Niharika Sravan\inst{8}
        \and
        Frank J. Masci\orcid{0000-0002-8532-9395}\fnmsep\inst{9}
        \and 
        Josiah Purdum\orcid{0000-0003-1227-3738}\fnmsep\inst{8}
      	}
   \institute{
                Aix Marseille Université, CNRS/IN2P3, CPPM, Marseille, France
          	    \and
         		Université de Lyon, Université Claude Bernard Lyon 1, CNRS/IN2P3, IP2I Lyon, F-69622, Villeurbanne, France
                \and
                Université Clermont Auvergne, CNRS/IN2P3, LPC, Clermont-Ferrand, France
                \and
                Institute of Astronomy and Kavli Institute for Cosmology, University of Cambridge, Madingley Road, Cambridge CB3 0HA, UK
                \and
                The Oskar Klein Centre for Cosmoparticle Physics, Department of Physics, Stockholm University, SE-10691 Stockholm, Sweden
                \and
                Sorbonne Université, CNRS/IN2P, LPNHE,    F-75005, Paris, France
                \and
                Université Montpellier, CNRS/IN2P3, LUPM,  F-34095, Montpellier, France
                \and
                Caltech Optical Observatories, California Institute of Technology, Pasadena, CA 91125, USA
                \and
                Infrared Processing and Analysis Center, M/S 100-22, 770 South Wilson Avenue, California Institute of Technology, Pasadena, CA 91125, USA
       	    }

   \date{}

  \abstract
     % Let's just do a single paragraph for abstract?
    {Measurements of the growth rate of structures at $z < 0.1$ with 
    peculiar velocity surveys have the potential of testing the validity of 
    general relativity on cosmic scales. In this work, we present 
    growth-rate measurements from realistic simulated sets of 
    type-Ia supernovae (\sns) from the Zwicky Transient Facility (ZTF). 
    We describe our simulation methodology, the light-curve fitting, and peculiar velocity 
    estimation. Using the maximum likelihood method, we derived constraints on $f\sigma_8$ using only ZTF \sn\ peculiar velocities.
    We carefully tested the method and we quantified biases due to selection effects (photometric detection, spectroscopic follow-up for typing) on several independent realizations.  
    We simulated the equivalent of 6 years of ZTF data, and considering 
    an unbiased spectroscopically typed sample at $z<0.06$, we obtained 
    unbiased estimates of $\fsig$ with an average uncertainty of 19\% precision. We also investigated the information gain in applying 
    bias correction methods. Our results validate our framework, which 
    can be used on real ZTF data. 
    } 
    

    \titlerunning{Growth-rate measurement with \sns\ using ZTF survey simulations}
    \authorrunning{B. Carreres et al. }
   \maketitle
%
%-------------------------------------------------------------------

\section{Introduction}

   The standard model of cosmology assumes that gravity is described at all scales by general relativity (GR) and its content is dominated by two exotic components, cold dark matter (CDM) and a dark energy component with the dynamics of a cosmological constant $\Lambda$. These components are required to explain the growth of structures and the acceleration of the expansion of the Universe. This flat $\Lambda$CDM+GR model has been successful in describing most, if not all, cosmological observations.
   
   The exact nature of dark energy remains unknown, and alternative models of gravity have been proposed to explain our observations without needing dark energy (see e.g., \citealt{cliftonModifiedGravityCosmology2012, zhaiEvaluationCosmologicalModels2017,ezquiagaDarkEnergyLight2018}). These models can predict the same background quantities 
   as the $\Lambda$CDM+GR model, such as the expansion rate $H(z)$ as a
   function of redshift $z$, but they can yield quite different
   predictions for quantities related to perturbations, such as 
   the linear growth rate of structures $f(z)$. In some of these
   models, the growth rate even becomes scale dependent. 
   To test whether our Universe is ruled by a $\Lambda$CDM+GR model 
   or some alternate gravity model, 
   not only do we need precise measurements of $H(z)$ from standard 
   candles (e.g., type-Ia supernovae) or standard rulers 
   (e.g., baryon acoustic oscillations) but also measurements of
   the growth of structures $f(z)$ with redshift-space
   distortions or of the amplitude of matter fluctuations with weak gravitational lensing. Ongoing and future cosmological surveys will constrain 
   the expansion rate to subpercent level precision and the growth 
   rate to a few percent which will allow us to test models of gravity.

   Most common measurements of the growth rate of structures are 
   based on the effect of redshift-space distortions  
   in the clustering of galaxies 
   \citep{guzzoTestNatureCosmic2008, 
   songReconstructingHistoryStructure2009}. 
   Peculiar velocities of galaxies modify  
   their cosmological redshift, such that
   when we estimate distances to these galaxies using their observed 
   redshifts, they are slightly misplaced relative to their true 
   comoving positions. The galaxy density field becomes 
   distorted in redshift space relative to real comoving space,
   and the two-point statistics of the galaxy density field 
   becomes anisotropic: the clustering along the line of sight is 
   enhanced relative to the clustering across the line of sight. 
   The amplitude of this anisotropy is proportional to the growth 
   rate $f$ and to the amplitude of matter fluctuations, commonly 
   described by the $\sigma_8$ parameter
   (the standard deviation of the matter field that has been top-hat smoothed on 
   scales of 8\hmpc).
   Clustering measurements of growth therefore usually quote the
   combination $f(z)\sigma_8(z)$.
   Several redshift-space distortion measurements have been performed in the past decade 
   by spectroscopic surveys including 
   WiggleZ \citep{blakeWiggleZDarkEnergy2011a}, 
   6dFGRS \citep{beutler6dFGalaxySurvey2012a}, 
   SDSS-II \citep{samushiaInterpretingLargescaleRedshiftspace2012}, 
   SDSS-MGS \citep{howlettClusteringSDSSMain2015},
   FastSound \citep{okumuraSubaruFMOSGalaxy2016},
   VIPERS \citep{pezzottaVIMOSPublicExtragalactic2017, delatorreVIMOSPublicExtragalactic2017}, 
   SDSS-III BOSS \citep{beutlerClusteringGalaxiesCompleted2017, griebClusteringGalaxiesCompleted2017, sanchezClusteringGalaxiesCompleted2017, satpathyClusteringGalaxiesCompleted2017},
   and more recently by SDSS-IV eBOSS 
   \citep{bautistaCompletedSDSSIVExtended2021, 
   gil-marinCompletedSDSSIVExtended2020,
   demattiaCompletedSDSSIVExtended2021, tamoneCompletedSDSSIVExtended2020, 
   houCompletedSDSSIVExtended2021, 
   neveuxCompletedSDSSIVExtended2020}.
   The latest measurements of $\fsig$ with redshift-space distortions 
   reach uncertainties of about 10 percent
   and currently no deviations from $\Lambda$CDM+GR 
   have been detected \citep{alamCompletedSDSSIVExtended2021}.
   
   Another method to measure $\fsig$ is to derive it from the 
   two-point statistics of direct peculiar velocity estimates for
   individual galaxies 
   \citep{gorskiCosmologicalVelocityCorrelations1989,
   straussDensityPeculiarVelocity1995}. 
   Peculiar velocities can be measured if both  
   redshifts and absolute distances can be estimated independently. 
   While spectroscopy provides precise redshifts, 
   distance estimates can be obtained using well-known correlations 
   such as the Tully-Fisher relation for spiral galaxies 
   (TF, \citealt{tullyNewMethodDetermining1977})
   or the Fundamental Plane for elliptical ones 
   (FP, \citealt{djorgovskiFundamentalPropertiesElliptical1987}).
   Such distances can be measured for galaxies at relatively low
   redshifts ($z<0.1$) since uncertainties quickly increase with
   redshift. 
   Current state-of-the-art samples of TF and FP distances include
   CosmicFlows4 \citep{tully_cosmicflows4_2022} and the SDSS-FP sample
   \citep{howlettSloanDigitalSky2022a}, both containing a few times $10^4$ 
   distance measurements. 
   The statistical properties of a sample of peculiar velocities 
   can be measured alone or in combination with an overlapping 
   galaxy density field, analogously to a multitracer analysis. 
   Several methods have been developed in the past years to 
   extract growth-rate measurements from sets of peculiar velocities: 
   The maximum-likelihood method, where velocity (and density) fields are assumed to be drawn from multivariate Gaussian 
    distributions
        \citep{johnson6dFGalaxySurvey2014, 
        hutererTestingLCDMLowest2017, 
        howlett2MTFVIMeasuring2017, 
        adamsJointGrowthrateMeasurements2020, 
        laiUsingPeculiarVelocity2023}. 
        This is the method we employ in this work;   
        The compressed two-point statistics such as two-point 
        correlation function, power spectrum, or average pair-wise 
        velocities \citep{nusserVelocitydensityCorrelationsCosmicflows32017, 
        dupuyEstimationLocalGrowth2019, 
        qinRedshiftspaceMomentumPower2019,
        turnerLocalMeasurementGrowth2022};
        The comparison between observed velocities to those reconstructed 
        from a density field \citep{davisLocalGravityLocal2011, carrickCosmologicalParametersComparison2015, boruahCosmicFlowsNearby2020,saidJointAnalysis6dFGS2020};
        The field level inference by evolving initial conditions or forward modeling method \citep{boruahReconstructingDarkMatter2021, prideaux-gheeFieldbasedPhysicalInference2023}

   Type-Ia supernovae (\sns) are well-known standardizable candles that have smaller intrinsic 
   scatter in standardized peak luminosity (of about 15 percent) than TF and FP relations (of about 40 percent), 
   so \sns\ can yield more precise peculiar velocities. 
   \sns\ have only been marginally used for growth-rate measurements (e.g., \citealt{boruahCosmicFlowsNearby2020}) 
   since most surveys only cover small parts of the sky or suffer from 
   being compilations of several different telescopes, which cover the sky inhomogeneously
   (e.g., \citealt{betouleImprovedCosmologicalConstraints2014, scolnicPantheonAnalysisFull2022}). 
   Photometric surveys with high cadence and large sky coverage, such 
   as the Zwicky Transient Facility 
   (ZTF, \citealt{grahamZwickyTransientFacility2019}) and the Rubin Observatory Legacy Survey of Space and Time (Rubin-LSST, \citealt{lsstsciencecollaborationLSSTScienceBook2009}) will provide a large and uniform sample of \sns\ that can be used 
   for peculiar velocity studies \citep{howlettMeasuringGrowthRate2017}.
   Combining peculiar velocities from \sns, Tully-Fisher and Fundamental Plane can set the best constraints on the growth-rate at $z<0.1$
   and will allow us to constrain alternatives to GR 
   \citep{kimComplementarityPeculiarVelocity2020, lyallTestingModifiedGravity2022}.
   
   In this work we study the possibility 
   for a first growth-rate measurement using uniquely \sn\ data from ZTF. 
   In preparation for the analysis of these data, we produced realistic simulations of ZTF \sn\ light-curves, including selection effects
   and instrumental noise, and we performed the analysis required to derive $f\sigma_8$, 
   based on the maximum likelihood method 
   \citep{johnson6dFGalaxySurvey2014, howlett2MTFVIMeasuring2017}.
   
   This article is organized as follows. 
   In Sect.~\ref{sec:simulations} we describe the pipeline to produce ZTF simulations of
   \sn\ observations. 
   In Sect.~\ref{sec:methodology} we present the method used to estimate $\fsig$. In
   Sect.~\ref{sec:results} we describe our main findings. In Sect.~\ref{sec:robtests} we consider variations of the baseline analysis. We finally conclude in Sect.~\ref{sec:conclusion}.

\section{ZTF simulations}
\label{sec:simulations} 

    This section describes our framework to produce realistic sets of simulated \sn\ light-curves from the Zwicky Transient Facility, including peculiar velocities and 
    multiple possible observational effects. 
    Our pipeline goes through the following steps:
        1) we extract host catalogs from a suitable N-body simulation;
        2) we generate \sn\ events with positions drawn from the host catalog and random dates;
        3) we generate their true light-curve parameters;
        4) we simulate ZTF-like light-curves based on real observations  (cadence, filters, noise);
        5) we introduce ZTF-like spectroscopic selection effects;
        6) we apply ZTF-like quality cuts on selected observations.
    Each of these steps are described in detail below. 
    This pipeline, named \snsim\footnote{\url{https://github.com/bastiencarreres/snsim}}, 
    was implemented in python language and is publicly available. 
    Another alternative software for producing ZTF simulations is \textsc{simsurvey}\footnote{\url{https://simsurvey.readthedocs.io/}} \citep{feindtSimsurveyEstimatingTransient2019}. 
    \textsc{simsurvey} was previously used to study the discovery rates of different transients before the start of the survey.
        
    \subsection{The N-body simulation}
    \label{sec:simulations:nbody} 

    To study the statistics of realistic nonlinear velocity fields, 
    we rely on velocities from halos found in matter-only N-body simulations 
    that are publicly available. 
    
    We used the OuterRim\footnote{\url{https://cosmology.alcf.anl.gov/}} cosmological simulation \citep{heitmannOuterRimSimulation2019} and focused on the snapshot at redshift $z = 0$. 
    The OuterRim simulation was widely used in recent cosmological measurements from the eBOSS 
    \citep{gil-marinClusteringSDSSIVExtended2018, 
    houClusteringSDSSIVExtended2018, 
    zarroukClusteringSDSSIVExtended2018, 
    avilaCompletedSDSSIVExtended2020, 
    rossiCompletedSDSSIVExtended2021, 
    smithCompletedSDSSIVExtended2020}. The OuterRim volume is a $(3~h^{-1} {\rm Gpc})^3$ cubic box.
    A total of 10,240$^3$ particles were evolved from initial conditions set at $z = 200$ 
    using the Zeldovich approximation and cosmological parameters displayed in Table~\ref{tab:cosmo_params}. 
    This corresponds to a particle mass $m_p = 1.85\times 10^9~h^{-1} M_\odot$. 
    
    Halos were defined using a friend-of-friends algorithm with a linking length of $b=0.168$, resulting in $1.9 \times 10^9$ halos with masses. 
     typically ranging from $m_{10\%} \sim 4 \times 10^{10} M_\odot$ (10th percentile) to $m_{90\%} \sim 4.6 \times 10^{11}M_\odot$ (90th percentile).
     
     In the rest of this work, we assign \sns\ to halo positions with 
     a probability that is independent of halo mass.

    \begin{table}
    \centering
    \caption{Cosmological parameters used in the OuterRim simulation. $H_0$ is given in km.s$^{-1}$.Mpc$^{-1}$}.
        \begin{tabular}{cccccccc} 
            \hline 
            \hline \\[-3ex]
            $H_0$ & $\omega_{\rm cdm}$ &  $\omega_{\rm b}$ & $n_s$ & $\sigma_8$ & $f$ & $\fsig$ \\ 
            \hline \\[-1.8ex] 
            $71.$ & $0.1109$ & $0.02258$ & $0.963$ & $0.800 $ & $0.478$ & $0.382$
        \end{tabular} 
    \label{tab:cosmo_params}
    \end{table}

    We have produced 27 realizations by selecting (1~$h^{-1}$Gpc)$^3$ subboxes.
    Each subbox encloses halos up to $z \sim 0.17$ when we place the observer at its center. 
    We show in Sect.~\ref{sec:simulations:ztf_selection_effects}
    that after all the selection effects we do not observe spectroscopically typed \sn\ above $z \sim 0.14$ in the ZTF survey.
    
    We did not attempt to populate halos with a realistic sample of galaxies, 
    even though potential effects can be introduced due to correlations of 
    supernova events with type of galaxy (passive or star-forming) 
    and how these galaxies connect to large-scale structures. 
    We leave this investigation for future work.

    \subsection{From N-body mocks to a survey-like host catalog}
    \label{sec:simulations:hosts_distances} 
    
    From the N-body simulation mocks, we convert spatial comoving coordinates of halos $(x, y, z)$ 
    to right-ascension (RA), declination (Dec) and redshift  $(z_\text{cos})$. 
    We first placed the observer at the center of each (1~$h^{-1}$Gpc)$^3$ subbox
    and computed distances of each halo to the observer. 

   To convert distances to cosmological redshifts $z_{\rm cos}$ we numerically invert the redshift-comoving distance relation, which for a flat $\Lambda$CDM universe is given by

    \begin{equation}
        r_\text{cos} \equiv r(z_{\rm cos}) = \frac{c}{H_0} \int_0^{z_{\rm  cos}} \frac{dz}{\sqrt{\Omega_m (1 + z)^3 + \Omega_\Lambda}},
        \label{eq:comodist}
    \end{equation}
    where $\Omega_m$ and $\Omega_\Lambda$ are respectively the density of matter and dark energy today. Since we assume a flat universe $\Omega_\Lambda = 1 - \Omega_m$.
    
    We then take into account the Doppler effect due to the peculiar velocity of the host and the velocity with respect to the Cosmic Microwave Background (CMB) frame to obtain the observed redshift $z_{\rm obs}$ as
    \begin{equation}
        (1 + z_{\rm obs}) = (1 + z_{\rm cos})(1 + z_{\rm p})(1 + z_\odot),
    \end{equation}
    where $z_{\rm cos}$ is the cosmological redshift, $z_{\rm p}$ the 
    red(or blue)shift due to the host peculiar velocity and $z_\odot$ is the shift due to the peculiar velocity of the Solar System with respect to CMB restframe. 
    We set $z_\odot = 0$ considering that it can be corrected using CMB measurement \citep{planckcollaborationPlanck2018Results2020a}.
    
    The expression for $z_{\rm p}$ derives from the relativistic Doppler effect due to the peculiar velocity and is given by
    \begin{equation}
        1 + z_{\rm p} = \frac{1 + \mathbf{v_p} \cdot \mathbf{\hat{n}} / c}{\sqrt{1 - \left(||\mathbf{v_p}||/c\right)^2}},
    \end{equation}
    where $\mathbf{v_p}$ is the 3-D peculiar velocity vector, 
    $c$ the speed of light, and $\mathbf{\hat{n}}$ is the unit vector pointing toward the \sn.
    The second-order term $(||\mathbf{v_p}||/ c)^2$ can be neglected, leading to
    \begin{equation}
        z_{\rm p} =\frac{\mathbf{v_p} \cdot \mathbf{\hat{n}}}{c}.
    \end{equation}
    This last expression allows us to compute the line-of-sight velocity from the Doppler shift.
    
    The relativistic beaming due to peculiar velocities change the luminosity distance $d_L$ \citep{huiCorrelatedFluctuationsLuminosity2006, davisEffectPeculiarVelocities2011} as

     \begin{equation}
        d_{L, \text{obs}} = (1+ z_{\rm p})^2 d_{L, \text{cos}},
        \label{eq:vp_distance_change}
    \end{equation}
    where $d_{L,\text{cos}} \equiv d_L(z_\text{cos}) = (1 + z_\text{cos}) r_\text{cos}$ is the cosmological luminosity distance. 
    The observed distance modulus is then given by
    \begin{align}
          \mu_{\rm obs} &= 5\log\left(\frac{d_{L,\text{obs}}}{10 \ {\rm pc}}\right) \\
                        &= 5\log\left((1 + z_{\rm p})^2 \frac{d_{L,{\rm cos}}}{10 \ {\rm pc}}\right)\\
                        &= \mu_{\rm cos} + 10\log\left(1 + z_{\rm p}\right).
                        \label{eq:muobs}
    \end{align}
    
\begin{figure}
        \centering
        \includegraphics[width=\columnwidth]{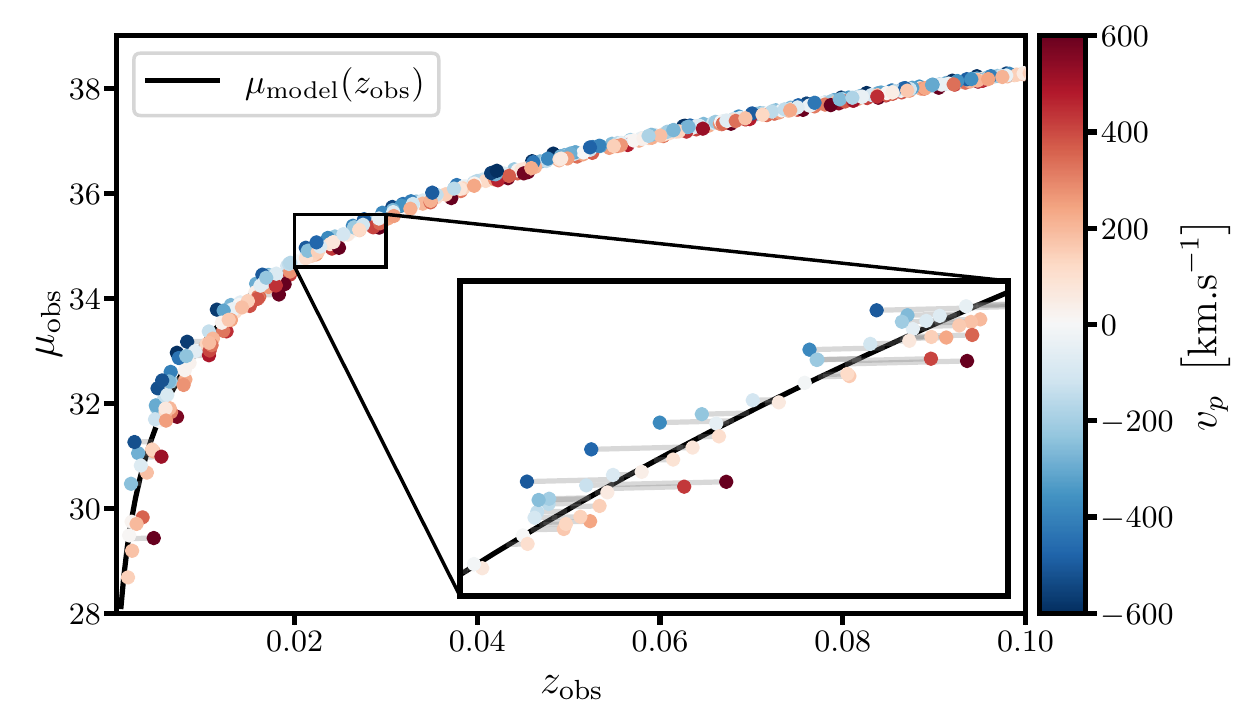}
        \caption{Toy model illustrating how peculiar velocities impact the observed 
        redshift $z_\text{obs}$ and the observed distance moduli $\mu_\text{obs}$.
        Radial peculiar velocities are randomly drawn from a Gaussian distribution with $\sigma_{\rm vp} = 300$ km/s. 
        The color of each point indicates the value of its radial peculiar velocities $v_p$. The effect on the x-axis is of first order on $v_p$ while the effect on the y-axis is of second order. }
        \label{fig:simple_sim}
    \end{figure}

    Figure~\ref{fig:simple_sim} illustrates these two effects on a random sample of distance indicators with Gaussian realizations of their peculiar velocities. 
    The effect on the observed redshifts is proportional to $(1+z_{\rm p})$ while 
    the effect on distance moduli is logarithmic, so subdominant (as seen in the 
    snippet of Fig.~\ref{fig:simple_sim}).
    
    \subsection{Generating \sn\ events}
    \label{sec:simulations:sn_lightcurves} 
    
    The next step is to generate  type-Ia supernova events.
    We used the latest estimates for the rate of explosions,
    \begin{equation}
        r_{v, P20}(z=0) = (2.35 \pm 0.24) \times 10^{-5} \text{Mpc}^{-3} \text{yr}^{-1},
        \label{eq:sn_rate}
    \end{equation}
    measured from ZTF data \citep{perleyZwickyTransientFacility2020}.
    We rescaled this rate to our fiducial $H_0$ value with 
        $r_v = r_{v, P20} \left(h/0.70\right)^3$.
    We also account for the time dilation at $z>0$ by scaling the rate by $1/(1+z)$.
    From this rate we computed the average number of \sns\ given the volume and the duration of our survey. We then drew the number of \sns\ in a given realization using a Poisson law.
    
    These \sn\ events were spatially assigned to the halos positions from 
    the N-body simulation. The velocity of the halos were also directly assigned 
    to their corresponding \sn. 
    This procedure neglects the velocity contribution from the relative velocity
    between the \sn\ and its host, which would simply add extra intrinsic scatter
    to their velocities.
    
    To generate the light-curves, we used the SALT2.4 model (\citealt{guySALT2UsingDistant2007, guySupernovaLegacySurvey2010}),
    which parameterized them by their stretch $x_1$, color $c$ and peak-magnitude 
    in the Bessel-B band $m_B$. 
    Using the Tripp relation \citep{tripptwopar1998}, the rest-frame magnitude in Bessel-B band for a given SN (indexed by the subscript $i$) is
    \begin{equation}
        M_{B, i}^* = M_B - \alpha x_{1, i} + \beta c_i + \sigma_{{\rm int}, i},
    \end{equation}
   where $\alpha$, $\beta$ and $M_B$ are common to all \sns\, and $x_{1, i}$, $c_i$ and the intrinsic scattering $\sigma_{{\rm int}, i}$ are randomly drawn from distributions described below.
    
    The absolute magnitude of \sns\ in Bessel-B band $M_B$, defined in the AB magnitude system, is fixed at the best-fit
    value of $-19.05$ (for $H_0 = 70 \ {\rm km.s}^{-1}{\rm .Mpc}^{-1}$) from \citealt{betouleImprovedCosmologicalConstraints2014}).
    We rescaled this $M_B$ value to our fiducial cosmology using
    \begin{equation}
        M_B = -19.05 + 5 \log\left(\frac{h}{0.7} \right).
    \end{equation}
    The $\alpha$ and $\beta$ parameters are also fixed  to best-fit values from \cite{betouleImprovedCosmologicalConstraints2014} that are $\alpha = 0.14$ and $\beta = 3.1$.
    The stretch parameter $x_1$ distribution is modeled using the redshift dependent two-Gaussian mixture from \citet{nicolasRedshiftEvolutionUnderlying2021}.
    The color parameter $c$ distribution follows the asymmetric model given by Table~1 of \citet{scolnicMEASURINGTYPEIA2016} for low-z (G10 model).
    The intrinsic scattering $\sigma_{\rm int}$ is drawn from a normal distribution with dispersion fixed to $\sigma_M = 0.12$. The values for the main input parameters are summarized in Table \ref{tab:snstandpar}.    
    \begin{table}
    \centering
    \caption{Input parameters of \sns\ standardization} 
        \begin{tabular}{cccc} 
            \hline 
            \hline \\[-3ex]
            $\alpha$ & $\beta$ &  $M_0$ & $\sigma_M$ \\ 
            \hline \\[-2.5ex] 
            $0.14$ & $3.1$ & $-19.019$ & $0.12$
        \end{tabular} 
    \label{tab:snstandpar}
    \end{table} 
    
    After generating $x_1$, $c$ and $\sigma_{\rm int}$ for each \sn, we compute their apparent magnitude given by
    \begin{equation}
        m_B = M_B^* + \mu_{\rm obs},
    \end{equation}   
    where $\mu_{\rm obs}$ is the observed distance modulus to the 
    \sn\ host (Eq.~\ref{eq:muobs}), which includes the peculiar velocity contribution.
    
    \subsection{Replicating ZTF observations} 
    \label{sec:simulations:ztf_obs}

    \begin{figure}
        \centering
        \includegraphics[width=\columnwidth]{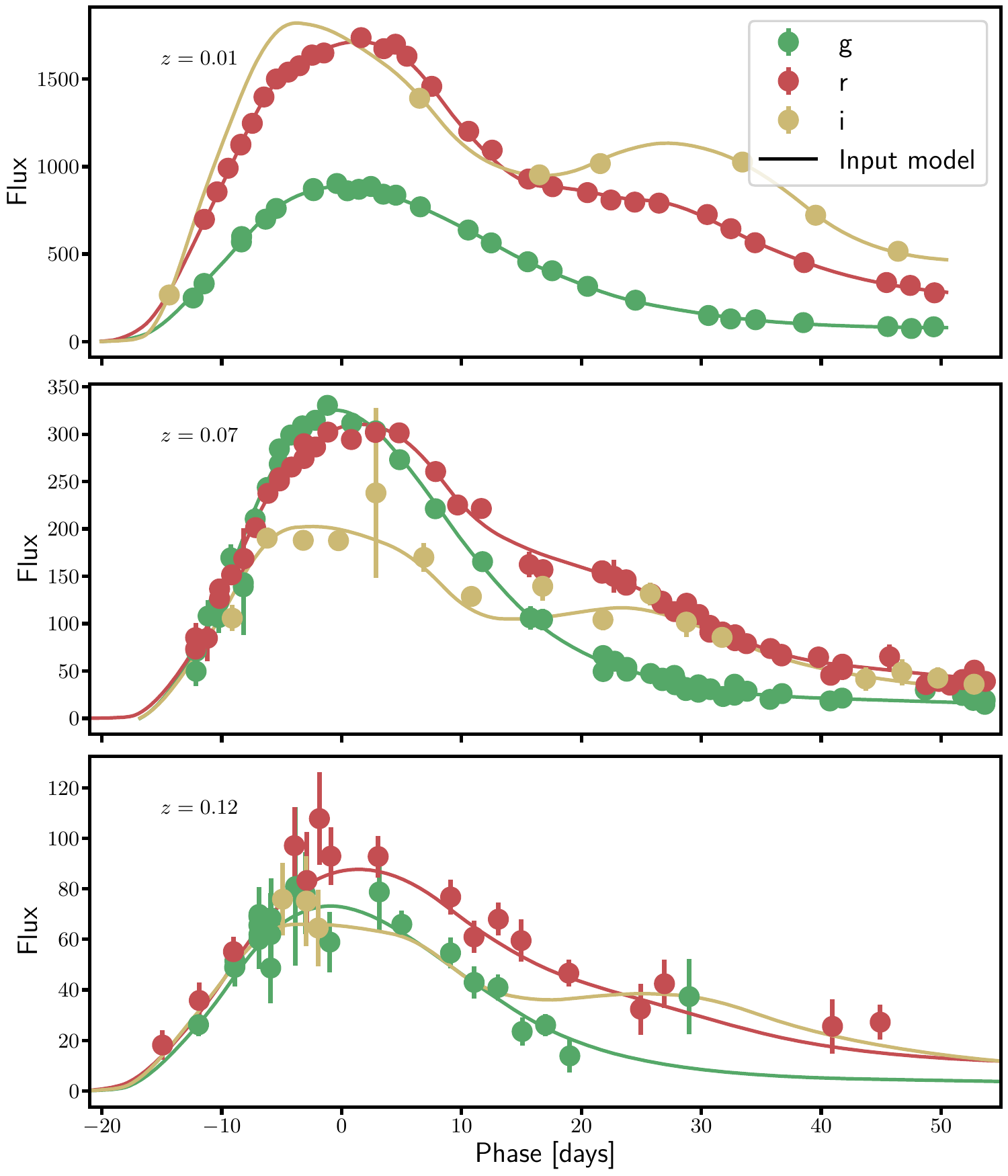}
        \caption{Examples of simulated ZTF light-curves of type-Ia supernovae at low, intermediate and high redshifts. Points with error bars are the simulated data, and the solid lines are the input light-curves.}
        \label{fig:LCsim}
    \end{figure}

    The Zwicky Transient Facility (ZTF) is conducting a photometric survey
    using the Samuel Oschin 48-inch (1.2-m) Schmidt Telescope at the 
    Palomar observatory \citep{grahamZwickyTransientFacility2019}. 
    It covers the entire northern visible sky in the g, r and i bands 
    with 30-second exposures pointing a fixed grid of fields with minimal 
    dithering \citep{bellm_survey_sched}.
    The ZTF camera hosts 16 charge coupled devices (CCD) containing a total of 0.6 gigapixels with an effective field of view of $\simeq 47~{\rm deg}^2$. 
    We use the focal plane dimensions, including inter-CCD gaps from \cite{dekanyZwickyTransientFacility2020} (Table 3). 
    
    The data processing pipelines are managed by Infrared Processing 
    and Analysis Center (IPAC, \cite{masciZwickyTransientFacility2019}), 
    which provides metadata tables. 
    In order to replicate ZTF observations we query these metadata using 
    the public code 
    \textsc{ztfquery}\footnote{\url{https://github.com/MickaelRigault/ztfquery}}. 
    
    The essential quantities of our simulations are:
        the dates of observations;
        the filters used;
        the limiting magnitude at 5-$\sigma$ of the observations $m_{5\sigma}$, which defines the magnitude at which the signal-to-noise ratio (S/N) is equal to 5;
        the CCD gain $G$ of each observation in units of electrons per analog-to-digital unit (ADU);
        the zero point ZP of each observation which gives the magnitude of an object that produces a flux of 1 ADU during an exposure.
    We also account for the Milky-Way dust extinction using the CCM89 model \citep{ccm89} and computing each object EBV from the \cite{schlegelMapsDustInfrared1998} dust map implemented in the pyhton package \textsc{sfdmap}\footnote{\url{https://github.com/kbarbary/sfdmap}}.
   
    We obtained true fluxes for each epoch of ZTF observations using \textsc{sncosmo} package\footnote{\url{https://sncosmo.readthedocs.io/}} \citep{barbary_sncosmo_2016}. 
    Since the flux  noise is dominated by the sky background at high-magnitude, we compute an effective sky noise using the limiting magnitude:
    \begin{equation}
        \sigma_{\rm sky} =  \frac{1}{5} 10^{-0.4 \left(m_{5\sigma} - \text{ZP}\right)}.
    \end{equation}

    A random noise is added to the true flux of each epoch. This noise is drawn from a Gaussian distribution with standard deviation given by
    \begin{equation}
        \sigma_F^2 = \frac{F}{G} + \sigma_{\rm sky}^2 +  \left(\frac{\ln(10)}{2.5}  F \right)^2 \sigma^2_\text{ZP},
    \end{equation}
    where we set $\sigma_\text{ZP} = 0.01$ as a calibration uncertainty. This calibration uncertainty will be refined in further works, together with taking into account host background, brighter-fatter effect, point spread function effects and calibration uniformity.

    Figure~\ref{fig:LCsim} displays examples of simulated 
    light-curves from ZTF SN Ia events, at three redshifts.
    We note that since i-band observations are not performed over the 
    whole sky, some supernovae have few or no i-band observations as the one 
    shown in the bottom panel of Fig.~\ref{fig:LCsim}.

     We simulated a 6-year data sample, similar to the full ZTF survey duration.
     We used observation metadata from 2018-06-19 to 2022-08-31, to follow realistic observing conditions. To obtain a 6-year simulated survey, we artificially increased the \sn\ rate.

    \subsection{ZTF selection from detection and spectroscopic typing}
    \label{sec:simulations:ztf_selection_effects}
    
    For cosmological analysis we require that detected objects are confirmed as type-Ia supernovae. The ZTF Bright Transient Survey (BTS) is a spectroscopic campaign to
    spectroscopically classify extragalactic transients brighter than 18.5 mag at peak brightness in either the g or r-filters 
    \citep{fremlingZwickyTransientFacility2020, perleyZwickyTransientFacility2020}. The BTS follow-up procedure requires stringent cuts, we describe their implementation in this section.

    Prior to spectroscopic selection we performed a photometric detection of sources by discarding \sn\ light-curves with less than two epochs with fluxes S/N above 5.
    In order to simulate the BTS spectroscopic selection effect, 
    we followed bullets 1 to 3 of the procedure presented in Sect. 2.3 of 
    \cite{perleyZwickyTransientFacility2020}: 
    1) prior to peak brightness, at least one observation with $-16.5 <t - t_{\rm peak} < -7.5$ days;
    2) around peak brightness, at least one observation  with $-7.5 < t - t_{\rm peak} < -2.5$ or $2.5 < t - t_{\rm peak} < 7.5$;
    3) after peak brightness, at least one observation with $7.5 < t - t_{\rm peak} < 16.5 $ or two observations, one within $2.5 < t - t_{\rm peak} < 7.5$ and one within $16.5 < t - t_{\rm peak} < 28.5$.
    Where $t_{\rm peak}$ is defined as the time where the light-curve reaches
    maximum flux (provided it has S/N above 5).
    
    At higher magnitude the spectroscopic efficiency drops. 
    To simulate this, we used the completeness given in Fig.~4 of 
    \citet{perleyZwickyTransientFacility2020} to randomly discard objects 
    as a function of their peak magnitude.
    
\begin{figure*}
        \centering
        \includegraphics[width=0.9\textwidth]{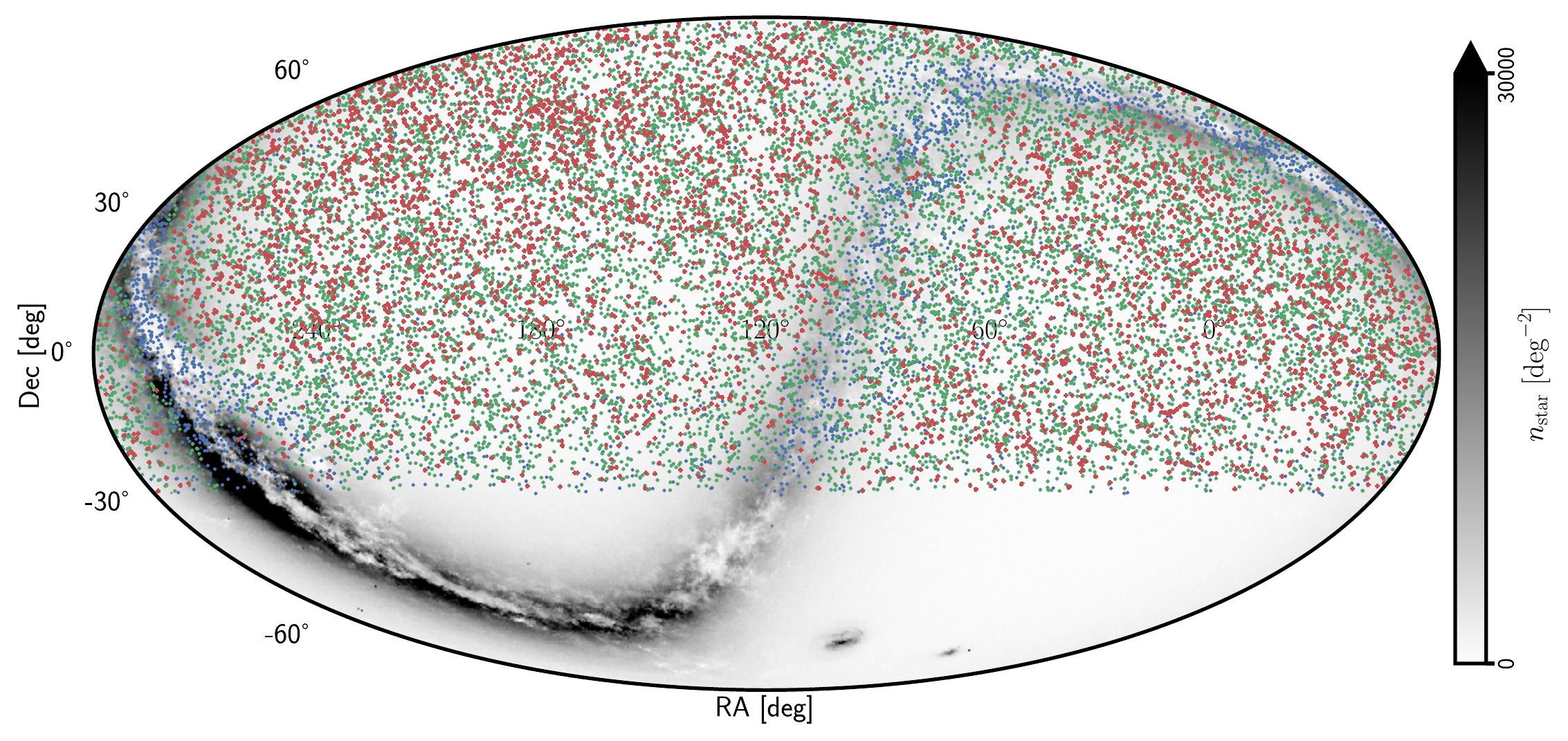}
        \includegraphics[width=0.49\textwidth]{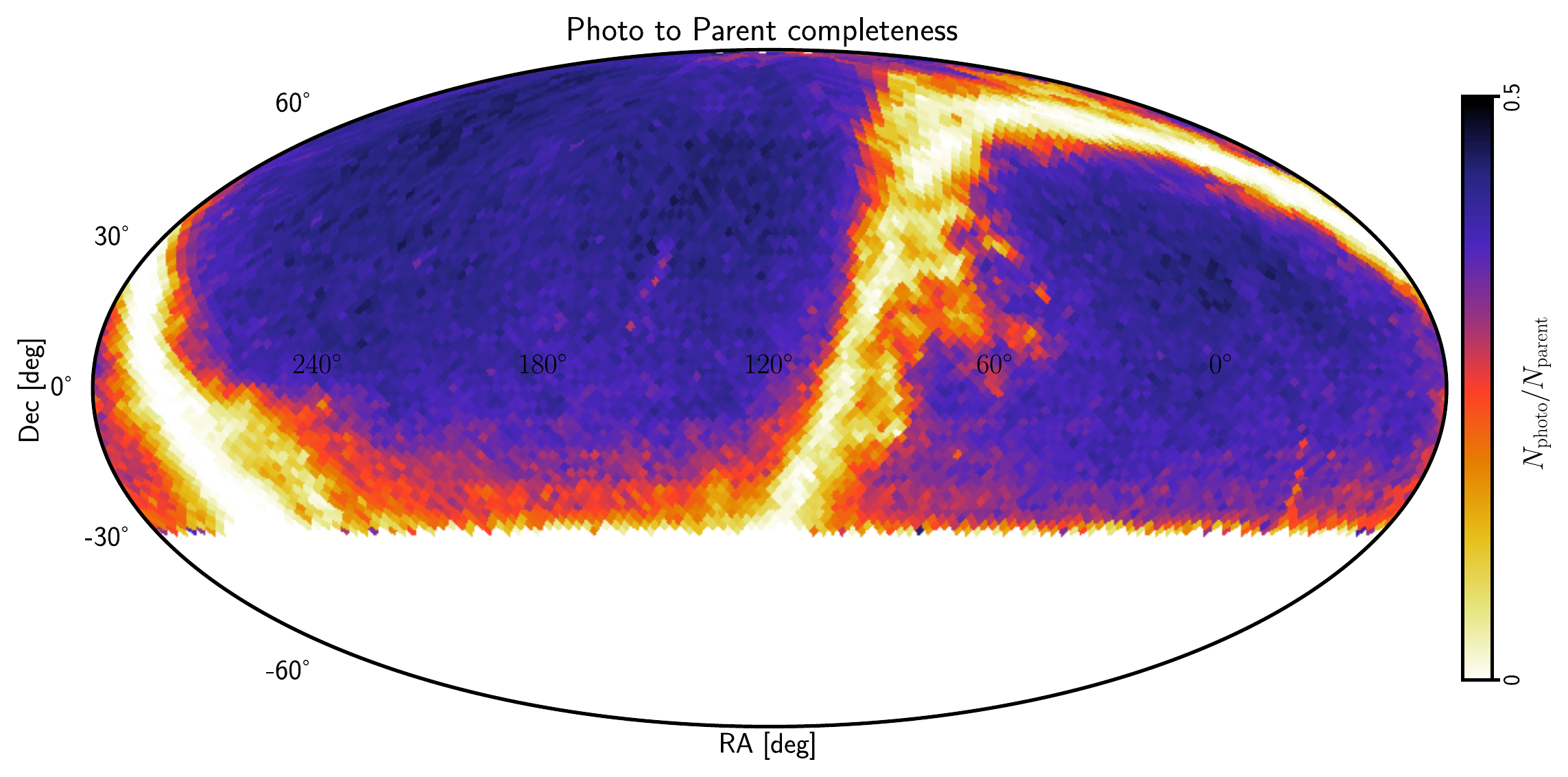}
        \includegraphics[width=0.49\textwidth]{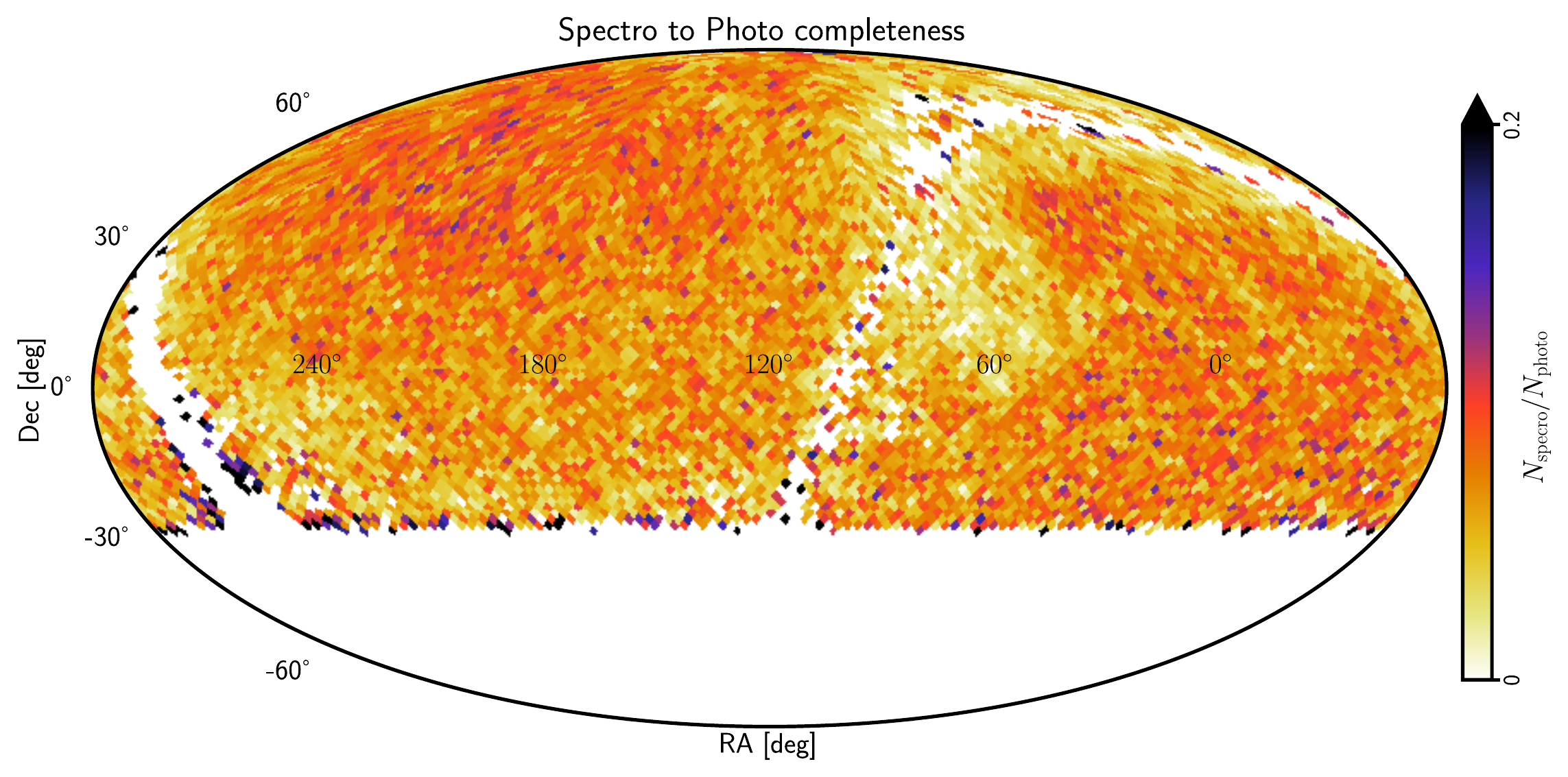}
        \caption{Angular distribution and completeness of the simulated ZTF \sn\ sample. \textit{Top}: Angular distribution of simulated ZTF type-Ia supernova from one mock realization of a \textit{6-year} program. 
        The parent sample of simulated \sns\ is shown in blue, those detected in photometry 
        are shown in green and those successfully typed
        with spectroscopy are shown in red.
        A map of stellar density from the Gaia satellite is shown
        in the background.
        \textit{Bottom}: Angular completeness of photometric (left) and spectroscopic samples (right).
        The photometric completeness is computed relative to the parent sample while 
        the spectroscopic one is relative to those detected in photometry, using the full redshift range of the simulation, that is $0 < z < 0.14$. 
        Different ranges are used for the color scales. 
        }
        \label{fig:angular_distribution}
    \end{figure*}

    Figure~\ref{fig:angular_distribution} shows the
    \sn\ angular distribution of a single mock realization of the 
    ZTF 6-year survey, before and after selection effects. 
    The parent sample (blue dots) uniformly covers the northern sky 
    at $\text{DEC} > -30$ deg, by construction.
    Photometry and spectroscopy preferentially select \sn\ events 
    outside the Galactic plane. 
    
    We estimated the sky coverage of each sample and defined 
    their completeness. 
    The sky coverage is calculated by assigning \sns\ of 27 mock 
    realizations to an angular mesh provided by the
    \textsc{healpix}\footnote{\url{http://healpix.sf.net}} 
    software \citep{gorskiHEALPixFrameworkHighResolution2005, 
    zoncaHealpyEqualArea2019}, which yields pixels of equal area. 
    Using 12,288 pixels ($n_\text{side}=32$), we estimated that the 
    parent sample covers uniformly
    an area of 31537.3~deg$^2$. The photometric and spectroscopic samples
    cover respectively 30698.0 and 28700.5~deg$^2$.
    Since photometric and spectroscopic samples are subsamples of the 
    parent sample, we can estimate their completeness
    by computing the ratio of the number of \sns\ in each sample to 
    the same number in the parent sample, in each angular pixel. 
    Bottom panels of Fig.~\ref{fig:angular_distribution}
    display two ratios: photometric to parent and spectroscopic to photometric. 
    While photometric completeness is nearly 40\% of the extra-Galactic sky, the spectroscopic completeness is 
    around 6 percent, mainly due to the magnitude cut imposed for follow-up. 
    Naturally these completeness values are dependent on the maximum 
    redshift of the simulation, that is $z=0.17$.
  
\begin{figure}
       \centering
       \includegraphics[width=\columnwidth]{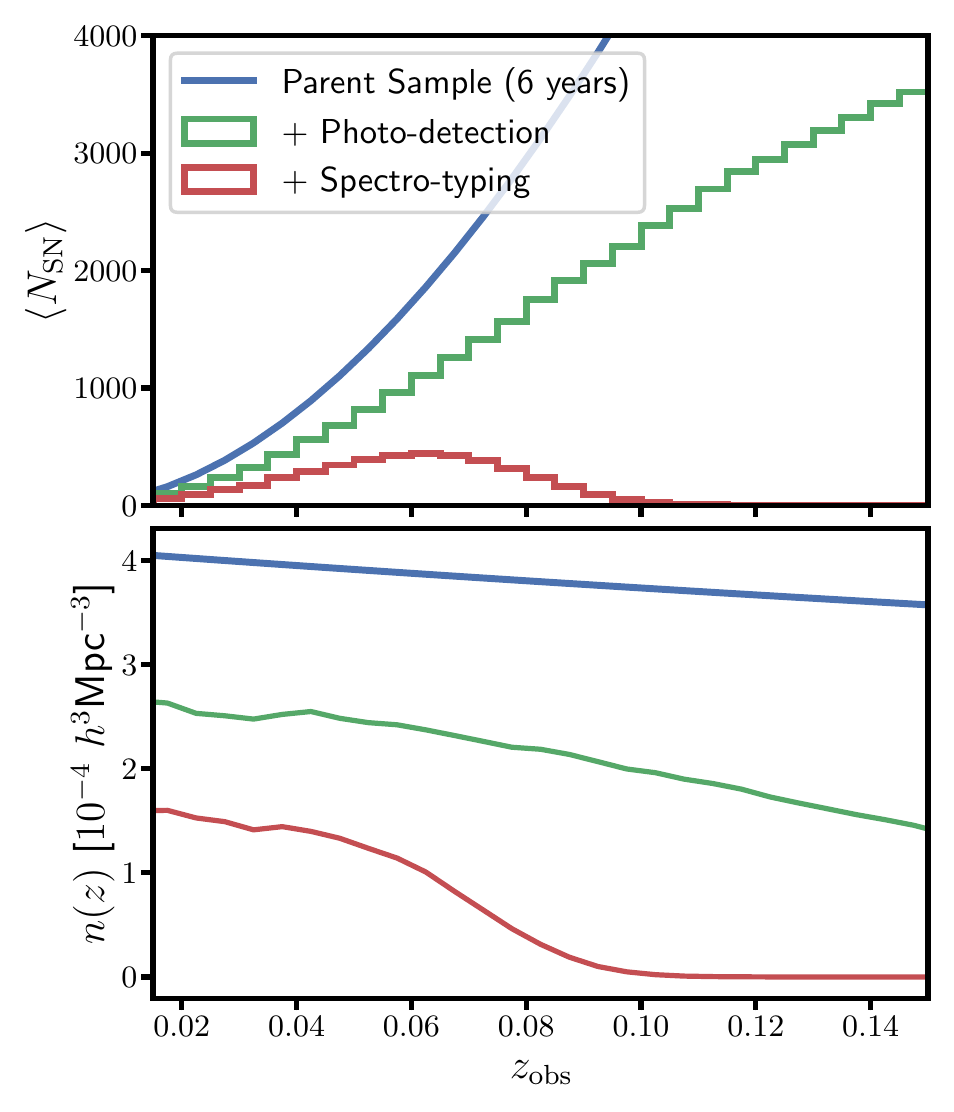}
       \caption{Number and density distributions of \sns\ with respect to redshift. The parent sample of simulated \sns\ is shown in blue, those detected 
       in photometry are shown in green and those successfully typed
       with spectroscopy are shown in red.\textit{Top panel}: Number of simulated ZTF type-Ia 
       supernova versus redshift per bins of  $\Delta z_{\rm obs} = 0.005$, 
       averaged over 27 independent mock realizations of a \textit{6-year} program. 
       \textit{Bottom panel}: Comoving number density of
       \sns\ versus redshift, calculated assuming the fiducial cosmology 
       from Table~\ref{tab:cosmo_params}. 
       }
       \label{fig:selection_effects}
    \end{figure}

    Figure~\ref{fig:selection_effects} displays the redshift distribution  
    of simulated \sns\ before and after selections caused by 
    photometric detection and spectroscopic follow-up.
    The top panel shows the absolute number of \sns\ per redshift bin 
    while the bottom panel shows the comoving number density $n(z)$ of \sns\ 
    as a function of redshift, for which 
    we see even more clearly the impact of the photometric detection and
    spectroscopic selection. 
    This density is the quantity of interest for clustering measurements.
    We computed these densities as the ratio of the number of \sns\ 
    in a redshift bin to its corresponding comoving volume. 
    The volume calculation assumes the fiducial cosmology from 
    Table~\ref{tab:cosmo_params}, in order to convert redshifts into distances,
    and the angular masks estimated above, shown in Fig.~\ref{fig:angular_distribution}.
    In Fig.~\ref{fig:selection_effects} we see that the rate of explosions 
    for the parent sample has a slight dependency on redshift due to 
    the time dilation factor $1/(1+z)$. 
    While the photometric sample extends to redshifts beyond $z=0.15$, 
    the density of the spectroscopic sample quickly drops beyond $z=0.06$.

    \subsection{Light-curve parameter adjustment and quality cuts}
    \label{sec:simulations:fits_and_quality_cuts}
    
    Each simulated \sn\ light-curve is fit 
    using the same framework (SALT2) that was used
    to generate them with the \textsc{sncosmo} package implementation. 
    For each light-curve, we fit for the stretch $x_1$, color $c$, 
    peak-magnitude $m_B$, and time of peak-brightness $t_0$.
    The redshift $z_\text{obs}$ of host galaxies is assumed to be provided by an external spectroscopic survey. We fixed it to its true value during the fit as we assume the error to be negligible.
    The covariance matrix $C_{{\rm SALT}, i}$ of the fit is defined by
    \begin{equation}
        C_{{\rm SALT}, i} = 
        \begin{pmatrix}
            \sigma^2_{m_B, i}      &  {\rm Cov}_{m_B x_1, i} & {\rm Cov}_{m_B c, i} \\
            {\rm Cov}_{m_B x_1, i} &   \sigma^2_{x_1, i}     & {\rm Cov}_{x_1 c, i} \\
            {\rm Cov}_{m_B c, i}   & {\rm Cov}_{x_1 c, i}    & \sigma^2_{c, i}
        \end{pmatrix},
        \label{eq:cov_salt}
    \end{equation}
    and will be used when fitting for standardization parameters in Sect.~\ref{sec:methodology:data_vector:hubble_diagram}.
     
    Some of our fits did not converge during the first attempt for at least two reasons.
    Firstly, some fits with excessive $\chi^2$ values correspond to light-curves with high S/N.
    To avoid this problem, we fit these light-curves by
    artificially increasing the flux uncertainties to help the minimizer to find the true minima. We then used these best-fit parameters as a first guess for a second iteration with the initial flux uncertainties.
    Secondly, some over-sampled light-curves show convergence problems, particularly when the oversampling occurs at the edges of the available phase range for the SALT2 model.
    When varying the peak-brightness time $t_0$ during the fit, it will also change the number of points considered in the fit. 
    Refitting those light-curves by discarding points $5$ days around the SALT2 model boundaries solved this issue.

    The final step in creating a full ZTF simulation of \sn\ light-curves
    is to apply quality cuts that are commonly done 
    for precision cosmological measurements. 
    We followed the procedure adopted on real data, as
    described in the first data release of the ZTF Type Ia Supernova survey \citep{dhawanZwickyTransientFacility2022}.
    Table~\ref{tab:cosmo_cuts} summarizes 
    the selections made on light-curves as well as
    the fraction of objects passing each criterion. 
    These fractions are obtained by averaging results from 
    27 independent mock realizations of the survey.
    The fraction of \sns\ with converged SALT2 fits is given in the 
    first row of Table~\ref{tab:cosmo_cuts}.
    We selected only best-fit models describing the data with 
    probability larger than 95 percent.
    In order to ensure a robust estimate of maximum brightness, 
    we selected light-curves containing at least three exposures 
    within 10 days of maximum brightness.
    We excluded any light-curve with best-fit stretch $|x_1|<3$ or
    color $|c|<0.3$.
    We also excluded light-curves for which the uncertainty in $t_0$
    or $x_1$ is larger than 1. 
    A final redshift cut $z_\text{obs}>0.02$ is meant to avoid supernovae velocities to be correlated within our local flow.
   
    \begin{table} 
    \centering
        \caption{Selection criteria applied to simulated ZTF type-Ia 
        supernova events to produce a cosmological sample.} 
        \label{tab:cosmo_cuts}

        \begin{tabular}{lcc}
            \hline 
            \hline \\[-2ex]
            \textbf{Cuts} & \textbf{Remains \%} & $\mathbf{\left<N \right>}$ \textbf{\sns} \\[1.ex]
            \hline \\[-1.8ex]
            SALT2 fit success &  88.7 & 3830 \\
            $P_{\rm fit} > 95 \%$ & 84.9 & 3664\\
            3 epochs with $|p| < 10 $ & 89.7 & 3873\\
            $|x_1| < 3$ & 89.5 & 3867  \\ 
            $|c| < 0.3$ &  88.8 & 3834  \\ 
            $\sigma_{t_0} < 1$ & 89.4 & 3862 \\
            $\sigma_{x_1} < 1$ & 89.3 & 3858  \\
            $z_{\rm obs} > 0.02$ & 97.9 & 4228\\[1.ex]
            \hline\\[-2ex]
            \textbf{All cuts} & {\bf 81.5} & {\bf 3520}\\
            \textbf{All cuts and $z < 0.06$} & {\bf 38.5} & {\bf 1660}
            \normalsize 
        \end{tabular} 
    \end{table} 

    After all cuts, the average number of SN Ia is 
    $\langle N\rangle \sim 3520$ for our 6-year sample.
    This number correspond to the spectroscopically 
    classified sample of \sns.

\section{Methodology}
\label{sec:methodology}

    In this section we introduce the methodology employed in this work to measure 
    the growth-rate of structures $\fsig$ from peculiar velocities 
    derived from a sample of standardized \sns. 
    We used the maximum-likelihood method which assumes that the peculiar velocity field is a Gaussian random field.

    The Gaussian likelihood is expressed as
    \begin{equation}
        \begin{split}    
         \mathcal{L}(\mathbf{p}, \mathbf{p}_{\rm HD}) =& (2 \pi)^{-\frac{n}{2}}|C(\mathbf{p}, \mathbf{p}_{\rm HD})|^{-\frac{1}{2}}  \\ &\times\exp\left[-\frac{1}{2}\mathbf{v}^T(\mathbf{p}_{\rm HD})C(\mathbf{p}, \mathbf{p}_{\rm HD})^{-1}\mathbf{v}(\mathbf{p}_{\rm HD})\right],
        \label{eq:likelihood}
        \end{split}
    \end{equation}
    where $\mathbf{v}$ is the data vector containing the sampled peculiar velocity field, $C(\mathbf{p}, \mathbf{p}_{\rm HD})$ is the covariance matrix describing correlations between velocities, $\mathbf{p}$ and $\mathbf{p}_{\rm HD}$ are vectors containing the parameters of the model. $\mathbf{p}_{\rm HD}$ refers to parameters of the Hubble diagram and \sn\ standardization and $\mathbf{p}$ refers to growth-rate related parameters including $\fsig$. We subsequently describe these parameters in detail in the following sections.

    Our methodology is based on works by 
    \citet{johnson6dFGalaxySurvey2014, howlett2MTFVIMeasuring2017, 
    adamsJointGrowthrateMeasurements2020, laiUsingPeculiarVelocity2023}, 
    who apply this methodology to samples of peculiar velocities derived 
    from Tully-Fisher and Fundamental Plane distances. 
    In this work we only applied this method to the velocity field, leaving
    the cross-correlation with a galaxy density field for future work.

    We start this section by describing the construction of the data vector
    $\mathbf{v}$ containing peculiar velocities (Sect.~\ref{sec:methodology:data_vector}), then the construction of the
    covariance matrix (Sect.~\ref{sec:methodology:covariance}).

    \subsection{The velocity data vector}
    \label{sec:methodology:data_vector}

    Peculiar velocities of \sn\ hosts can be extracted via 
    the residuals with respect to the Hubble diagram, 
    as illustrated in Figs.~\ref{fig:simple_sim} and \ref{fig:pecveleffect}.
    Here we present how we fit for the Hubble diagram parameters $\mathbf{p}_{\rm HD}$ .

    \subsubsection{Hubble diagram and \sn\ standardization}
\label{sec:methodology:data_vector:hubble_diagram}
    
\begin{figure}
        \centering
    \includegraphics[width=0.9\columnwidth]{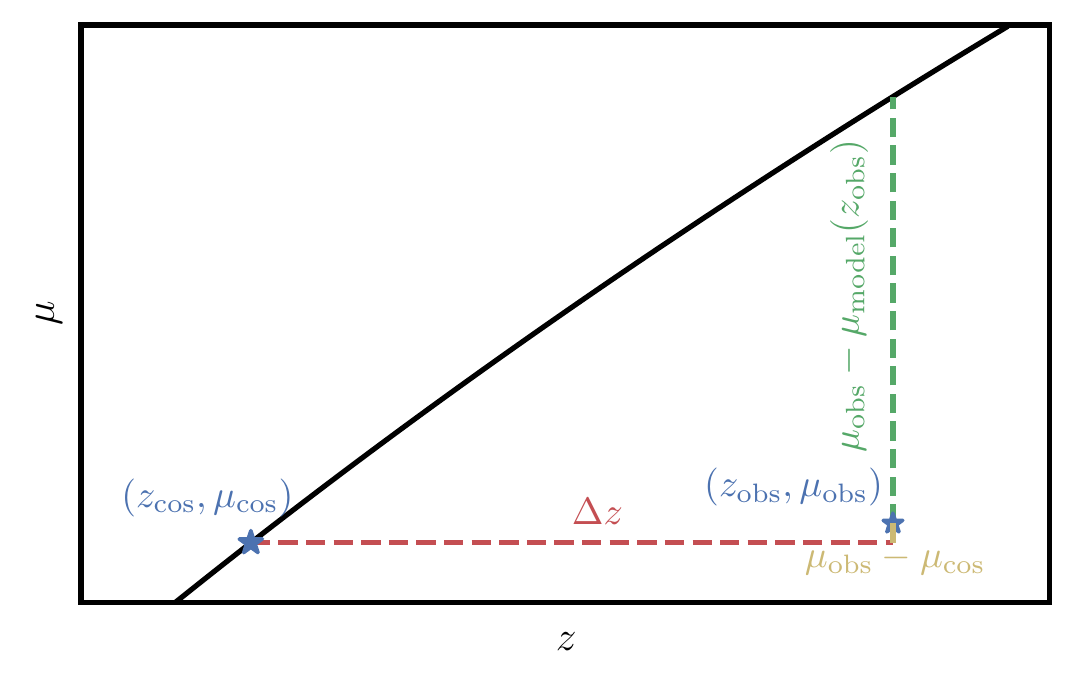}
        \caption{Representation of the different effects of peculiar velocities on the Hubble diagram for a single SN. The red dotted line shows the Doppler shift, the yellow line shows the relativistic beaming effect and the green dotted line shows the Hubble residual.}
        \label{fig:pecveleffect}
    \end{figure}

    Usually the fit of the Hubble diagram varies background cosmological parameters (e.g., $\Omega_m$) as well as standardization parameters ($\alpha$, $\beta$ and $M_0$). However, given the small span of redshift of ZTF, we cannot constrain $\Omega_m$. 
    We can either fix the Hubble diagram by assuming true input cosmological parameters of the simulation or assume low-redshift linear relation. 
    In this work we decided to fix the cosmology to simulation input, thus our Hubble diagram parameters are $\mathbf{p}_{\rm HD} = \{\alpha, \beta, M_0, \sigma_M\}$ and we only fit for the \sn\ standardization expressed by the Tripp relation:
    \begin{equation}
        \mu_{{\rm obs}, i}(\mathbf{p}_{\rm HD}) = m_{B, i} - (M_B - \alpha x_{1, i} + \beta c_i).
        \label{eq:tripp}
    \end{equation}
    As shown on Figs. \ref{fig:simple_sim}  and \ref{fig:pecveleffect}, the quantity of interest in order to derive peculiar 
    velocities from \sns\ is the Hubble diagram residuals which is the difference between the observed distance modulus and the model distance modulus evaluated at the observed redshift $z_{\rm obs}$. Hubble residuals are given by
    \begin{equation}
        \Delta \mu_i(\mathbf{p}_{\rm HD}) = \mu_{{\rm obs}, i}(\mathbf{p}_{\rm HD}) - \mu_\text{model}(z_{{\rm obs}, i}),
        \label{eq:dmu}
    \end{equation}
    and their uncertainties by
    \begin{equation}
        \sigma^2_{\mu, i}(\mathbf{p}_{\rm HD}) = \mathbf{A}_{i}(\mathbf{p}_{\rm HD})^T C_{{\rm SALT}, i} \mathbf{A}_{i}(\mathbf{p}_{\rm HD}) + \sigma_M^2,
    \end{equation}
    where the vector $A_i$ is given by
    \begin{equation}
        \mathbf{A}_i(\mathbf{p}_{\rm HD}) = 
        \begin{pmatrix}
            1\\
            \alpha \\
            - \beta
        \end{pmatrix},
    \end{equation}
    and the covariance $C_{{\rm SALT}, i}$ is written in Eq.~\ref{eq:cov_salt}. 
    
    \subsubsection{Peculiar velocities from Hubble residuals}
    \label{sec:methodology:data_vector:pecvel}

    From Eq. 15 of \cite{huiCorrelatedFluctuationsLuminosity2006}  we can show that a 
    first-order expansion of Hubble diagram residuals \eqref{eq:dmu} with respect to peculiar velocities gives the estimator (see Appendix \ref{ap:vestimator:derivation}):     
    \begin{equation}
        \hat{v}_i(\mathbf{p}_{\rm HD}) = - \frac{\ln(10) c}{5} \left(\frac{(1 + z_i)c}{H(z_i) r(z_i)} - 1 \right)^{-1} \Delta \mu_i(\mathbf{p}_{\rm HD}).
        \label{eq:vest}
    \end{equation}
    Although the derivation of the estimator gives that $z_i$ is evaluated as the cosmological redshift $z_i = z_{\mathrm{cos},i}$, as stated in \cite{huiCorrelatedFluctuationsLuminosity2006} replacing it by $z_\mathrm{obs}$ is only a second order approximation.
    Since we do not have access to cosmological redshifts, we used the observed redshift $z_i = z_{{\rm obs}, i}$ to estimate velocities. However, the estimator is valid in a regime where the peculiar redshift $z_p$ is small enough compared to the cosmological redshift $z_\mathrm{cos}$. This leads to small biases on velocity estimates, in particular for nearby galaxies with high velocities. We further discuss this point in Appendix \ref{ap:vestimator:diffest}.
    This estimator is used in \cite{johnson6dFGalaxySurvey2014} but
    throughout the literature other variants of this estimator including more approximations have been used. 
    In Appendix \ref{ap:vestimator:diffest}, we also compare performances of 
    different estimators. 
    We concluded that the bias is small for all of them but we choose to use (\ref{eq:vest}) since 
    it is the least biased. 
    However this estimator depends on the cosmological model used, and using cosmological parameters that differ from the true ones will result in a bias of the velocity estimator. 
    In Appendix \ref{ap:vestimator:estcosmo} we evaluate this bias as a function of $\Omega_m$ and we conclude that it is negligible.
    
    We compute the uncertainty on our velocity estimations as
    \begin{equation}
        \sigma_{\hat{v}, i}(\mathbf{p}_{\rm HD})=\frac{\ln(10)c}{5} \left(\frac{(1 + z_{{\rm obs}, i})c}{H(z_{{\rm obs}, i})r(z_{{\rm obs}, i})} - 1 \right)^{-1} \sigma_{\mu, i}(\mathbf{p}_{\rm HD}).
    \end{equation}
    
    In the low-redshift limit, this error grow quasi-linearly with redshift:
    \begin{equation}
        \sigma_{\hat{v}} \simeq 1400 \left( \frac{z}{0.1} \right)  \left( \frac{\sigma_\mu}{0.1} \right)\ {\rm km} \ {\rm s}^{-1}.
        \label{eq:verrorlin}
    \end{equation}
    All biases that we underline in the previous paragraph are below $\sim 5\%$ for a typical velocity of $v\sim 300$~km.s$^{-1}$. 
       
    \subsection{The covariance matrix}
    \label{sec:methodology:covariance}

    The modeling of the statistical properties of our sample of 
    peculiar velocities is made through the covariance matrix 
    $C(\mathbf{p}, \mathbf{p}_{\rm HD})$ of the Gaussian likelihood from Eq.~\ref{eq:likelihood}. 
    This covariance can be decomposed in two parts: an analytical part, 
    which depends on a theoretical modeling of the large-scale cosmological
    correlations of velocities, and an observational part which accounts
    for observational uncertainties. 

    \subsubsection{Modeling large-scale cosmological correlations}
    
    The analytical part of the covariance $C^{vv}$ will depend on 
    the growth-rate parameter $\fsig$, and other nuisance parameters, 
    which will be varied altogether when maximizing the likelihood.
    Each element of the covariance matrix $C_{ij}^{vv} \equiv \langle v_i v_j^* \rangle$ is defined as the correlation function of 
    the radial velocity field at two positions $\mathbf{r}_i$ and $\mathbf{r}_j$. 
    This correlation can be written as 
    an inverse Fourier transform of the velocity-velocity correlations in 
    Fourier space. The radial component $v_i$ of the three-dimensional velocity field $\mathbf{v}(\mathbf{r})$ at a position $\mathbf{r}_i$ 
    can be written as:
    \begin{equation}
        v_i = \hat{\mathbf{r}}_i \cdot \mathbf{v}(\mathbf{r}_i)
        = \int \frac{\text{d}^3k}{(2\pi)^3} e^{i \mathbf{k}\cdot \mathbf{r}_i} 
            \hat{\mathbf{r}}_i \cdot \mathbf{v}(\mathbf{k}).
    \end{equation}
    The covariance is therefore:
    \begin{multline}
        C^{vv}_{ij} = \langle v_i v_j^* \rangle \\
         = \iint \frac{\text{d}^3k}{(2\pi)^3} \frac{\text{d}^3k'}{(2\pi)^3} 
         e^{i (\mathbf{k}\cdot \mathbf{r}_i - \mathbf{k}'\cdot \mathbf{r}_j)}
         \Bigl\langle (\hat{\mathbf{r}}_i \cdot \mathbf{v}  (\mathbf{k} ) )
                      (\hat{\mathbf{r}}_j \cdot \mathbf{v}^*(\mathbf{k}') )
         \Bigr\rangle.
         \label{eq:covariance_0}
    \end{multline}

    We can write the expression (\ref{eq:covariance_0}) using the velocity-divergence $\theta(\mathbf{r})$, a scalar field defined as 
    \begin{equation}
        \mathbf{\nabla} \cdot \mathbf{v}(\mathbf{r}, a) \equiv   - aH(a) f(a) \theta(\mathbf{r}, a)
        \label{eq:velocity_divergence}
    \end{equation}
    where $a=1/(1+z)$ is the scale factor, $H(a)$ is the Hubble 
    rate and $f(a)$ is the growth-rate. 

    In linear theory, valid on large scales, it is common to 
    consider that the velocity field is irrotational, in which case we can write in Fourier space:
    \begin{equation}
        \mathbf{v}(\mathbf{k}, a) = -i a f(a) H(a) \theta(\mathbf{k}, a) \frac{\hat{\mathbf{k}}}{k}.
    \end{equation}

    Defining $\mu \equiv \hat{\mathbf{r}}\cdot \hat{\mathbf{k}}$
    and the velocity-divergence auto power spectrum as $\langle \theta(\mathbf{k}) \theta^*(\mathbf{k}') \rangle \equiv (2\pi)^3 \delta_D(\mathbf{k}-\mathbf{k}')P_{\theta \theta}(k)$, we
    can simplify 
    Eq.~\ref{eq:covariance_0} to
    \begin{equation}
        C^{vv}_{ij} 
         = \int \frac{\text{d}^3k}{(2\pi)^3} 
         e^{i \mathbf{k}\cdot (\mathbf{r}_i -\mathbf{r}_j)} (af(a)H(a))^2 \frac{\mu_i \mu_j}{k^2} P_{\theta \theta}(k).
         \label{eq:covariance_1}
    \end{equation}

    The resulting formula has been computed analytically and 
    used in previous work 
    \citep{abate_peculiar_2008,
    johnson6dFGalaxySurvey2014, 
    howlett2MTFVIMeasuring2017}. Eq.~\ref{eq:covariance_1} can be written as
    \begin{equation}
        C_{ij}^{vv} = \frac{\left(a H f\right)^2}{2 \pi^2} \int_0^{+\infty} P_{\theta\theta}(k) W_{ij}(k; \mathbf{r}_i, \mathbf{r}_j)dk.
        \label{eq:covariance_2}
    \end{equation}
    The window function $W_{ij}$ is given in 
    \cite{maPeculiarVelocityField2011} (Appendix A) as
    \begin{align}
        W_{ij}(k; \mathbf{r}_i, \mathbf{r}_j) &=\int \frac{d\Omega_k}{4 \pi} e^{i\mathbf{k}(\mathbf{r}_i - \mathbf{r}_j)}\mu_i \mu_j \\
        \begin{split}
            &= \frac{1}{3} \left[ j_0\left(k r_{ij}\right) 
                             - 2j_2\left(k r_{ij}\right) 
                         \right] \cos (\alpha_{ij}) \\
            & \ \ + \frac{1}{r_{ij}^2} j_2\left(k r_{ij}\right)r_i r_j \sin^2(\alpha_{ij}),
        \end{split}
    \end{align}
    where $\alpha_{ij}$ is the angle between $\hat{\mathbf{r}}_i$ and $\hat{\mathbf{r}}_j$, 
    $r_{ij} \equiv |\mathbf{r}_j - \mathbf{r}_j|$,
    and $j_0(x)$ and $j_2(x)$ are the zeroth and second order 
    spherical Bessel functions.

    Since $P_{\theta\theta}$ is computed using a fiducial $(\sigma_8)_{\rm fid}$ we normalize it. To be more explicit we also introduce $f_{\rm fid}$:
    \begin{equation}
        C_{ij}^{vv} = \frac{(a H)^2}{2\pi^2}\frac{(\fsig)^2}{(\fsig)^2_{\rm fid}} \int_0^{+\infty} f_{\rm fid}^2P_{\theta\theta}(k) W_{ij}(k; \mathbf{r}_i, \mathbf{r}_j)dk.
        \label{eq:covariance_3}
    \end{equation}
    
    We account for the impact of using positions in redshift space, which are themselves affected by peculiar velocities, by 
    including the following empirical damping function based on N-body simulations \citep{kodaArePeculiarVelocity2014}:
    \begin{equation}
        D_u(k) = \frac{\sin\left(k\sigma_u\right)}{k\sigma_u},
    \end{equation}
    where $\sigma_u \sim 15$\hmpc\ can be fit as a free parameter.
    Equation~\ref{eq:covariance_3} then becomes at $z=0$:
    \begin{equation}
        C_{ij}^{vv} = \frac{H_0^2}{2\pi^2}\frac{(\fsig)^2}{(\fsig)^2_{\rm fid}} \int_0^{+\infty} f_{\rm fid}^2P_{\theta\theta}(k)D_u^2(k) W_{ij}(k; \mathbf{r}_i, \mathbf{r}_j) {\rm d}k.
        \label{eq:covariance_sigmau}
    \end{equation}
    
    \citet{damExploringRedshiftspacePeculiar2021} explore
    analytically the impact of redshift-space distortions on the velocity-divergence
    power spectrum and we leave its implementation for future work.

    \subsubsection{Numerical considerations}
    \label{sec:methodology:covariance:numerical}

    Since the integrals in Eq.~\ref{eq:covariance_sigmau} are computed 
    numerically in practice, we need to impose 
    integration limits $k_\text{min}$ and $k_\text{max}$. The lower integration bound is imposed by the N-body simulation size as $k_{\rm min} = 2\pi/L = 2.1\times 10^{-3}$~\ihmpc. We chose the value of $k_{\rm max}$ such that the integrals converge for any
    configuration of pairs of velocity tracers.
    Previous works \citep{johnson6dFGalaxySurvey2014, howlett2MTFVIMeasuring2017} have chosen low values for $k_\text{max}$ (0.1 or 0.2~\ihmpc) in order 
    to avoid the nonlinear clustering on small scales. We observed that for such $k_{\rm max}$ values, the power spectrum integral does not fully converge.This has to be mitigated with the presence of the damping term $D_u$ that will reduce the variation of convergence due to the choice of $k_{\rm max}$. Therefore, we took 
    $k_{\rm max} = 1$~\ihmpc as the higher integration bound. 
    In Appendix \ref{ap:pwconv} we check that integrals involving the power
    spectrum have correctly converged for all $\sigma_u$ values.
  
    The covariance matrix $C^{vv}_{ij}$ has $N^2$ coefficient where $N$ is
    the number of peculiar velocity measurements. This matrix can become 
    prohibitively large if we want to invert it multiple times for each 
    evaluation of the likelihood. While the dependency with $\fsig$ can be
    factored out of the covariance, that is not the case for the $\sigma_u$ 
    parameter for which the full covariance has to be recomputed. Similarly to \cite{howlett_2mtf_2017}, we precomputed matrices with $\sigma_u \in [0, 50]$ \hmpc, using $\Delta \sigma_u = 0.02$ \hmpc. We then interpolated each matrix coefficient as a function of $\sigma_u$.

    When the number of halos at a given redshift range is too small, 
    more than one \sn\ can be associated to the same halo. 
    Since these \sns\ then have exactly the same position, 
    the covariance matrix becomes noninvertible. 
    We note that this happens rarely, at most for four pairs in our mocks.
    Thus, we decided to consider them as one unique data point with an 
    averaged velocity:
    \begin{equation}
        \hat{v}_\mathrm{eff} = \frac{\sum \hat{v}_i w_i}{\sum w_i},
    \end{equation}
    where the weights $w_i$ are $w_i = (\sigma_{\hat{v},i})^{-2}$. 
    
    \subsubsection{The velocity-divergence power spectrum}
    \label{sec:methodology:covariance:velocity_power_spectrum}

    To compute the cosmological part of the covariance matrix 
    (Eq.~\ref{eq:covariance_sigmau}) we need a model for the 
    velocity-divergence power spectrum $P_{\theta\theta}(k)$.
    In this work, we tested three different choices detailed below:
    the linear theory, an empirical model based on simulations and
    a perturbation theory model. 
    
    In linear theory, the continuity equation states that the velocity-divergence field $\theta$ is equal to the matter overdensity field $\delta$. 
    Therefore, the velocity divergence power spectrum is the same as the density power spectrum 
    $P^{\rm lin}_{\theta\theta}(k) = P^{\rm lin}_{\delta\delta}(k)$.
    The linear matter power spectrum $P^\text{lin}_{\delta\delta}(k)$ 
    is computed using the Boltzmann solver 
    \textsc{camb}\footnote{\url{https://camb.info}} \citep{lewisEfficientComputationCosmic2000} 
    using the cosmological parameters from Table \ref{tab:cosmo_params}. 
   
    It is well known that linear theory fails to describe the density field on small scales, typically for $k > 0.1$\ihmpc.
    \cite{bel_accurate_2019} constructed an empirical model for the nonlinear $P_{\theta\theta}(k)$ using the following parametrization:
    \begin{equation}
        P_{\theta\theta}^\textrm{non-lin}(k) = P_{\theta\theta}^\textrm{lin}(k) \exp\left[-k\left(a_1+a_2 k + a_3 k^2\right)\right],
    \end{equation}
    where the coefficients $a_i$ depend on $\sigma_8$ and are obtained from a fit to an N-body simulation:
    \begin{equation}
        \left\{\begin{matrix}a_1 &=& -0.817 + 3.198 \sigma_8 \\
               a_2 &=& 0.877 - 4.191\sigma_8 \\
               a_3 &=& -1.199 + 4.629\sigma_8\end{matrix}\right. .
    \end{equation}
    we checked that changing this $\sigma_8$ has negligible effect.
    
    Alternatively, we can use a model based on 
    regularized perturbation theory (RegPT, \citealt{taruya_regpt_2012})
    computed up to 2-loop expansion.
    We used a publicly available implementation of the RegPT\footnote{\url{https://github.com/adematti/pyregpt}}. 
    
   \begin{figure}
       \centering
       \includegraphics[width=\columnwidth]{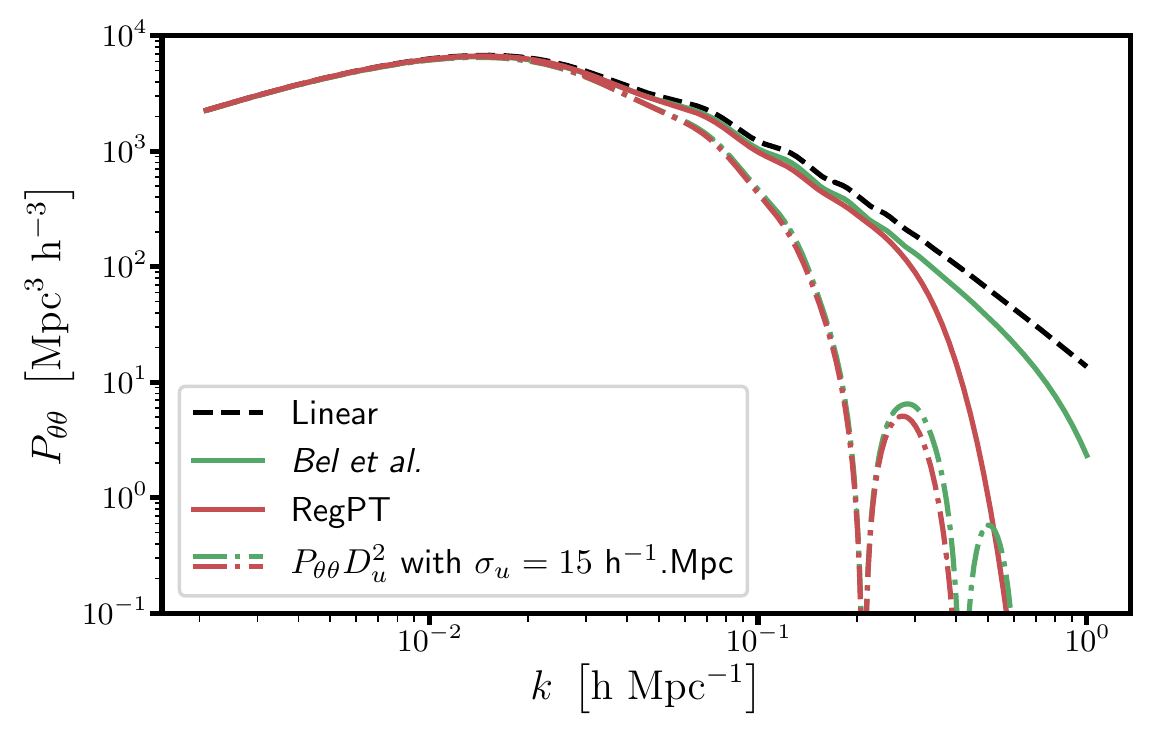}
       \caption{Models for the isotropic velocity divergence power spectrum of matter at $z=0$: linear theory (black dashed), an empirical model based on fits to N-body simulations from \citet{bel_accurate_2019} (green), and 
       a beyond first-order perturbation theory model from \citet{taruya_regpt_2012} (red). For these two models, we show the power spectrum with a damping function using $\sigma_u = 15$ \hmpc\ (dashed lines). Nonlinearities become important at $k>0.1$\ihmpc.}
       \label{fig:power_spectra}
   \end{figure}
   
   Figure~\ref{fig:power_spectra} compares the different models for 
   $P_{\theta\theta}(k)$ used in this work: the linear one and the two nonlinear models. 
   As expected, most differences are seen on small scales, when $k>0.1$\,\ihmpc.
   In Sect.~\ref{sec:robtests} we study the impact of the choice of model for the measurement of $\fsig$ from peculiar velocity data.

    \subsubsection{Including observational uncertainties}
    \label{sec:methodology:covariance:obs_uncertainties}

    The covariance matrix $C(\mathbf{p},\mathbf{p}_{\rm HD})$ has to also take into account random motions on very small scales and observational uncertainties. The random motions are modeled by 
    a diagonal term of velocity dispersion $\sigma_v$. We also assumed that 
    uncertainties on estimated velocities $\sigma_{\hat{v}, i}$ are 
    uncorrelated. 
    We checked in Appendix \ref{ap:vestimator:estgauss} that our 
    estimated velocities follow a Gaussian distribution.
    The parameter vector is $\mathbf{p} = \{\fsig, \sigma_v, \sigma_u  \}$ and the expression of the covariance matrix is
    \begin{equation}
         C_{ij}(\mathbf{p},\mathbf{p}_{\rm HD}) 
         = C_{ij}^{vv}(f\sigma_8, \sigma_u)  + \left[ \sigma_v^2 + \sigma_{\hat{v},i}^2(\mathbf{p}_{\rm HD})\right] \delta^K_{ij},
    \end{equation}
    where $C_{ij}^{vv}(\fsig, \sigma_u)$ is given by 
    Eq.~\ref{eq:covariance_sigmau} and $\delta^K_{ij}$ is the Kronecker delta.

    \subsection{Likelihood exploration}
    \label{sec:methodology:likelihood_exploration}

    Using the data vector of peculiar velocities from Sect.~\ref{sec:methodology:data_vector} and  the covariance matrix from Sect.~\ref{sec:methodology:covariance}, 
    we proceeded to explore the likelihood (Eq.~\ref{eq:likelihood})
    in order to constrain the growth-rate $\fsig$, the
    velocity dispersion $\sigma_v$ and the damping term $\sigma_u$. 
    
    We used the gradient descent algorithm  \textsc{iminuit}\footnote{\url{https://iminuit.readthedocs.io/}} \citep{iminuit} to 
    find the maximum likelihood. 
    Errors are given as symmetric errors computed with \textsc{Hesse} function or asymmetric errors using the \textsc{Minos} function. 
    In order to check these uncertainties evaluations and more generally the 
    likelihood profile, we used a Markov Chain Monte-Carlo (MCMC) algorithm implemented in the \textsc{emcee}\footnote{\url{https://emcee.readthedocs.io/}} package.
    We fund excellent agreement between both evaluations so we mostly used the
    faster maximization by \textsc{iminuit}, unless stated otherwise. 

\section{Results}
\label{sec:results}
    In this section, we present measurements of $\fsig$ from our simulated sets of ZTF \sns. As described in Sect. \ref{sec:simulations:ztf_selection_effects}, the ZTF observation strategy, particularly the spectroscopic follow-up of transients for classification, 
    introduces strong selection effects, which can lead to 
    biases in peculiar velocities and therefore on our estimates 
    of $\fsig$. We defined a sample with limited selection effects 
    and showed that we obtain unbiased results on $\fsig$. 
    To help identify effects on the $\fsig$ fit we performed it in three different configurations, with increasing complexity: in the first, we fit $\mathbf{p}$ parameters using the true input velocities (i.e., no Hubble diagram fit); in the second we fit $\mathbf{p}$ parameters using estimated velocities but still fixing $\mathbf{p}_{\rm HD}$ to input values; finally we fit $\mathbf{p}$ and $\mathbf{p}_{\rm HD}$ parameters simultaneously. This last configuration is our baseline choice.
    We detail our findings in the following.
    
    \subsection{Selection effects on Hubble residuals and velocities}
    \label{sec:results:effects_selection}    

    \begin{figure}
        Effect of the selection bias versus redshift.
        \centering
        \includegraphics[width=\columnwidth]{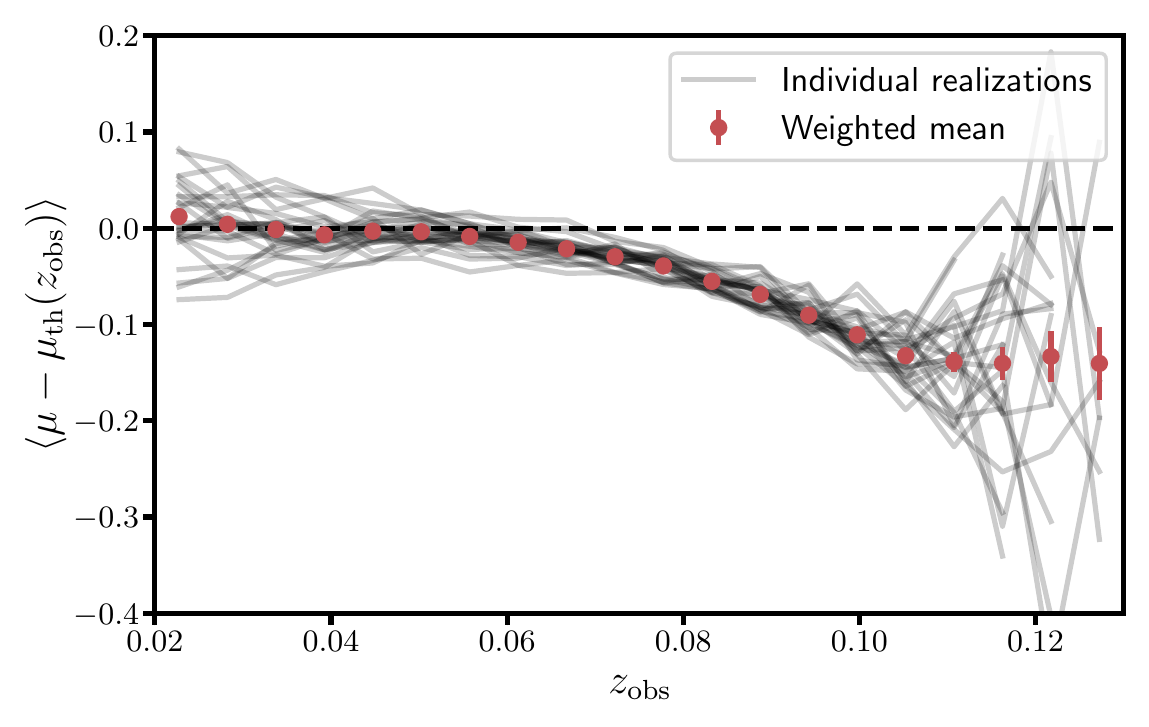}
        \includegraphics[width=\columnwidth]{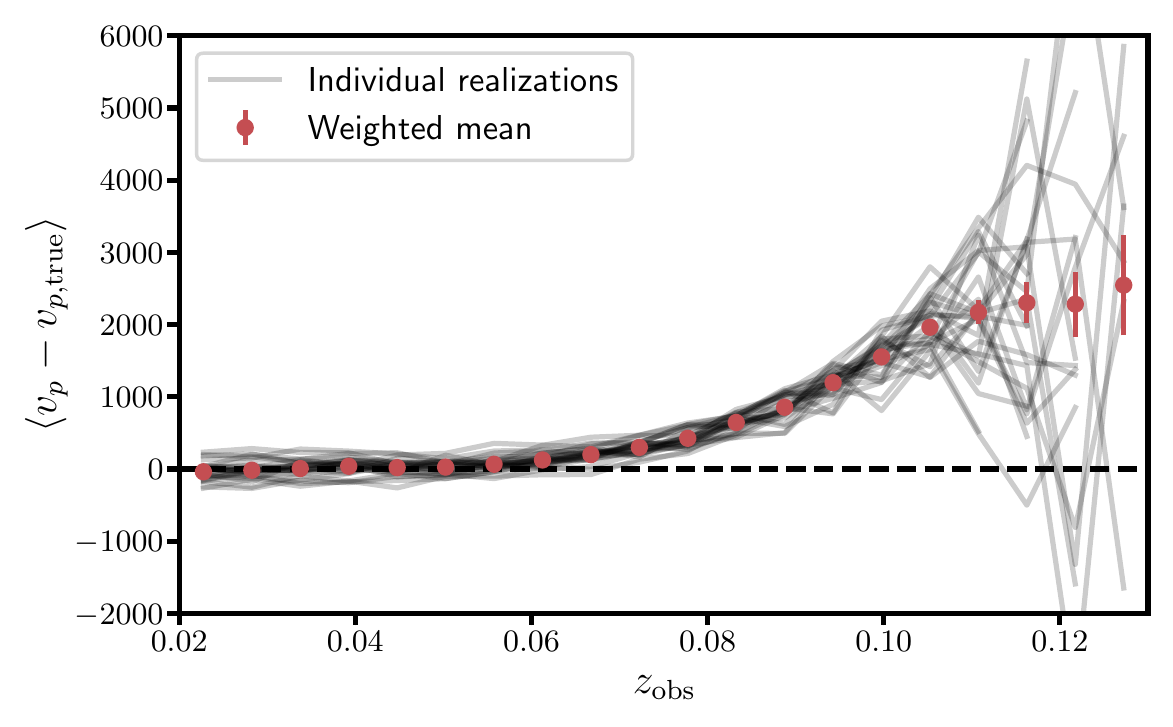}
        \caption{Hubble residuals and estimated velocities versus redshift.\textit{Top panel} : Hubble residuals of the 27 mocks. The gray lines represent each mock, red points are the weighted means taken within each redshift bin. \textit{Bottom panel}: same for the estimated peculiar velocities.}
        \label{fig:HDres}
    \end{figure}

    We started by assuming the true Tripp relation (Eq.~\ref{eq:tripp}) used
    to build our simulation (i.e., fixing the $\mathbf{p}_{\rm HD}$ to value of Table \ref{tab:snstandpar}) in order 
    to standardize the simulated \sns\ and obtain distance moduli $\mu$. 
    In other words, we do not perform a fit for these values just yet as we would do in the real analysis (see next section). 
    With this procedure we can disentangle different sources of 
    biases to the final analysis. 
    
    Figure~\ref{fig:HDres} shows the residuals to the Hubble diagram 
    and how these residuals translate to biases in the estimated
    velocities. 
    The analysis was performed on our 27 mock realizations. 
    We can see that at redshifts above $z \sim 0.06$, 
    the Hubble diagram residuals become increasingly biased, reaching $\Delta_\mu \simeq -0.13$ at $z \sim 0.12$.  This is a manifestation of a selection effect (or Malmquist bias), mainly caused by the spectroscopic follow-up for typing the 
    transients (see Sect.~\ref{sec:simulations:ztf_selection_effects}). 
    We converted Hubble residuals relative to the true input 
    Hubble diagram into peculiar velocities using Eq.~\ref{eq:vest}. 
    The bottom panel of Fig.~\ref{fig:HDres} displays the 
    comparison between estimated velocities and the true input radial
    velocities of the simulation. 
    We can see that the selection bias 
    simply translates to a fake outflow above 
    $z_{\rm obs}\sim 0.06$. However, the true peculiar velocity distribution itself is not biased by these selection effects. As we said previously peculiar velocities have only a negligible effect on \sn\ magnitude, thus they are not affected by the sample bias that is mostly a magnitude cut.
    
    We can see in Fig. \ref{fig:HDres} that at $z_{\rm obs} = 0.02$ the mean residuals seem to have a small positive bias. This effect is due to the fact that a sharp cut in $z_{obs}$ leads to an asymmetric cut in velocity space: there are more hosts with higher $z_{\rm cos}$ and negative velocities that contaminate the redshift bin than lower redshift hosts with positive velocities. We checked that replacing $z_{\rm obs}$ by $z_{\rm cos}$ removes this positive bias. Nonetheless, this bias at the low-redshift end does not impact our growth rate measurements (see next section).

    \subsection{Growth-rate measurement forecast from the complete sample ($z<0.06$)}   
    \label{sec:results:forecast006}
  
\begin{figure}
            \centering\includegraphics[width=\columnwidth]{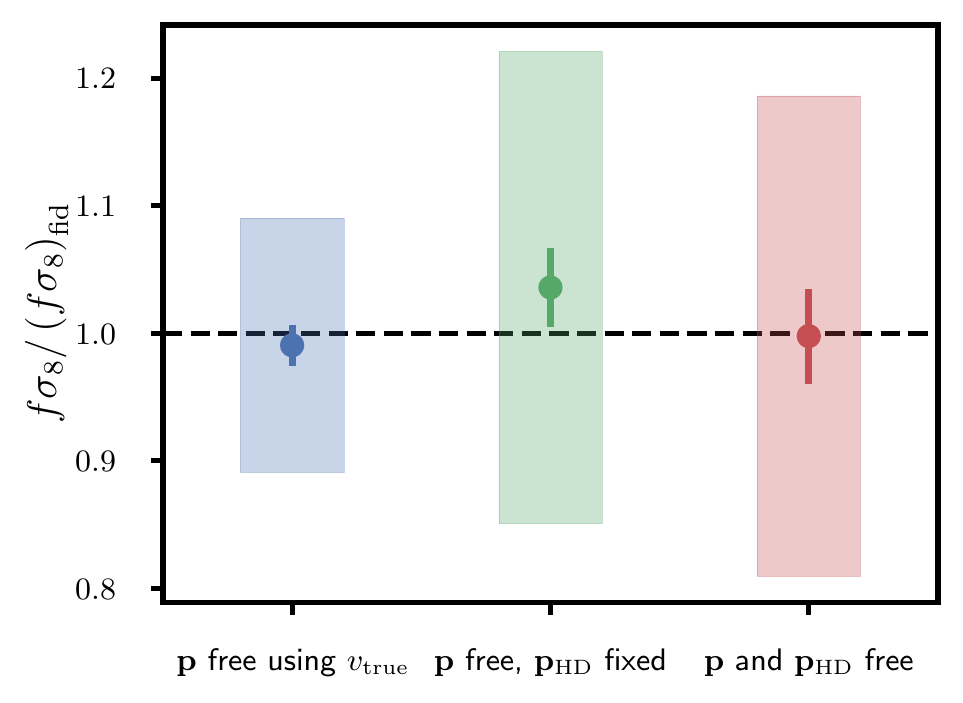}
            \caption{Best fit $\fsig$ for the ``complete sample'' within a redshift range of $z_{\rm obs} \in [0.02, 0.06]$. The points with errorbars show the mean obtained with our 27 mocks, the colored boxes show the averages of uncertainties.}
            \label{fig:fs8z006}
        \end{figure}

        Since Hubble residuals, and hence velocities, become strongly biased with increasing redshift, we decided to cut our sample at $z_{\rm obs} = 0.06$ where the sample bias remains below $\sim -0.01$ mag on $\mu$. With this cut we are left with $\langle N \rangle \sim 1660$ \sns\ at redshift $z < 0.06$. We performed the measurement of the growth rate $\fsig$ for the three types of analyses mentioned earlier. Results are summarized in Fig.~\ref{fig:fs8z006}  and commented below.  
        
        We first fit parameters $\mathbf{p}$ using the true peculiar velocities from the simulation. The point on the \textit{left} in Fig. \ref{fig:fs8z006} shows the 
        estimation of $\fsig$ from these fits for 27 mock realizations. 
        We obtained $\left<\fsig / (\fsig)_{\rm fid}\right> = 0.991 \pm 0.016$ with an averaged uncertainty\footnote{The averaged uncertainty on $\fsig$ is computed on the 27 mocks as $\sqrt{\left<\sigma_{\fsig}^2\right>}/ (\fsig)_{\rm fid}$} of $0.100$. This nonbiased result show that the model of covariance is a good description of our data. This result also set the minimum error that we can achieve with our sample if we access a perfect measurement of each velocity.
        
        We then performed our fit using estimated velocities but fixing $\mathbf{p}_{\rm HD}$ to input values. With these noisy velocity measurements, we observed that we do not have enough constraining power to measure $\sigma_u$. Due to positive degeneracy between the high-value of $\sigma_u$ and $\fsig$ we then over-estimated $\sigma_u$ and $f\sigma_8$. 
        We note here that the large values of $\sigma_u$ much above the scale of redshift-space distortions are not physical.
        To overcome this problem, we imposed a Gaussian prior on $\sigma_u$. This prior is centered on the value of $\mu_{p(\sigma_u)} = 15$\hmpc\ with a scale of $\sigma_{p(\sigma_u)} = 50$ percent. We chose this prior value near what has been observed in \cite{kodaArePeculiarVelocity2014}. This is also very close to what we obtain in our fit using true velocities ($\left<\sigma_u  \right> = 14.6\pm0.5$ \hmpc)
        In Sect.~\ref{sec:results:psigu} we further discuss the impact of this choice of prior. We recover $\left<\fsig / (\fsig)_{\rm fid}\right> = 1.036 \pm 0.031$ with a bias smaller than $2 \sigma$.  However this bias is undetectable for one single mock realization since the average uncertainty is $0.185$.
        This is shown in the \textit{middle} of Fig.~\ref{fig:fs8z006}.
        
        We then proceeded to a global fit letting all parameters free ($\mathbf{p}$ and $\mathbf{p}_{\rm HD}$) as one would ideally do in the analysis. Using the 6-year dataset, we verified that the $\fsig$ value is not biased, as can be seen in Fig.~\ref{fig:fs8z006}.
        \begin{figure}
            \centering        
            \includegraphics[width=\columnwidth]{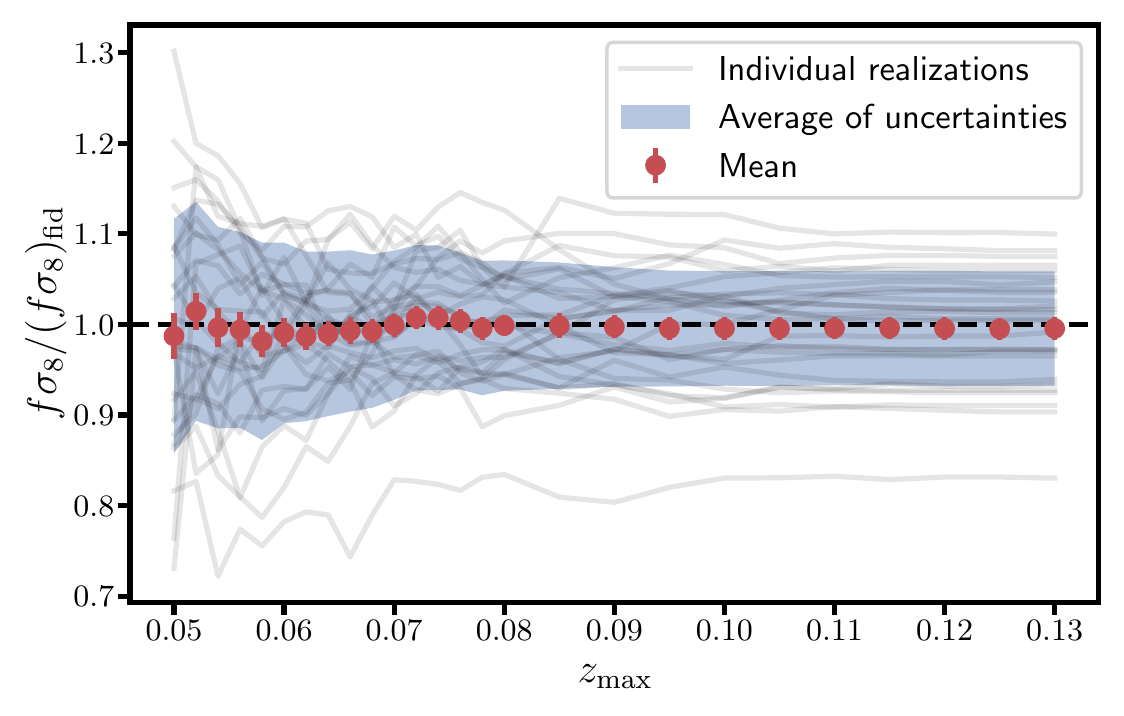}
            \includegraphics[width=\columnwidth]{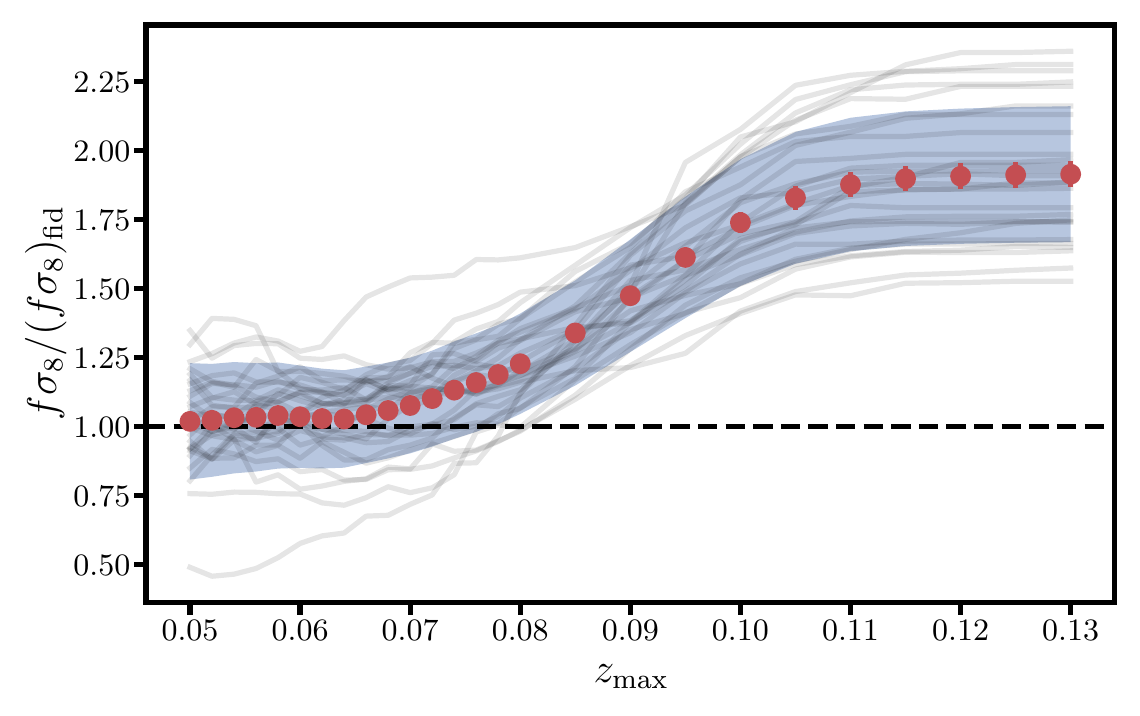}
            \includegraphics[width=\columnwidth]{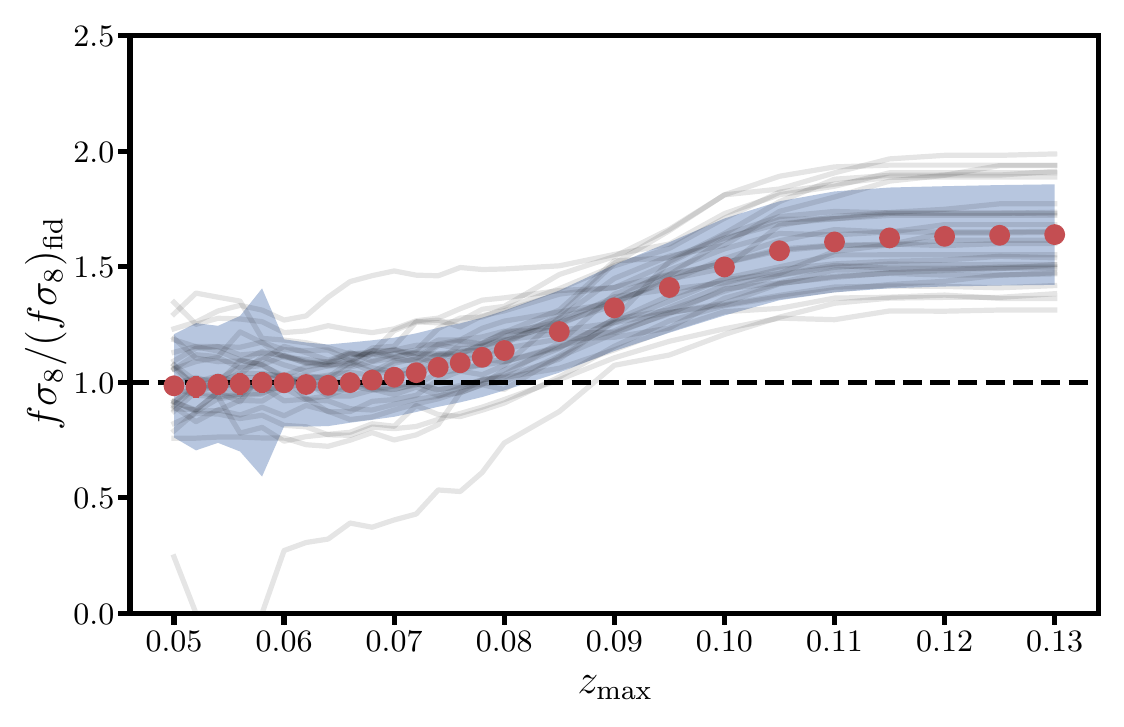}
    
            \caption{$\fsig$ fit results versus redshift cut upper bound $z_{\rm max}$.
            \textit{Top panel} : fit results using $v_{\rm true}$. \textit{Mid panel} : fit results using $\hat{v}$ with  $\mathbf{p}$ free and  $\mathbf{p}_{\rm HD}$ fixed. \textit{Bottom panel} : fit results using $\hat{v}$ with $\mathbf{p}$ and $\mathbf{p}_{\rm HD}$ free.}
            \label{fig:fs8bias}
        \end{figure}
        However, on the \textit{bottom panel} of Fig.~\ref{fig:fs8bias} we can notice that one of our 27 mocks has low fit values of $\fsig$ and that its fit became unstable for the smallest redshift range leading to null value of $\fsig$ with extremely large errors. We checked that this mock does not give abnormal values of $\fsig$ for the two previous fits (using true velocities and fixing $\mathbf{p}_\mathrm{HD}$), and that removing it from our sample does not affect significantly our results. Since its minimum is reported as valid by \textsc{Minuit} we kept it in our main results. We obtained
        \begin{ceqn}
        \begin{equation}             \left<\fsig / (\fsig)_{\rm fid}\right> = 0.998 \pm  0.037,
        \label{eq:fs8_baseline_value}
        \end{equation}
        \end{ceqn}
       and an average uncertainty of
        \begin{ceqn}
        \begin{equation}             \sqrt{\left<\sigma_{\fsig}^2\right>} / (\fsig)_{\rm fid} =0.188.
        \label{eq:fs8_baseline_error}
        \end{equation}
        \end{ceqn}

        \begin{table}
        \centering
        \caption{Results obtained for parameters $p \in {\mathbf{p}, \mathbf{p}_{\rm HD}}$ on our 27 realizations of ZTF 6-year \sn\ survey when considering the redshift range $z \in [0.02, 0.06]$.} 
            \begin{tabular}{cccc} 
                \hline 
                \hline\\[-2.5ex]
                Parameter & $p_\text{true}$ & $\langle p \rangle$ & $\sqrt{\left<\sigma_p^2\right>}$ \\[0.8ex]
                \hline 
                $\fsig / (\fsig)_{\rm fid}$ & 1.0 & $0.998 \pm  0.037$ &  $0.188$ \\
                $\alpha$ & 0.14 & $0.1356 \pm 0.0006$ &  $0.004$\\ 
                $\beta$ & 3.1 & $ 3.054 \pm 0.006$ & $0.04$ \\
                $M_0$ & -19.019 & $-19.027 \pm 0.002$ & $0.014$\\
                $\sigma_M$ & 0.12 & $0.1196 \pm 0.0008$ & $0.004 $\\
                $\sigma_u$ & - &  $14.1 \pm 0.5$ & $6.4$ \\
                $\sigma_v$ & - &   $168 \pm 20$ & $186$ \\
              \hline 
                \hline 
            \end{tabular} 
        \label{tab:results2}
        \end{table} 

\begin{figure*}
        \centering\includegraphics[width=0.9\textwidth]{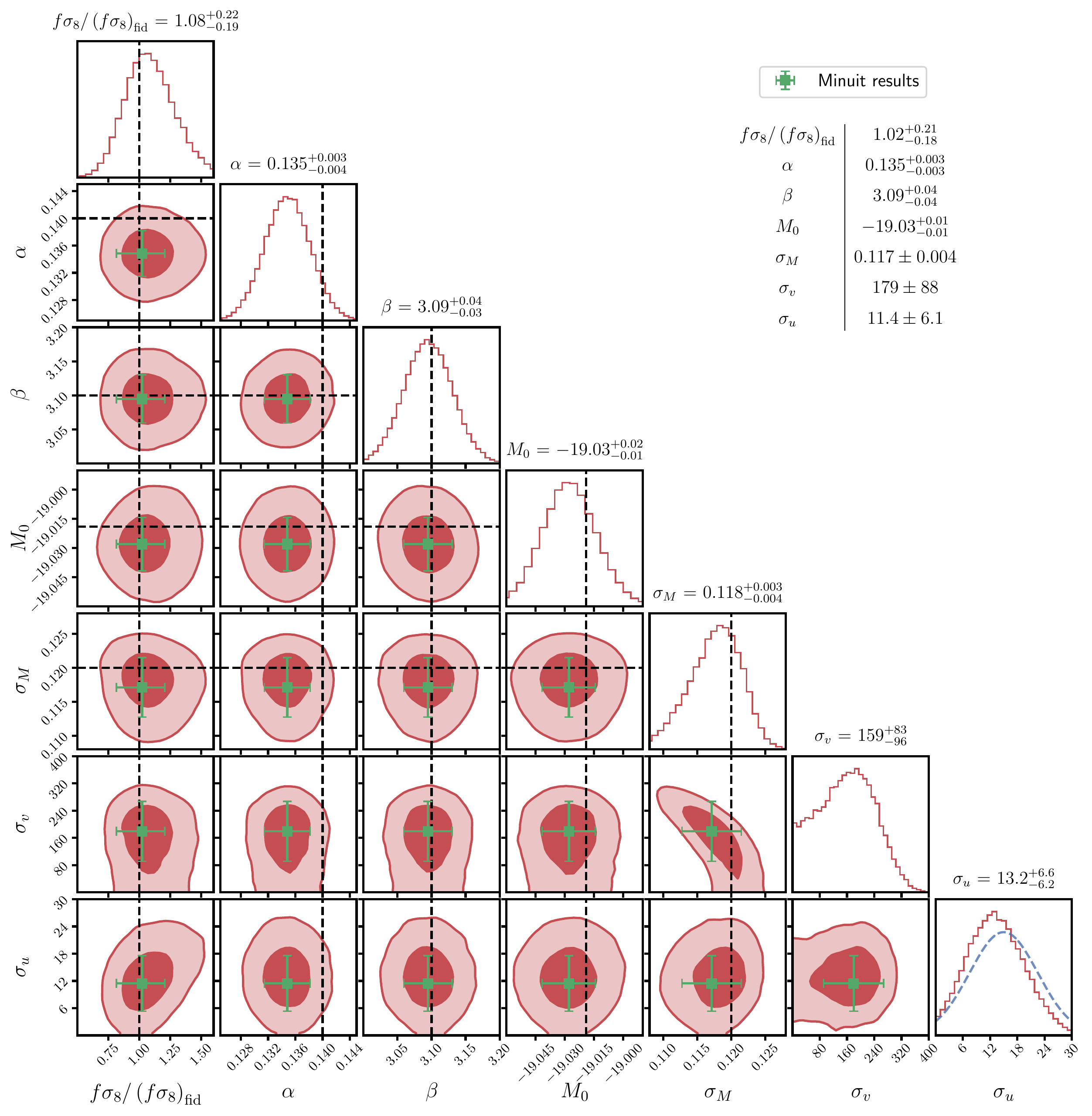}
        \caption{Posterior distributions for the joint fit (Hubble diagram and growth-rate parameters) of a single mock realization of the ZTF 6-year \sn\ program.
        The red contours show 1 and 2-$\sigma$ levels, the dotted black lines are the true values, the dotted blue line represents the prior on $\sigma_u$ and the green square show the \textsc{minuit} results.}
        \label{fig:mcmc6yearsfull}
    \end{figure*}

        We summarized our results on the 27 mocks in Table~\ref{tab:results2}. We see that $M_0$ is $\sim0.01$ mag lower than the input value, this is what dominated the sample selection bias study in Sect.~\ref{sec:results:effects_selection}. The standardization parameters $\alpha$ and $\beta$ are, on average, biased with respect to the input values. We have checked with a Hubble diagram fit on simulation outputs on the parent sample (before any selection) that we retrieved the true input parameters. We concluded that these biases come from selection effects but have negligible impact on $\fsig$ since we do not observe a strong correlation between these parameters. The \sn\ intrinsic scattering $\sigma_M$ is retrieved with good precision.
        To check the likelihood profile, we also ran a MCMC for one of our mocks. The corresponding
        posterior distributions are presented in Fig.~\ref{fig:mcmc6yearsfull}.
        The asymmetric \textsc{Minos} errors and MCMC chains analysis reveal a slightly larger upper error-bar for $\fsig$. This can be explained by the degeneracy of $\fsig$ with $\sigma_u$. We also observed the correlation between the intrinsic scattering $\sigma_M$ and the velocity noise parameter $\sigma_v$. We can also note that $\sigma_u$ constraints are dominated by our prior. 
    
    \subsection{Comparison with previous measurements}

    \begin{figure}
        \centering 
        \includegraphics[width=\columnwidth]{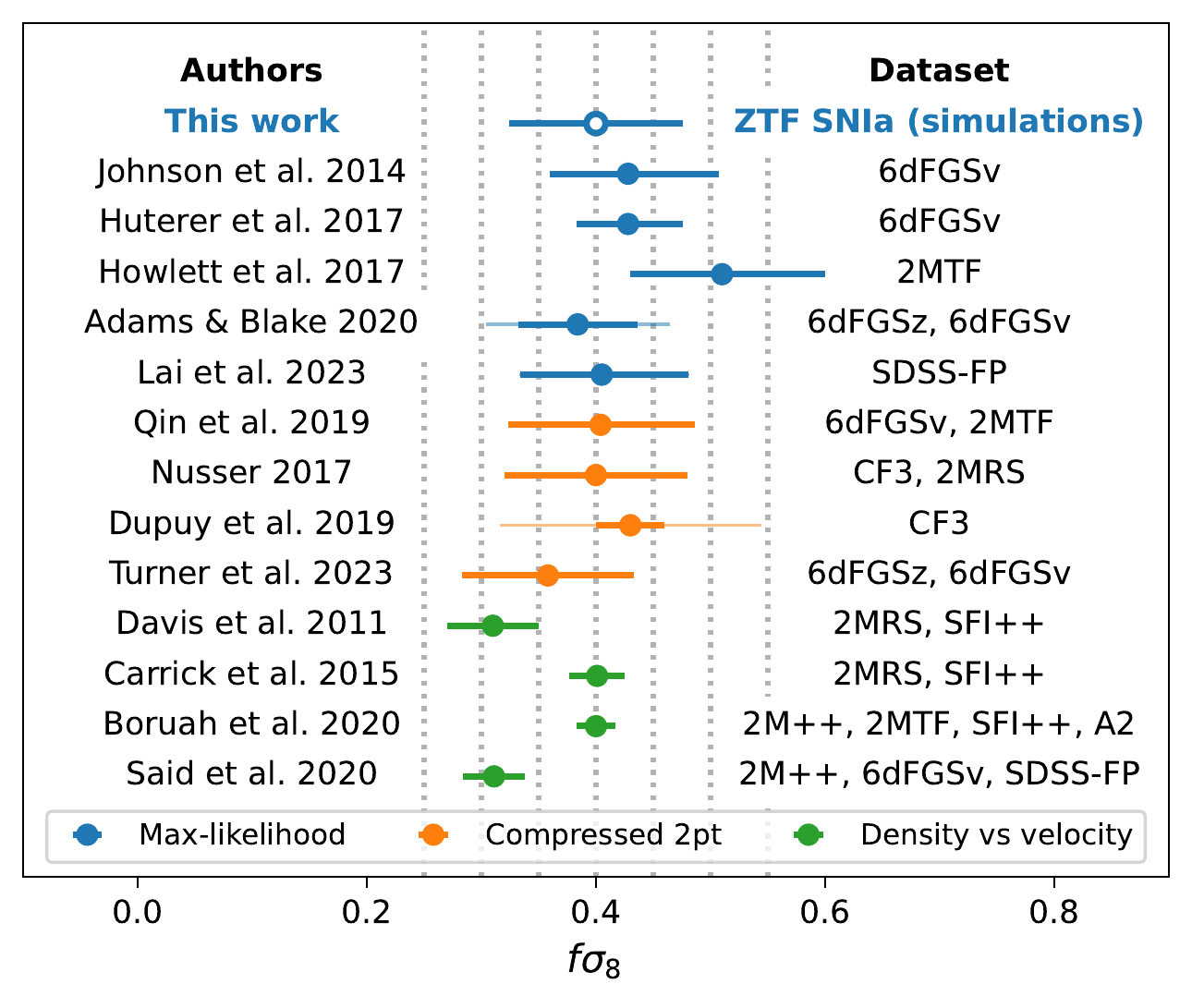}
        \caption{Measurements of the growth-rate of structures $f\sigma_8$ from peculiar velocity and galaxy survey data. 
         Error bars with lighter shade are those including quoted systematic errors (except for Dupuy et al. 2019, 
         where the extra contribution is from cosmic variance). 
         Our prediction is from Eq.~\ref{eq:fs8_baseline_error} and 
         only considers the spectroscopically classified sample of ZTF \sns\ between $0.02 < z< 0.06$. }
        \label{fig:previous_fs8}
    \end{figure}

    Our baseline analysis of the ZTF 6-years \sn\ sample yields an uncertainty of 19\% on the growth 
    rate of structures $\fsig$ (Eq.~\ref{eq:fs8_baseline_error}) when considering 
    the $\langle N \rangle \sim 1660$ \sns\ distributed over more than 28k deg$^2$ and $0.02 < z < 0.06$. We considered only velocity-velocity correlations in this work. It is interesting to compare our predictions to past measurements of $\fsig$ using peculiar velocity data. 

    Figure~\ref{fig:previous_fs8} compares our estimate from ZTF simulations to previous measurements, all of them using Tully-Fisher or Fundamental Plane distances (except \citealt{johnson6dFGalaxySurvey2014, boruahCosmicFlowsNearby2020} who include \sns\ from several compilations). 
    Most datasets also include a galaxy survey in order to cross-correlate
    density and velocity field, so their information source is larger 
    than in our case where we just use velocities. 
    We also make the distinction between methods (maximum-likelihood, compressed 2-pt statistics and reconstruction-based) since in 
    principle they do not use the same amount of information from 
    the data. 

    Our result has similar constraining power to those measurements that use only peculiar velocity data (and not density) and the similar methodology as ours, such as \citet{johnson6dFGalaxySurvey2014,howlett2MTFVIMeasuring2017},
    who obtain 15 and 16\%respectively. 
    \citet{johnson6dFGalaxySurvey2014} used 8896 FP distances from the 6dFGS between $0 < z < 0.05$ (southern sky only) and 303 \sns\ (heterogeneously distributed over the full sky). 
    \citet{howlett2MTFVIMeasuring2017} used 2062 TF distances between $0.002 < z < 0.03$, which is half the span in our sample.
    There are slight differences in analysis choices such as the values for $k_\text{max}$ in the evaluation of the model, or the assignment 
    of tracers to a mesh, which we do not use. 
    
    We expect constraints to improve when combining ZTF \sns\ with an overlapping galaxy survey. A great candidate is the DESI Bright Galaxy Survey \citep{hahnDESIBrightGalaxy2022} which has large 
    area and redshift overlap with ZTF. 
    
    \section{Robustness tests and alternative forecasts}
    \label{sec:robtests}

    In this section we study how our results are affected when we vary the prior on $\sigma_u$ or when we change the power spectrum model. We also present the expected precision for 30 months of data as well as when we include information beyond our complete sample cut at $z=0.06$. 

    \subsection{Impact of $\sigma_u$ prior parameters on $\fsig$}
    \label{sec:results:psigu}
    
    In our analysis, the $\sigma_u$ parameter is responsible for 
    a loss of the constraining power and its degeneracy with $\fsig$ can lead to biased results. 
    Some recent work \citep{laiUsingPeculiarVelocity2023,howlett_sloan_2022} proposed to fix this parameter 
    with a simulation-based value. In our analysis, we chose to use a Gaussian prior. Here we evaluate the impact 
    of this choice on the estimated $\fsig$.

\begin{figure}[ht]
        \centering
        \includegraphics[width=0.49\textwidth]{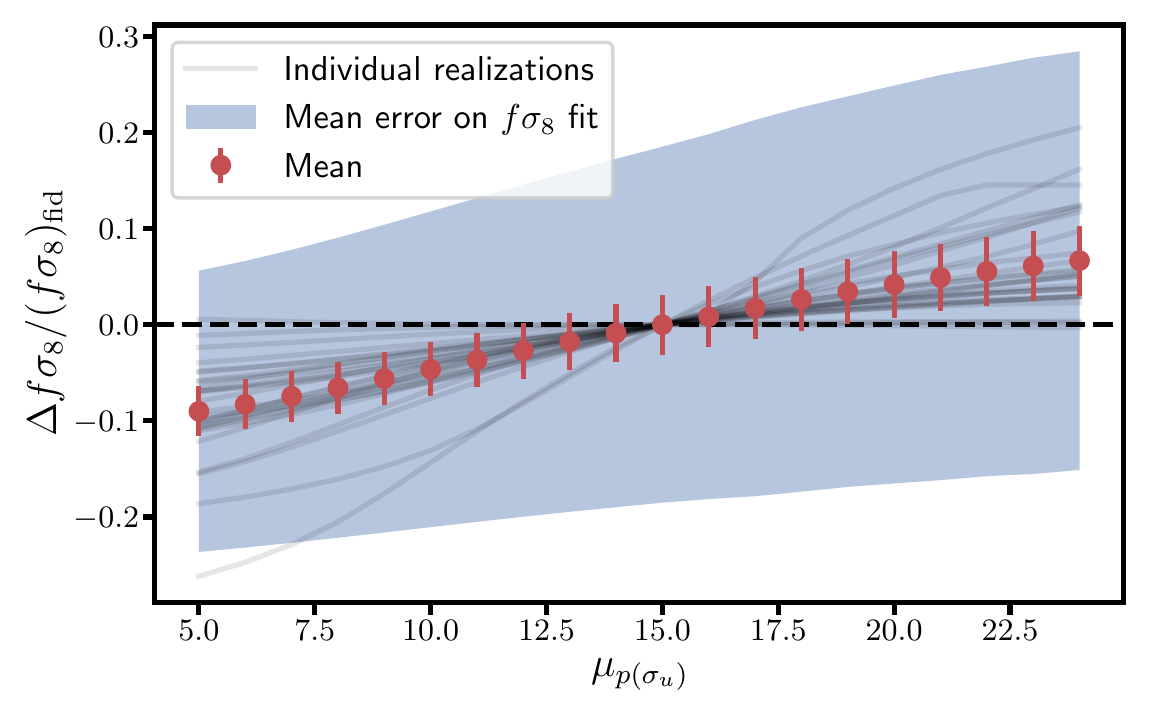}
        \includegraphics[width=0.49\textwidth]{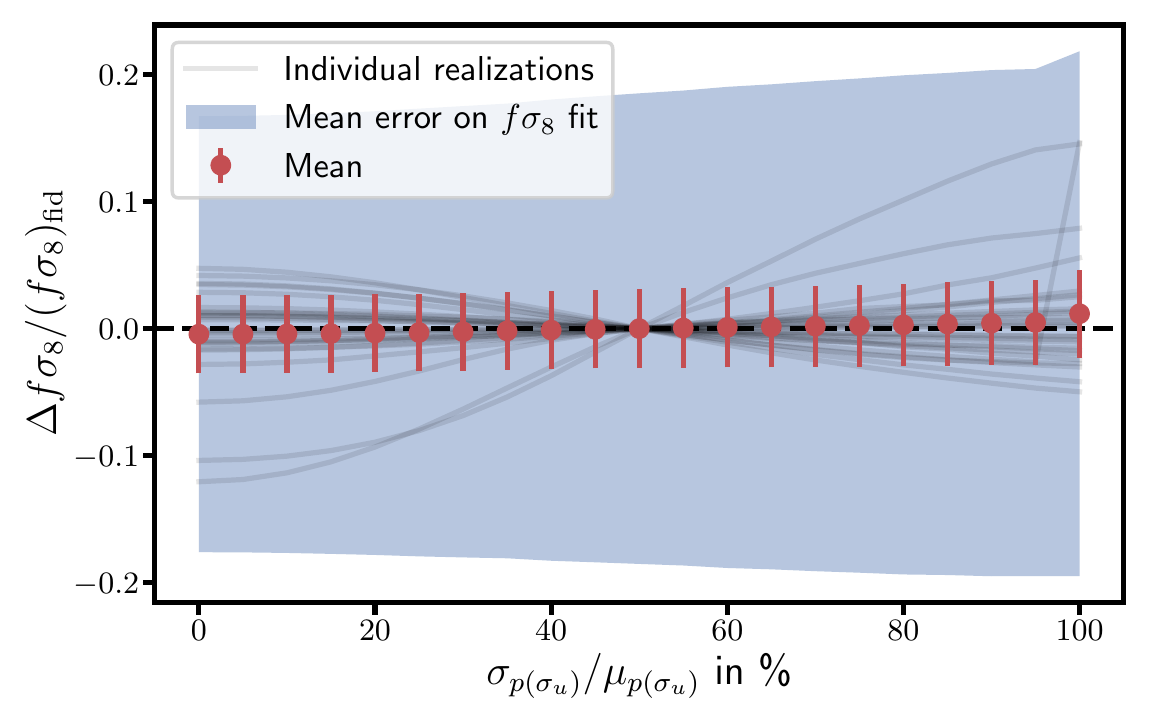}
        \caption{Effect of the $\sigma_u$ prior on the $\fsig$ fit results.
        \textit{Top panel}: Evolution of the difference between $f\sigma_8$ fit value and our baseline ($\sigma_{p(\sigma_u)} = 50 \%$) in function of the central value of the Gaussian prior on $\sigma_u$. The width of the prior is fixed to $\sigma_{p(\sigma_u)} = 10$\hmpc.
        \textit{Bottom panel}: Evolution of the difference between $f\sigma_8$ fit value and our baseline ($\sigma_{p(\sigma_u)} = 50 \%$). The central value of the prior is fixed to $\mu_{p(\sigma_u)} = 15$\hmpc. We use the estimated velocities of the \textit{6-year} sample}
        \label{fig:sigu_prior}
    \end{figure}

    The \textit{top panel} of Fig.~\ref{fig:sigu_prior} shows the evolution of the $\fsig$ result as a function of the central value of the Gaussian $\sigma_u$ prior. Between a central value of 5 to 25~\hmpc, we get a variation of $\fsig$ from $\sim -9$ to $\sim +6$ percent with respect to our baseline fit. 
    In \cite{kodaArePeculiarVelocity2014}, $\sigma_u$ is found, using N-body simulations, within the range of $[13, 15]$~\hmpc. In this range we found a less than $\sim 2 \%$ variation of $f\sigma_8$ . This result has to be mitigated since \cite{howlett2MTFVIMeasuring2017} found a value of $\sigma_u = 6.75^{+1.74}_{-6.75}$\hmpc\ on 2MTF data and \cite{laiUsingPeculiarVelocity2023} found that a $\sigma_u$ within $[19, 23]$~\hmpc\ better match their mocks.
    The bottom of Fig.~\ref{fig:sigu_prior} shows the evolution of the $f\sigma_8$ result as a function of the scale of the Gaussian $\sigma_u$ prior.
    We found that our results are insensitive to the width of this prior.
    
    \subsection{Effect of the power spectrum model}
        The power spectrum from \cite{bel_accurate_2019} used in the previous section comes from a fit on N-body simulations. We checked the impact of this choice using the RegPT power-spectrum model.
        We performed the same fit as in Sect.~\ref{sec:results:forecast006} using true velocities but including all our \sns\ up to $z = 0.13$ since the true velocities are not biased. We found with our 27 mocks a mean difference of $\left< \Delta\fsig\right>/ (\fsig)_{\rm fid} = (-1.5 \pm 4.9) \times 10^{-4}$. This difference is negligible. This is expected since the overall integral of these two power-spectra only differs by less than $\sim 1\%$ with $\sigma_u = 15$ \hmpc.
    
        We also performed a fit using the linear power spectrum. We found a difference of $\left<\Delta\fsig \right> / (\fsig)_{\rm fid} = (-2.7 \pm 0.06)\times 10^{-2}$. This difference comes from the fact that the linear power spectrum overestimates the power on small scales resulting in a variance overestimation. However this bias is small and nondetectable compared to the expected uncertainty of one realization.

    \subsection{ZTF DR2 forecast}
        \label{sec:robtests:dr2forecast}
        The next data release for ZTF (DR2) is expected to publish supernovae lightcurves for a survey of thirty-months. We simulated this sample.
        The statistics available after all our cuts is on average $\left<N \right> \sim 775$ \sns\ for our 27 mocks.
        From the fit with $\mathbf{p_\mathrm{HD}}$ fixed we obtained $\left<f\sigma_8 / (f\sigma_8)_{\rm fid} \right> = 0.968 \pm 0.046$, with an average uncertainty on $\fsig$ of 0.246. This result is compatible with the fiducial value.
        When fitting with $\mathbf{p}$ and $\mathbf{p_\mathrm{HD}}$ free, we obtained on average $\left<f\sigma_8 / (f\sigma_8)_{\rm fid} \right> = 0.923 \pm 0.051$, at $\sim 1.5\sigma$ from the fiducial value. This small bias can be explained by the fact that, as seen in the third paragraph of Sect.~\ref{sec:results:forecast006}, this fit can become unstable, due to a combination of low number of SNIa and the large number of free parameters. For the DR2 samples, a larger fraction of realizations yield excessively low values for $f\sigma_8$. This would indicate that the distribution of $f\sigma_8$ becomes non-Gaussian at such low number density of tracers. For a work on real data, this point would need further investigation. However, this bias is still negligible compared to the averaged uncertainty on $\fsig$ of about 0.255. Since the number of \sns\ is approximately divided by $\sim 2$ within the same volume we observe the expected $\sqrt{N}$ law as $0.255 \sim 0.188 \times \sqrt{2}$.
        
    \subsection{Impact of a bias correction beyond $z = 0.06$}
    \label{sec:robtests:fakedebias}
    With our cut at $z=0.06$, we are left with $\langle N \rangle \sim 1660$ \sns. This number represents only half of the spectroscopically typed sample. In Fig.~\ref{fig:fs8bias} we show $\fsig$ results as a function of the redshift upper bound $z_{\rm max}$. In the \textit{middle} and \textit{bottom} panel of Fig.~\ref{fig:fs8bias}, we show the impact of this sample bias on the $\fsig$ fit compared to the fit using true velocities of the \textit{top pannel}. After $z > 0.06$ the bias begins to grow up to $\sim 180$ percent at $z=0.12$.

    This bias cannot be corrected simply by using the methods introduced in \cite{betouleImprovedCosmologicalConstraints2014} or in \cite{kesslerCorrectingTypeIa2017}. A correction in redshift bins will shift the velocity of an entire redshift shell with the same factor, leading to fake velocity correlation.
    
    Although we do not propose a method to actually make the correction, we explored how much we could improve the $\fsig$ measurement. In what follows, we assume that we can perfectly correct for selection biases and construct an unbiased sample following two steps. First we take the velocity uncertainties $\sigma_{\hat{ v}, i}$ of the selected sample. Then, we draw new velocities from a Gaussian distribution centered on $v_{\text{true},i}$ with standard deviation of $\sigma_{\hat{ v}, i}$.

\begin{figure}
        \centering        
        \includegraphics[width=\columnwidth]{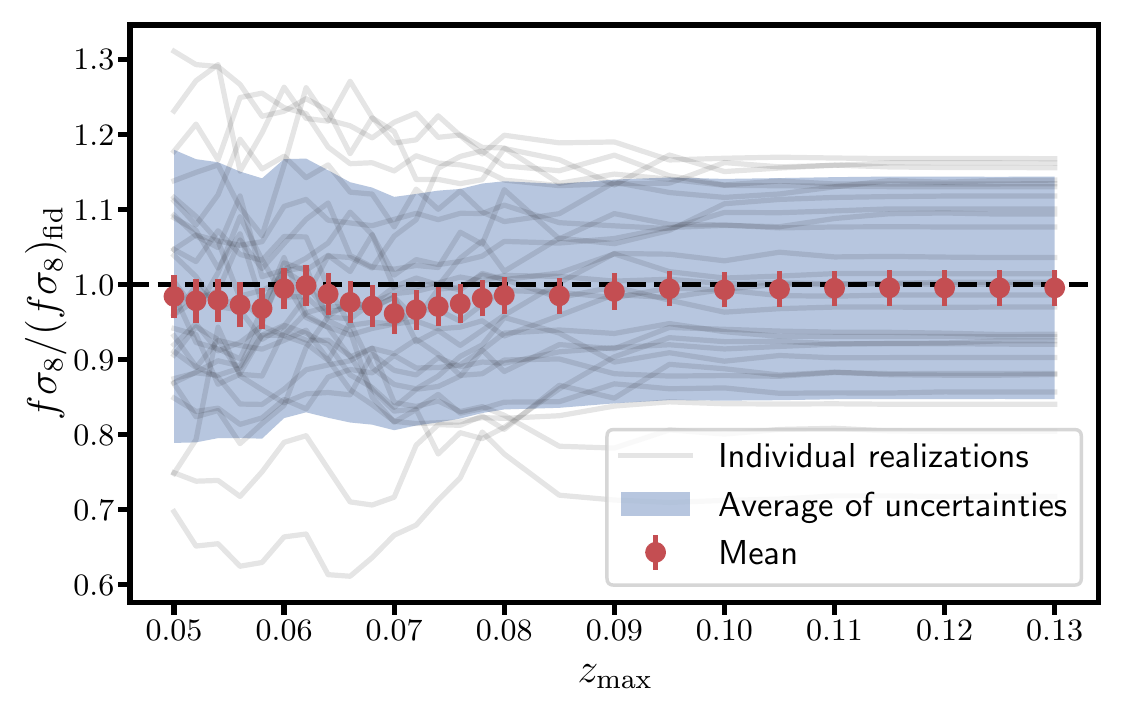}
        \includegraphics[width=\columnwidth]{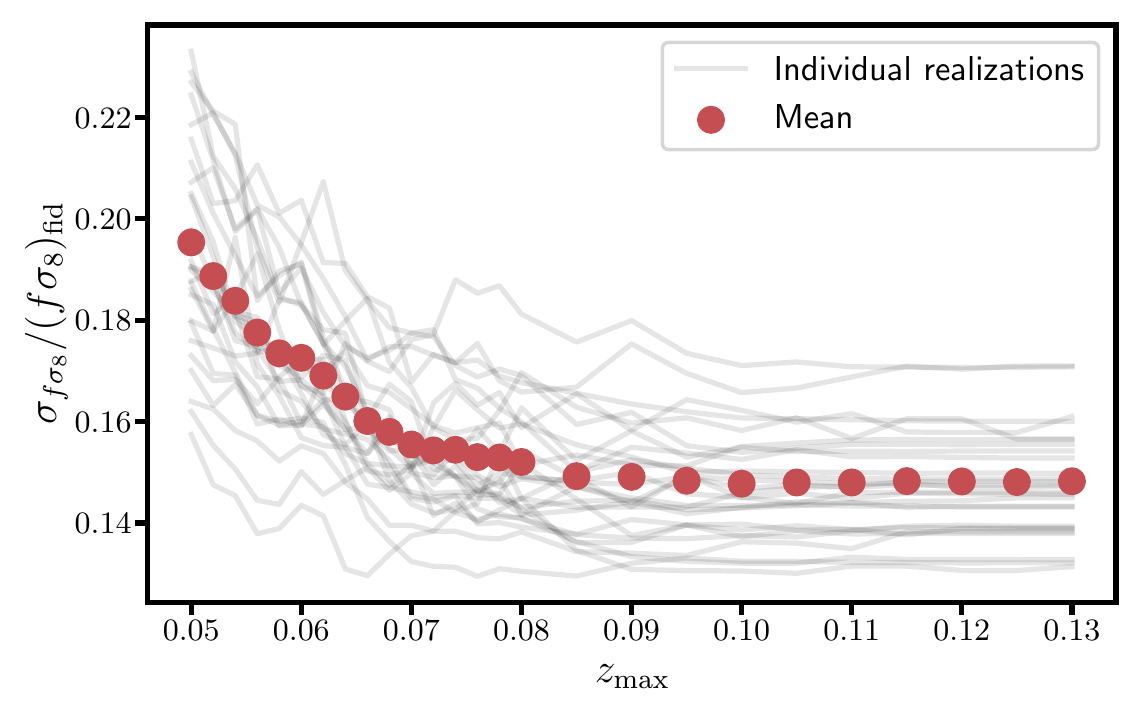}
        \caption{$\fsig$ constraints as a function of the upper bound in the redshift range $z_\text{max}$, using an artificially perfect correction for the bias on the
        velocities. The \textit{top panel} shows the best-fit values and the \textit{bottom panel} its uncertainties. 
        Gray lines represent each mock, red points are average of 27 realizations.}
        \label{fig:fakedebias}
    \end{figure}

    With this new sample of artificially corrected \sn\ velocities, 
    we fit for $\fsig$ extending the range in redshift. 
    Results are shown in Fig.~\ref{fig:fakedebias}. 
    We obtained for $z<0.06$ ($\left< N_{SN}\right> = 1660$) a result of $\left<f\sigma_8 / (f\sigma_8)_{\rm fid} \right> =  0.994 \pm 0.027$, with an averaged uncertainty of 0.173 and for $z<0.13$ ($\left< N_{SN}\right> = 3520)$ (i.e, a statistic multiplied by $\sim 2$) a result of $\left<f\sigma_8 / (f\sigma_8)_{\rm fid} \right> = 0.995 \pm 0.024$, with an averaged uncertainty of 0.148.
    We see that the uncertainties on $\fsig$ only decrease from 17 to 15 percent when including \sns\ with redshifts $z_{\rm obs} > 0.06$.
    This can be explained by two effects: 
    Firstly, when going deeper in redshift, the observed volume increases faster than the number of \sns\, which at some point also decreases due to sample selection. Thus the density decreases quickly, as we shown on bottom panel of Fig.~\ref{fig:selection_effects}. Secondly, the errors on peculiar velocities increase with redshift as stated in Eq.~\ref{eq:verrorlin}. These two effects result in a reduced constraining power from \sns\ at $z>0.06$. 

\section{Conclusions}
\label{sec:conclusion}
    
    In this paper we have presented detailed simulations of ZTF \sn\ samples equivalent to 6 years of data. We used these simulations to study the measurement of the growth rate of structures $\fsig$
    from the clustering of \sn\ peculiar velocities. 
    
    Our simulations aim to faithfully reproduce real observations, including several features: peculiar velocities and positions of host galaxies drawn from an N-body simulation, lightcurve sampling using actual ZTF metadata, realistic fluxes and uncertainties, and selection effects from photometric detection and spectroscopic follow-up for typing. 
   We used the SALT2 model to adjust \sn\ light-curve parameters on the measurements and applied quality cuts to reproduce cosmological samples. 
    
    We have then presented the methodology proposed to derive peculiar velocities from \sn\ distances and measure the growth rate. 
    We used the commonly employed maximum-likelihood method, which assumes the peculiar velocity field to be a multivariate Gaussian random field. The covariance matrix used in the likelihood is a function of the growth rate parameter plus \sn\ standardization and nuisance parameters. Our baseline choice of analysis fits for all 
    parameters at once. We showed that all our results are robust against variations of the main assumptions of our analysis. 

    Our simulations showed that selection effects, mainly the one imposed by the spectroscopic typing, create a bias in distance estimates at $z_{\rm obs} > 0.06$. Biases in distances translate to biases in 
    the estimates of peculiar velocities and thus on the measurements of the growth rate.

    We defined an unbiased sample of \sns\ by considering only those at $z_{\rm obs} < 0.06$ with which we derive constraints on $\fsig$. Using the equivalent of 6-years of ZTF data and our baseline analysis settings, we showed that we can obtain unbiased estimates of $\fsig$ with a $19\%$ precision. 
    This precision is comparable to previous measurements on data using peculiar velocity samples derived from the Fundamental Plane or the Tully-Fisher relations. 
    Our result showcases the great potential of using \sn\ distances alone for growth-rate measurements.

    Since selection effects significantly reduce the \sn\ sample size, 
    we investigated the gain in applying a bias correction to \sns\ at 
    $z_{\rm obs} > 0.06$. Assuming an artificially perfect bias correction
    and using the full available redshift range of spectroscopically typed \sns\, our constraints on $\fsig$ reduce from 17 to 15\%. 
    This small improvement is mainly due to the rapid decrease in comoving density of tracers between $0.06 < z < 0.10$ and the increase in the velocity uncertainties due to intrinsic scatter of \sn\ peak brightness.  
    Significant improvement could be expected by using a photometrically typed sample of \sns\, which is larger and push the decline of the comoving density to a higher redshift. We leave this investigation for future work. 

    The work of this paper sets the basis for the measurement of the growth rate with real ZTF data. 
    The same methodologies can be applied to \sn\ samples from the Vera Rubin Observatory, where spectroscopic follow-up cannot be performed 
    and measurements will rely on photometric typing.

\begin{acknowledgements}
 Simulation logs are based on observations obtained with the Samuel Oschin Telescope 48-inch and the 60-inch Telescope at the Palomar Observatory as part of the Zwicky Transient Facility project. ZTF is supported by the National Science Foundation under Grants No. AST-1440341 and AST-2034437 and a collaboration including current partners Caltech, IPAC, the Weizmann Institute of Science, the Oskar Klein Center at Stockholm University, the University of Maryland, Deutsches Elektronen-Synchrotron and Humboldt University, the TANGO Consortium of Taiwan, the University of Wisconsin at Milwaukee, Trinity College Dublin, Lawrence Livermore National Laboratories, IN2P3, University of Warwick, Ruhr University Bochum, Northwestern University and former partners the University of Washington, Los Alamos National Laboratories, and Lawrence Berkeley National Laboratories. Operations are conducted by COO, IPAC, and UW.

The project leading to this publication has received funding from 
Excellence Initiative of Aix-Marseille University - A*MIDEX, 
a French ``Investissements d'Avenir'' program (AMX-20-CE-02 - DARKUNI).

Some of the results in this paper have been derived using the healpy and HEALPix packages.

\end{acknowledgements}


\bibliographystyle{aa}
\bibliography{MyBib.bib}

\begin{thebibliography}{94}
\expandafter\ifx\csname natexlab\endcsname\relax\def\natexlab#1{#1}\fi

\bibitem[{Abate {et~al.}(2008)Abate, Bridle, Teodoro, Warren, \&
  Hendry}]{abate_peculiar_2008}
Abate, A., Bridle, S., Teodoro, L. F.~A., Warren, M.~S., \& Hendry, M. 2008,
  Monthly Notices of the Royal Astronomical Society, 389, 1739

\bibitem[{Adams \& Blake(2020)}]{adamsJointGrowthrateMeasurements2020}
Adams, C. \& Blake, C. 2020, Monthly Notices of the Royal Astronomical Society,
  494, 3275

\bibitem[{Alam {et~al.}(2021)Alam, Aubert, Avila, Balland, Bautista, Bershady,
  Bizyaev, Blanton, Bolton, Bovy, Brinkmann, Brownstein, Burtin, Chabanier,
  Chapman, Choi, Chuang, Comparat, Cousinou, Cuceu, Dawson, {de la Torre}, {de
  Mattia}, Agathe, {des Bourboux}, Escoffier, Etourneau, Farr, {Font-Ribera},
  Frinchaboy, Fromenteau, {Gil-Mar{\'i}n}, Le~Goff, {Gonzalez-Morales},
  {Gonzalez-Perez}, Grabowski, Guy, Hawken, Hou, Kong, Parker, Klaene, Kneib,
  Lin, Long, Lyke, {de la Macorra}, Martini, Masters, Mohammad, Moon, Mueller,
  {Mu{\~n}oz-Guti{\'e}rrez}, Myers, Nadathur, Neveux, Newman, Noterdaeme,
  Oravetz, Oravetz, {Palanque-Delabrouille}, Pan, Paviot, Percival,
  {P{\'e}rez-R{\`a}fols}, Petitjean, Pieri, Prakash, Raichoor, Ravoux, Rezaie,
  Rich, Ross, Rossi, Ruggeri, {Ruhlmann-Kleider}, S{\'a}nchez, S{\'a}nchez,
  {S{\'a}nchez-Gallego}, Sayres, Schneider, Seo, Shafieloo, Slosar, Smith,
  Stermer, Tamone, Tinker, Tojeiro, {Vargas-Maga{\~n}a}, Variu, Wang, Weaver,
  Weijmans, Y{\`e}che, Zarrouk, Zhao, Zhao, \&
  Zheng}]{alamCompletedSDSSIVExtended2021}
Alam, S., Aubert, M., Avila, S., {et~al.} 2021, Physical Review D, 103, 083533

\bibitem[{Avila {et~al.}(2020)Avila, {Gonzalez-Perez}, Mohammad, {de~Mattia},
  Zhao, Raichoor, Tamone, Alam, Bautista, Bianchi, Burtin, Chapman, Chuang,
  Comparat, Dawson, Divers, {du~Mas~des~Bourboux}, {Gil-Marin}, Mueller, Habib,
  Heitmann, {Ruhlmann-Kleider}, Padilla, Percival, Ross, Seo, Schneider, \&
  Zhao}]{avilaCompletedSDSSIVExtended2020}
Avila, S., {Gonzalez-Perez}, V., Mohammad, F.~G., {et~al.} 2020, Monthly
  Notices of the Royal Astronomical Society, 499, 5486

\bibitem[{Barbary {et~al.}(2016)Barbary, Bailey, Barentsen, Barclay, Biswas,
  Boone, Craig, Feindt, Friesen, Goldstein, Jha, Jones, Mondon,
  Papadogiannakis, Perrefort, Pierel, Rodney, Rose, Saunders, Sipőcz,
  Sofiatti, Thomas, van Santen, Vincenzi, Wang, \&
  Wood-Vasey}]{barbary_sncosmo_2016}
Barbary, K., Bailey, S., Barentsen, G., {et~al.} 2016, {SNCosmo}

\bibitem[{Bautista {et~al.}(2021)Bautista, Paviot, Vargas~Maga{\~n}a, {de la
  Torre}, Fromenteau, {Gil-Mar{\'i}n}, Ross, Burtin, Dawson, Hou, Kneib, {de
  Mattia}, Percival, Rossi, Tojeiro, Zhao, Zhao, Alam, Brownstein, Chapman,
  Choi, Chuang, Escoffier, {de la Macorra}, {du Mas des Bourboux}, Mohammad,
  Moon, M{\"u}ller, Nadathur, Newman, Schneider, Seo, \&
  Wang}]{bautistaCompletedSDSSIVExtended2021}
Bautista, J.~E., Paviot, R., Vargas~Maga{\~n}a, M., {et~al.} 2021, Monthly
  Notices of the Royal Astronomical Society, 500, 736

\bibitem[{Bel {et~al.}(2019)Bel, Pezzotta, Carbone, Sefusatti, \&
  Guzzo}]{bel_accurate_2019}
Bel, J., Pezzotta, A., Carbone, C., Sefusatti, E., \& Guzzo, L. 2019, Astronomy
  \& Astrophysics, 622, A109, arXiv: 1809.09338

\bibitem[{{Bellm} {et~al.}(2019){Bellm}, {Kulkarni}, {Barlow}, {Feindt},
  {Graham}, {Goobar}, {Kupfer}, {Ngeow}, {Nugent}, {Ofek}, {Prince}, {Riddle},
  {Walters}, \& {Ye}}]{bellm_survey_sched}
{Bellm}, E.~C., {Kulkarni}, S.~R., {Barlow}, T., {et~al.} 2019, \pasp, 131,
  068003

\bibitem[{Betoule {et~al.}(2014)Betoule, Kessler, Guy, Mosher, Hardin, Biswas,
  Astier, {El-Hage}, Konig, Kuhlmann, Marriner, Pain, Regnault, Balland,
  Bassett, Brown, Campbell, Carlberg, {Cellier-Holzem}, Cinabro, Conley,
  D'Andrea, DePoy, Doi, Ellis, Fabbro, Filippenko, Foley, Frieman, Fouchez,
  Galbany, Goobar, Gupta, Hill, Hlozek, Hogan, Hook, Howell, Jha, Le~Guillou,
  Leloudas, Lidman, Marshall, M{\"o}ller, Mour{\~a}o, Neveu, Nichol, Olmstead,
  {Palanque-Delabrouille}, Perlmutter, Prieto, Pritchet, Richmond, Riess,
  {Ruhlmann-Kleider}, Sako, Schahmaneche, Schneider, Smith, Sollerman,
  Sullivan, Walton, \& Wheeler}]{betouleImprovedCosmologicalConstraints2014}
Betoule, M., Kessler, R., Guy, J., {et~al.} 2014, Astronomy and Astrophysics,
  568, A22

\bibitem[{Beutler {et~al.}(2012)Beutler, Blake, Colless, Jones,
  {Staveley-Smith}, Poole, Campbell, Parker, Saunders, \&
  Watson}]{beutler6dFGalaxySurvey2012a}
Beutler, F., Blake, C., Colless, M., {et~al.} 2012, Monthly Notices of the
  Royal Astronomical Society, 423, 3430

\bibitem[{Beutler {et~al.}(2017)Beutler, Seo, Saito, Chuang, Cuesta,
  Eisenstein, {Gil-Mar{\'i}n}, Grieb, Hand, Kitaura, Modi, Nichol, Olmstead,
  Percival, Prada, S{\'a}nchez, {Rodriguez-Torres}, Ross, Ross, Schneider,
  Tinker, Tojeiro, \&
  {Vargas-Maga{\~n}a}}]{beutlerClusteringGalaxiesCompleted2017}
Beutler, F., Seo, H.-J., Saito, S., {et~al.} 2017, Monthly Notices of the Royal
  Astronomical Society, 466, 2242

\bibitem[{Blake {et~al.}(2011)Blake, Brough, Colless, Contreras, Couch, Croom,
  Davis, Drinkwater, Forster, Gilbank, Gladders, Glazebrook, Jelliffe, Jurek,
  Li, Madore, Martin, Pimbblet, Poole, Pracy, Sharp, Wisnioski, Woods, Wyder,
  \& Yee}]{blakeWiggleZDarkEnergy2011a}
Blake, C., Brough, S., Colless, M., {et~al.} 2011, Monthly Notices of the Royal
  Astronomical Society, 415, 2876

\bibitem[{Boruah {et~al.}(2020)Boruah, Hudson, \&
  Lavaux}]{boruahCosmicFlowsNearby2020}
Boruah, S.~S., Hudson, M.~J., \& Lavaux, G. 2020, Monthly Notices of the Royal
  Astronomical Society, 498, 2703

\bibitem[{Boruah {et~al.}(2021)Boruah, Lavaux, \&
  Hudson}]{boruahReconstructingDarkMatter2021}
Boruah, S.~S., Lavaux, G., \& Hudson, M.~J. 2021, arXiv:2111.15535 [astro-ph]

\bibitem[{{Cardelli} {et~al.}(1989){Cardelli}, {Clayton}, \& {Mathis}}]{ccm89}
{Cardelli}, J.~A., {Clayton}, G.~C., \& {Mathis}, J.~S. 1989, \apj, 345, 245

\bibitem[{Carrick {et~al.}(2015)Carrick, Turnbull, Lavaux, \&
  Hudson}]{carrickCosmologicalParametersComparison2015}
Carrick, J., Turnbull, S.~J., Lavaux, G., \& Hudson, M.~J. 2015, Monthly
  Notices of the Royal Astronomical Society, 450, 317

\bibitem[{Clifton {et~al.}(2012)Clifton, Ferreira, Padilla, \&
  Skordis}]{cliftonModifiedGravityCosmology2012}
Clifton, T., Ferreira, P.~G., Padilla, A., \& Skordis, C. 2012, Physics
  Reports, 513, 1

\bibitem[{Dam {et~al.}(2021)Dam, Bolejko, \&
  Lewis}]{damExploringRedshiftspacePeculiar2021}
Dam, L., Bolejko, K., \& Lewis, G.~F. 2021, Journal of Cosmology and
  Astroparticle Physics, 2021, 018

\bibitem[{Davis {et~al.}(2011{\natexlab{a}})Davis, Nusser, Masters, Springob,
  Huchra, \& Lemson}]{davisLocalGravityLocal2011}
Davis, M., Nusser, A., Masters, K.~L., {et~al.} 2011{\natexlab{a}}, Monthly
  Notices of the Royal Astronomical Society, 413, 2906

\bibitem[{Davis {et~al.}(2011{\natexlab{b}})Davis, Hui, Frieman, Haugb{\o}lle,
  Kessler, Sinclair, Sollerman, Bassett, Marriner, M{\"o}rtsell, Nichol,
  Richmond, Sako, Schneider, \& Smith}]{davisEffectPeculiarVelocities2011}
Davis, T.~M., Hui, L., Frieman, J.~A., {et~al.} 2011{\natexlab{b}}, The
  Astrophysical Journal, 741, 67

\bibitem[{{de la Torre} {et~al.}(2017){de la Torre}, Jullo, Giocoli, Pezzotta,
  Bel, Granett, Guzzo, Garilli, Scodeggio, Bolzonella, Abbas, Adami, Bottini,
  Cappi, Cucciati, Davidzon, Franzetti, Fritz, Iovino, Krywult, Le~Brun,
  Le~F{\`e}vre, Maccagni, Ma{\l}ek, Marulli, Polletta, Pollo, Tasca, Tojeiro,
  Vergani, Zanichelli, Arnouts, Branchini, Coupon, De~Lucia, Ilbert, Moutard,
  Moscardini, Peacock, Metcalf, Prada, \&
  Yepes}]{delatorreVIMOSPublicExtragalactic2017}
{de la Torre}, S., Jullo, E., Giocoli, C., {et~al.} 2017, Astronomy and
  Astrophysics, 608, A44

\bibitem[{{de Mattia} {et~al.}(2021){de Mattia}, {Ruhlmann-Kleider}, Raichoor,
  Ross, Tamone, Zhao, Alam, Avila, Burtin, Bautista, Beutler, Brinkmann,
  Brownstein, Chapman, Chuang, Comparat, {du Mas des Bourboux}, Dawson, {de la
  Macorra}, {Gil-Mar{\'i}n}, {Gonzalez-Perez}, Gorgoni, Hou, Kong, Lin,
  Nadathur, Newman, Mueller, Percival, Rezaie, Rossi, Schneider, Tiwari, Vivek,
  Wang, \& Zhao}]{demattiaCompletedSDSSIVExtended2021}
{de Mattia}, A., {Ruhlmann-Kleider}, V., Raichoor, A., {et~al.} 2021, Monthly
  Notices of the Royal Astronomical Society, 501, 5616

\bibitem[{Dekany {et~al.}(2020)Dekany, Smith, Riddle, Feeney, Porter, Hale,
  Zolkower, Belicki, Kaye, Henning, Walters, Cromer, Delacroix, Rodriguez,
  Reiley, Mao, Hover, Murphy, Burruss, Baker, Kowalski, Reif, Mueller, Bellm,
  Graham, \& Kulkarni}]{dekanyZwickyTransientFacility2020}
Dekany, R., Smith, R.~M., Riddle, R., {et~al.} 2020, Publications of the
  Astronomical Society of the Pacific, 132, 038001

\bibitem[{Dembinski \& et~al.(2020)}]{iminuit}
Dembinski, H. \& et~al., P.~O. 2020

\bibitem[{Dhawan {et~al.}(2022)Dhawan, Goobar, Smith, Johansson, Rigault,
  Nordin, Biswas, Goldstein, Nugent, Kim, Miller, Graham, Medford, Kasliwal,
  Kulkarni, Duev, Bellm, Rosnet, Riddle, \&
  Sollerman}]{dhawanZwickyTransientFacility2022}
Dhawan, S., Goobar, A., Smith, M., {et~al.} 2022, Monthly Notices of the Royal
  Astronomical Society, 510, 2228

\bibitem[{Djorgovski \&
  Davis(1987)}]{djorgovskiFundamentalPropertiesElliptical1987}
Djorgovski, S. \& Davis, M. 1987, The Astrophysical Journal, 313, 59

\bibitem[{Dupuy {et~al.}(2019)Dupuy, Courtois, \&
  Kubik}]{dupuyEstimationLocalGrowth2019}
Dupuy, A., Courtois, H.~M., \& Kubik, B. 2019, Monthly Notices of the Royal
  Astronomical Society, 486, 440

\bibitem[{Ezquiaga \& Zumalac{\'a}rregui(2018)}]{ezquiagaDarkEnergyLight2018}
Ezquiaga, J.~M. \& Zumalac{\'a}rregui, M. 2018, Frontiers in Astronomy and
  Space Sciences, 5, 44

\bibitem[{Feindt {et~al.}(2019)Feindt, Nordin, Rigault, Brinnel, Dhawan,
  Goobar, \& Kowalski}]{feindtSimsurveyEstimatingTransient2019}
Feindt, U., Nordin, J., Rigault, M., {et~al.} 2019, Journal of Cosmology and
  Astroparticle Physics, 10, 005

\bibitem[{Fremling {et~al.}(2020)Fremling, Miller, Sharma, Dugas, Perley,
  Taggart, Sollerman, Goobar, Graham, Neill, Nordin, Rigault, Walters,
  Andreoni, Bagdasaryan, Belicki, Cannella, Bellm, Cenko, De, Dekany,
  Frederick, Golkhou, Graham, Helou, Ho, Kasliwal, Kupfer, Laher, Mahabal,
  Masci, Riddle, Rusholme, Schulze, Shupe, Smith, {van Velzen}, Yan, Yao,
  Zhuang, \& Kulkarni}]{fremlingZwickyTransientFacility2020}
Fremling, C., Miller, A.~A., Sharma, Y., {et~al.} 2020, The Astrophysical
  Journal, 895, 32

\bibitem[{{Gil-Mar{\'i}n} {et~al.}(2020){Gil-Mar{\'i}n}, Bautista, Paviot,
  {Vargas-Maga{\~n}a}, {de la Torre}, Fromenteau, Alam, {\'A}vila, Burtin,
  Chuang, Dawson, Hou, {de Mattia}, Mohammad, M{\"u}ller, Nadathur, Neveux,
  Percival, Raichoor, Rezaie, Ross, Rossi, {Ruhlmann-Kleider}, Smith, Tamone,
  Tinker, Tojeiro, Wang, Zhao, Zhao, Brinkmann, Brownstein, Choi, Escoffier,
  {de la Macorra}, Moon, Newman, Schneider, Seo, \&
  Vivek}]{gil-marinCompletedSDSSIVExtended2020}
{Gil-Mar{\'i}n}, H., Bautista, J.~E., Paviot, R., {et~al.} 2020, Monthly
  Notices of the Royal Astronomical Society, 498, 2492

\bibitem[{{Gil-Mar{\'i}n} {et~al.}(2018){Gil-Mar{\'i}n}, Guy, Zarrouk, Burtin,
  Chuang, Percival, Ross, Ruggeri, Tojerio, Zhao, Wang, Bautista, Hou,
  S{\'a}nchez, P{\^a}ris, Baumgarten, Brownstein, Dawson, Eftekharzadeh,
  {Gonz{\'a}lez-P{\'e}rez}, Habib, Heitmann, Myers, Rossi, Schneider, Seo,
  Tinker, \& Zhao}]{gil-marinClusteringSDSSIVExtended2018}
{Gil-Mar{\'i}n}, H., Guy, J., Zarrouk, P., {et~al.} 2018, Monthly Notices of
  the Royal Astronomical Society, 477, 1604

\bibitem[{Gorski {et~al.}(1989)Gorski, Davis, Strauss, White, \&
  Yahil}]{gorskiCosmologicalVelocityCorrelations1989}
Gorski, K.~M., Davis, M., Strauss, M.~A., White, S. D.~M., \& Yahil, A. 1989,
  The Astrophysical Journal, 344, 1

\bibitem[{G{\'o}rski {et~al.}(2005)G{\'o}rski, Hivon, Banday, Wandelt, Hansen,
  Reinecke, \& Bartelmann}]{gorskiHEALPixFrameworkHighResolution2005}
G{\'o}rski, K.~M., Hivon, E., Banday, A.~J., {et~al.} 2005, The Astrophysical
  Journal, 622, 759

\bibitem[{Graham {et~al.}(2019)Graham, Kulkarni, Bellm, Adams, Barbarino,
  Blagorodnova, Bodewits, Bolin, Brady, Cenko, Chang, Coughlin, De, Eadie,
  Farnham, Feindt, Franckowiak, Fremling, Gezari, Ghosh, Goldstein, Golkhou,
  Goobar, Ho, Huppenkothen, Ivezi{\'c}, Jones, Juric, Kaplan, Kasliwal, Kelley,
  Kupfer, Lee, Lin, Lunnan, Mahabal, Miller, Ngeow, Nugent, Ofek, Prince,
  Rauch, {van Roestel}, Schulze, Singer, Sollerman, Taddia, Yan, Ye, Yu,
  Barlow, Bauer, Beck, Belicki, Biswas, Brinnel, Brooke, Bue, Bulla, Burruss,
  Connolly, Cromer, Cunningham, Dekany, Delacroix, Desai, Duev, Feeney, Flynn,
  Frederick, {Gal-Yam}, Giomi, Groom, Hacopians, Hale, Helou, Henning, Hover,
  Hillenbrand, Howell, Hung, Imel, Ip, Jackson, Kaspi, Kaye, Kowalski, Kramer,
  Kuhn, Landry, Laher, Mao, Masci, Monkewitz, Murphy, Nordin, Patterson,
  Penprase, Porter, Rebbapragada, Reiley, Riddle, Rigault, Rodriguez, Rusholme,
  {van Santen}, Shupe, Smith, Soumagnac, Stein, Surace, Szkody, Terek,
  Van~Sistine, {van Velzen}, Vestrand, Walters, Ward, Zhang, \&
  Zolkower}]{grahamZwickyTransientFacility2019}
Graham, M.~J., Kulkarni, S.~R., Bellm, E.~C., {et~al.} 2019, Publications of
  the Astronomical Society of the Pacific, 131, 078001

\bibitem[{Grieb {et~al.}(2017)Grieb, S{\'a}nchez, {Salazar-Albornoz},
  Scoccimarro, Crocce, Dalla~Vecchia, Montesano, {Gil-Mar{\'i}n}, Ross,
  Beutler, {Rodr{\'i}guez-Torres}, Chuang, Prada, Kitaura, Cuesta, Eisenstein,
  Percival, {Vargas-Maga{\~n}a}, Tinker, Tojeiro, Brownstein, Maraston, Nichol,
  Olmstead, Samushia, Seo, Streblyanska, \&
  Zhao}]{griebClusteringGalaxiesCompleted2017}
Grieb, J.~N., S{\'a}nchez, A.~G., {Salazar-Albornoz}, S., {et~al.} 2017,
  Monthly Notices of the Royal Astronomical Society, 467, 2085

\bibitem[{Guy {et~al.}(2007)Guy, Astier, Baumont, Hardin, Pain, Regnault, Basa,
  Carlberg, Conley, Fabbro, Fouchez, Hook, Howell, Perrett, Pritchet, Rich,
  Sullivan, Antilogus, Aubourg, Bazin, Bronder, Filiol,
  {Palanque-Delabrouille}, Ripoche, \&
  {Ruhlmann-Kleider}}]{guySALT2UsingDistant2007}
Guy, J., Astier, P., Baumont, S., {et~al.} 2007, Astronomy and Astrophysics,
  466, 11

\bibitem[{Guy {et~al.}(2010)Guy, Sullivan, Conley, Regnault, Astier, Balland,
  Basa, Carlberg, Fouchez, Hardin, Hook, Howell, Pain, {Palanque-Delabrouille},
  Perrett, Pritchet, Rich, {Ruhlmann-Kleider}, Balam, Baumont, Ellis, Fabbro,
  Fakhouri, Fourmanoit, {Gonz{\'a}lez-Gait{\'a}n}, Graham, Hsiao, Kronborg,
  Lidman, Mourao, Perlmutter, Ripoche, Suzuki, \&
  Walker}]{guySupernovaLegacySurvey2010}
Guy, J., Sullivan, M., Conley, A., {et~al.} 2010, Astronomy and Astrophysics,
  523, 7

\bibitem[{Guzzo {et~al.}(2008)Guzzo, Pierleoni, Meneux, Branchini, F{\`e}vre,
  Marinoni, Garilli, Blaizot, Lucia, Pollo, McCracken, Bottini, Brun, Maccagni,
  Picat, Scaramella, Scodeggio, Tresse, Vettolani, Zanichelli, Adami, Arnouts,
  Bardelli, Bolzonella, Bongiorno, Cappi, Charlot, Ciliegi, Contini, Cucciati,
  de~la Torre, Dolag, Foucaud, Franzetti, Gavignaud, Ilbert, Iovino,
  Lamareille, Marano, Mazure, Memeo, Merighi, Moscardini, Paltani, Pell{\`o},
  {Perez-Montero}, Pozzetti, Radovich, Vergani, Zamorani, \&
  Zucca}]{guzzoTestNatureCosmic2008}
Guzzo, L., Pierleoni, M., Meneux, B., {et~al.} 2008, Nature, 451, 541

\bibitem[{Hahn {et~al.}(2022)Hahn, Wilson, {Ruiz-Macias}, Cole, Weinberg,
  Moustakas, Kremin, Tinker, Smith, Wechsler, Ahlen, Alam, Bailey, Brooks,
  Cooper, Davis, Dawson, Dey, Dey, Eftekharzadeh, Eisenstein, Fanning,
  {Forero-Romero}, Frenk, Gazta{\~n}aga, Gontcho, Guy, Honscheid, Ishak,
  Juneau, Kehoe, Kisner, Lan, Landriau, Guillou, Levi, Magneville, Martini,
  Meisner, Myers, Nie, Norberg, {Palanque-Delabrouille}, Percival, Poppett,
  Prada, Raichoor, Ross, Safonova, Saulder, Schlafly, Schlegel, {Sierra-Porta},
  Tarle, Weaver, Y{\`e}che, Zarrouk, Zhou, Zhou, \&
  Zou}]{hahnDESIBrightGalaxy2022}
Hahn, C., Wilson, M.~J., {Ruiz-Macias}, O., {et~al.} 2022, {{DESI Bright Galaxy
  Survey}}: {{Final Target Selection}}, {{Design}}, and {{Validation}}

\bibitem[{Heitmann {et~al.}(2019)Heitmann, Finkel, Pope, Morozov, Frontiere,
  Habib, Rangel, Uram, Korytov, Child, Flender, Insley, \&
  Rizzi}]{heitmannOuterRimSimulation2019}
Heitmann, K., Finkel, H., Pope, A., {et~al.} 2019, The Astrophysical Journal
  Supplement Series, 245, 16

\bibitem[{Hou {et~al.}(2021)Hou, S{\'a}nchez, Ross, Smith, Neveux, Bautista,
  Burtin, Zhao, Scoccimarro, Dawson, {de Mattia}, {de la Macorra}, {du Mas des
  Bourboux}, Eisenstein, {Gil-Mar{\'i}n}, Lyke, Mohammad, Mueller, Percival,
  Rossi, Vargas~Maga{\~n}a, Zarrouk, Zhao, Brinkmann, Brownstein, Chuang,
  Myers, Newman, Schneider, \& Vivek}]{houCompletedSDSSIVExtended2021}
Hou, J., S{\'a}nchez, A.~G., Ross, A.~J., {et~al.} 2021, Monthly Notices of the
  Royal Astronomical Society, 500, 1201

\bibitem[{Hou {et~al.}(2018)Hou, S{\'a}nchez, Scoccimarro, {Salazar-Albornoz},
  Burtin, {Gil-Mar{\'i}n}, Percival, Ruggeri, Zarrouk, Zhao, Bautista,
  Brinkmann, Brownstein, Dawson, Devi, Myers, Habib, Heitmann, Tojeiro, Rossi,
  Schneider, Seo, \& Wang}]{houClusteringSDSSIVExtended2018}
Hou, J., S{\'a}nchez, A.~G., Scoccimarro, R., {et~al.} 2018, Monthly Notices of
  the Royal Astronomical Society, 480, 2521

\bibitem[{Howlett {et~al.}(2017{\natexlab{a}})Howlett, Robotham, Lagos, \&
  Kim}]{howlettMeasuringGrowthRate2017}
Howlett, C., Robotham, A. S.~G., Lagos, C. D.~P., \& Kim, A.~G.
  2017{\natexlab{a}}, The Astrophysical Journal, 847, 128

\bibitem[{Howlett {et~al.}(2017{\natexlab{b}})Howlett, Robotham, Lagos, \&
  Kim}]{howlett_measuring_2017}
Howlett, C., Robotham, A. S.~G., Lagos, C. D.~P., \& Kim, A.~G.
  2017{\natexlab{b}}, The Astrophysical Journal, 847, 128

\bibitem[{Howlett {et~al.}(2015)Howlett, Ross, Samushia, Percival, \&
  Manera}]{howlettClusteringSDSSMain2015}
Howlett, C., Ross, A.~J., Samushia, L., Percival, W.~J., \& Manera, M. 2015,
  Monthly Notices of the Royal Astronomical Society, 449, 848

\bibitem[{Howlett {et~al.}(2022)Howlett, Said, Lucey, Colless, Qin, Lai, Tully,
  \& Davis}]{howlettSloanDigitalSky2022a}
Howlett, C., Said, K., Lucey, J.~R., {et~al.} 2022, Monthly Notices of the
  Royal Astronomical Society, 515, 953

\bibitem[{{Howlett} {et~al.}(2022){Howlett}, {Said}, {Lucey}, {Colless}, {Qin},
  {Lai}, {Tully}, \& {Davis}}]{howlett_sloan_2022}
{Howlett}, C., {Said}, K., {Lucey}, J.~R., {et~al.} 2022, \mnras, 515, 953

\bibitem[{Howlett {et~al.}(2017{\natexlab{c}})Howlett, {Staveley-Smith}, Elahi,
  Hong, Jarrett, Jones, Koribalski, Macri, Masters, \&
  Springob}]{howlett2MTFVIMeasuring2017}
Howlett, C., {Staveley-Smith}, L., Elahi, P.~J., {et~al.} 2017{\natexlab{c}},
  Monthly Notices of the Royal Astronomical Society, 471, 3135

\bibitem[{Howlett {et~al.}(2017{\natexlab{d}})Howlett, Staveley-Smith, Elahi,
  Hong, Jarrett, Jones, Koribalski, Macri, Masters, \&
  Springob}]{howlett_2mtf_2017}
Howlett, C., Staveley-Smith, L., Elahi, P.~J., {et~al.} 2017{\natexlab{d}},
  Monthly Notices of the Royal Astronomical Society, 471, 3135, arXiv:
  1706.05130

\bibitem[{Hui \& Greene(2006)}]{huiCorrelatedFluctuationsLuminosity2006}
Hui, L. \& Greene, P.~B. 2006, Physical Review D, 73, 123526

\bibitem[{Huterer {et~al.}(2017{\natexlab{a}})Huterer, Shafer, Scolnic, \&
  Schmidt}]{huterer_testing_2017}
Huterer, D., Shafer, D.~L., Scolnic, D., \& Schmidt, F. 2017{\natexlab{a}},
  Journal of Cosmology and Astroparticle Physics, 2017, 015, arXiv: 1611.09862

\bibitem[{Huterer {et~al.}(2017{\natexlab{b}})Huterer, Shafer, Scolnic, \&
  Schmidt}]{hutererTestingLCDMLowest2017}
Huterer, D., Shafer, D.~L., Scolnic, D.~M., \& Schmidt, F. 2017{\natexlab{b}},
  Journal of Cosmology and Astroparticle Physics, 05, 015

\bibitem[{Johnson {et~al.}(2014)Johnson, Blake, Koda, Ma, Colless, Crocce,
  Davis, Jones, Magoulas, Lucey, Mould, Scrimgeour, \&
  Springob}]{johnson6dFGalaxySurvey2014}
Johnson, A., Blake, C., Koda, J., {et~al.} 2014, Monthly Notices of the Royal
  Astronomical Society, 444, 3926

\bibitem[{Kessler \& Scolnic(2017)}]{kesslerCorrectingTypeIa2017}
Kessler, R. \& Scolnic, D. 2017, The Astrophysical Journal, 836, 56

\bibitem[{Kim \& Linder(2020)}]{kimComplementarityPeculiarVelocity2020}
Kim, A.~G. \& Linder, E.~V. 2020, Physical Review D, 101, 023516

\bibitem[{Koda {et~al.}(2014)Koda, Blake, Davis, Magoulas, Springob,
  Scrimgeour, Johnson, Poole, \&
  {Staveley-Smith}}]{kodaArePeculiarVelocity2014}
Koda, J., Blake, C., Davis, T., {et~al.} 2014, Monthly Notices of the Royal
  Astronomical Society, 445, 4267

\bibitem[{Lai {et~al.}(2023)Lai, Howlett, \&
  Davis}]{laiUsingPeculiarVelocity2023}
Lai, Y., Howlett, C., \& Davis, T.~M. 2023, Monthly Notices of the Royal
  Astronomical Society, 518, 1840

\bibitem[{Lewis {et~al.}(2000)Lewis, Challinor, \&
  Lasenby}]{lewisEfficientComputationCosmic2000}
Lewis, A., Challinor, A., \& Lasenby, A. 2000, The Astrophysical Journal, 538,
  473

\bibitem[{{LSST Science Collaboration} {et~al.}(2009){LSST Science
  Collaboration}, Abell, Allison, Anderson, Andrew, Angel, Armus, Arnett,
  Asztalos, Axelrod, Bailey, Ballantyne, Bankert, Barkhouse, Barr, Barrientos,
  Barth, Bartlett, Becker, Becla, Beers, Bernstein, Biswas, Blanton, Bloom,
  Bochanski, Boeshaar, Borne, Bradac, Brandt, Bridge, Brown, Brunner, Bullock,
  Burgasser, Burge, Burke, Cargile, Chandrasekharan, Chartas, Chesley, Chu,
  Cinabro, Claire, Claver, Clowe, Connolly, Cook, Cooke, Cooray, Covey,
  Culliton, {de Jong}, {de Vries}, Debattista, Delgado, Dell'Antonio, Dhital,
  Di~Stefano, Dickinson, Dilday, Djorgovski, Dobler, Donalek,
  {Dubois-Felsmann}, Durech, Eliasdottir, Eracleous, Eyer, Falco, Fan,
  Fassnacht, Ferguson, Fernandez, Fields, Finkbeiner, Figueroa, Fox, Francke,
  Frank, Frieman, Fromenteau, Furqan, Galaz, {Gal-Yam}, Garnavich, Gawiser,
  Geary, Gee, Gibson, Gilmore, Grace, Green, Gressler, Grillmair, Habib,
  Haggerty, Hamuy, Harris, Hawley, Heavens, Hebb, Henry, Hileman, Hilton,
  Hoadley, Holberg, Holman, Howell, Infante, Ivezic, Jacoby, Jain, {R},
  {Jedicke}, Jee, Garrett~Jernigan, Jha, Johnston, Jones, Juric, Kaasalainen,
  {Styliani}, {Kafka}, Kahn, Kaib, Kalirai, Kantor, Kasliwal, Keeton, Kessler,
  Knezevic, Kowalski, Krabbendam, Krughoff, Kulkarni, Kuhlman, Lacy, Lepine,
  Liang, Lien, Lira, Long, Lorenz, Lotz, Lupton, Lutz, Macri, Mahabal,
  Mandelbaum, Marshall, May, McGehee, Meadows, Meert, Milani, Miller, Miller,
  Mills, Minniti, Monet, Mukadam, Nakar, Neill, Newman, Nikolaev, Nordby,
  O'Connor, Oguri, Oliver, Olivier, Olsen, Olsen, Olszewski, Oluseyi, Padilla,
  Parker, Pepper, Peterson, Petry, Pinto, Pizagno, Popescu, Prsa, Radcka,
  Raddick, Rasmussen, Rau, Rho, Rhoads, Richards, Ridgway, Robertson, Roskar,
  Saha, Sarajedini, Scannapieco, Schalk, Schindler, Schmidt, Schmidt,
  Schneider, Schumacher, Scranton, Sebag, Seppala, Shemmer, Simon, Sivertz,
  Smith, Allyn~Smith, Smith, Spitz, Stanford, Stassun, Strader, Strauss,
  Stubbs, Sweeney, Szalay, Szkody, Takada, Thorman, Trilling, Trimble, Tyson,
  Van~Berg, Vanden~Berk, VanderPlas, Verde, Vrsnak, Walkowicz, Wandelt, Wang,
  Wang, Warner, Wechsler, West, Wiecha, Williams, Willman, Wittman, Wolff,
  {Wood-Vasey}, Wozniak, Young, Zentner, \&
  Zhan}]{lsstsciencecollaborationLSSTScienceBook2009}
{LSST Science Collaboration}, Abell, P.~A., Allison, J., {et~al.} 2009, {{LSST
  Science Book}}, {{Version}} 2.0

\bibitem[{Lyall {et~al.}(2022)Lyall, Blake, Turner, Ruggeri, \&
  Winther}]{lyallTestingModifiedGravity2022}
Lyall, S., Blake, C., Turner, R., Ruggeri, R., \& Winther, H. 2022, Testing
  Modified Gravity Scenarios with Direct Peculiar Velocities

\bibitem[{Ma {et~al.}(2011)Ma, Gordon, \&
  Feldman}]{maPeculiarVelocityField2011}
Ma, Y.-Z., Gordon, C., \& Feldman, H.~A. 2011, Physical Review D, 83, 103002

\bibitem[{Masci {et~al.}(2019)Masci, Laher, Rusholme, Shupe, Groom, Surace,
  Jackson, Monkewitz, Beck, Flynn, Terek, Landry, Hacopians, Desai, Howell,
  Brooke, Imel, Wachter, Ye, Lin, Cenko, Cunningham, Rebbapragada, Bue, Miller,
  Mahabal, Bellm, Patterson, Juri{\'c}, Golkhou, Ofek, Walters, Graham,
  Kasliwal, Dekany, Kupfer, Burdge, Cannella, Barlow, Van~Sistine, Giomi,
  Fremling, Blagorodnova, Levitan, Riddle, Smith, Helou, Prince, \&
  Kulkarni}]{masciZwickyTransientFacility2019}
Masci, F.~J., Laher, R.~R., Rusholme, B., {et~al.} 2019, Publications of the
  Astronomical Society of the Pacific, 131, 018003

\bibitem[{Neveux {et~al.}(2020)Neveux, Burtin, {de Mattia}, Smith, Ross, Hou,
  Bautista, Brinkmann, Chuang, Dawson, {Gil-Mar{\'i}n}, Lyke, {de la Macorra},
  {du Mas des Bourboux}, Mohammad, M{\"u}ller, Myers, Newman, Percival, Rossi,
  Schneider, Vivek, Zarrouk, Zhao, \& Zhao}]{neveuxCompletedSDSSIVExtended2020}
Neveux, R., Burtin, E., {de Mattia}, A., {et~al.} 2020, Monthly Notices of the
  Royal Astronomical Society, 499, 210

\bibitem[{Nicolas {et~al.}(2021)Nicolas, Rigault, Copin, Graziani, Aldering,
  Briday, Kim, Nordin, Perlmutter, \&
  Smith}]{nicolasRedshiftEvolutionUnderlying2021}
Nicolas, N., Rigault, M., Copin, Y., {et~al.} 2021, Astronomy and Astrophysics,
  649, A74

\bibitem[{Nusser(2017)}]{nusserVelocitydensityCorrelationsCosmicflows32017}
Nusser, A. 2017, Monthly Notices of the Royal Astronomical Society, 470, 445

\bibitem[{Okumura {et~al.}(2016)Okumura, Hikage, Totani, Tonegawa, Okada,
  Glazebrook, Blake, Ferreira, More, Taruya, Tsujikawa, Akiyama, Dalton, Goto,
  Ishikawa, Iwamuro, Matsubara, Nishimichi, Ohta, Shimizu, Takahashi, Takato,
  Tamura, Yabe, \& Yoshida}]{okumuraSubaruFMOSGalaxy2016}
Okumura, T., Hikage, C., Totani, T., {et~al.} 2016, Publications of the
  Astronomical Society of Japan, 68

\bibitem[{Perley {et~al.}(2020)Perley, Fremling, Sollerman, Miller, Dahiwale,
  Sharma, Bellm, Biswas, Brink, Bruch, De, Dekany, Drake, Duev, Filippenko,
  {Gal-Yam}, Goobar, Graham, Graham, Ho, Irani, Kasliwal, Kim, Kulkarni,
  Mahabal, Masci, Modak, Neill, Nordin, Riddle, Soumagnac, Strotjohann,
  Schulze, Taggart, Tzanidakis, Walters, \&
  Yan}]{perleyZwickyTransientFacility2020}
Perley, D.~A., Fremling, C., Sollerman, J., {et~al.} 2020, The Astrophysical
  Journal, 904, 35

\bibitem[{Pezzotta {et~al.}(2017)Pezzotta, {de la Torre}, Bel, Granett, Guzzo,
  Peacock, Garilli, Scodeggio, Bolzonella, Abbas, Adami, Bottini, Cappi,
  Cucciati, Davidzon, Franzetti, Fritz, Iovino, Krywult, Le~Brun, Le~F{\`e}vre,
  Maccagni, Ma{\l}ek, Marulli, Polletta, Pollo, Tasca, Tojeiro, Vergani,
  Zanichelli, Arnouts, Branchini, Coupon, De~Lucia, Koda, Ilbert, Mohammad,
  Moutard, \& Moscardini}]{pezzottaVIMOSPublicExtragalactic2017}
Pezzotta, A., {de la Torre}, S., Bel, J., {et~al.} 2017, Astronomy and
  Astrophysics, 604, A33

\bibitem[{{Planck Collaboration} {et~al.}(2020{\natexlab{a}}){Planck
  Collaboration}, Aghanim, Akrami, Arroja, Ashdown, Aumont, Baccigalupi,
  Ballardini, Banday, Barreiro, Bartolo, Basak, Battye, Benabed, Bernard,
  Bersanelli, Bielewicz, Bock, Bond, Borrill, Bouchet, Boulanger, Bucher,
  Burigana, Butler, Calabrese, Cardoso, Carron, Casaponsa, Challinor, Chiang,
  Colombo, Combet, Contreras, Crill, Cuttaia, {de Bernardis}, {de Zotti},
  Delabrouille, Delouis, D{\'e}sert, Di~Valentino, Dickinson, Diego, Donzelli,
  Dor{\'e}, Douspis, Ducout, Dupac, Efstathiou, Elsner, En{\ss}lin, Eriksen,
  Falgarone, Fantaye, Fergusson, {Fernandez-Cobos}, Finelli, Forastieri,
  Frailis, Franceschi, Frolov, Galeotta, Galli, Ganga, {G{\'e}nova-Santos},
  Gerbino, Ghosh, {Gonz{\'a}lez-Nuevo}, G{\'o}rski, Gratton, Gruppuso,
  Gudmundsson, Hamann, Handley, Hansen, Helou, Herranz, Hildebrandt, Hivon,
  Huang, Jaffe, Jones, Karakci, Keih{\"a}nen, Keskitalo, Kiiveri, Kim, Kisner,
  Knox, Krachmalnicoff, Kunz, {Kurki-Suonio}, Lagache, Lamarre, Langer,
  Lasenby, Lattanzi, Lawrence, Le~Jeune, Leahy, Lesgourgues, Levrier, Lewis,
  Liguori, Lilje, Lilley, Lindholm, {L{\'o}pez-Caniego}, Lubin, Ma,
  {Mac{\'i}as-P{\'e}rez}, Maggio, Maino, Mandolesi, Mangilli,
  {Marcos-Caballero}, Maris, Martin, Martinelli, {Mart{\'i}nez-Gonz{\'a}lez},
  Matarrese, Mauri, McEwen, Meerburg, Meinhold, Melchiorri, Mennella,
  Migliaccio, Millea, Mitra, {Miville-Desch{\^e}nes}, Molinari, Moneti,
  Montier, Morgante, Moss, Mottet, M{\"u}nchmeyer, Natoli,
  {N{\o}rgaard-Nielsen}, Oxborrow, Pagano, Paoletti, Partridge, Patanchon,
  Pearson, Peel, Peiris, Perrotta, Pettorino, Piacentini, Polastri, Polenta,
  Puget, Rachen, Reinecke, Remazeilles, Renault, Renzi, Rocha, Rosset, Roudier,
  {Rubi{\~n}o-Mart{\'i}n}, {Ruiz-Granados}, Salvati, Sandri, Savelainen, Scott,
  Shellard, Shiraishi, Sirignano, Sirri, Spencer, Sunyaev, {Suur-Uski}, Tauber,
  Tavagnacco, Tenti, Terenzi, Toffolatti, Tomasi, Trombetti, Valiviita,
  Van~Tent, Vibert, Vielva, Villa, Vittorio, Wandelt, Wehus, White, White,
  Zacchei, \& Zonca}]{planckcollaborationPlanck2018Results2020a}
{Planck Collaboration}, Aghanim, N., Akrami, Y., {et~al.} 2020{\natexlab{a}},
  Astronomy and Astrophysics, 641, A1

\bibitem[{{Planck Collaboration} {et~al.}(2020{\natexlab{b}}){Planck
  Collaboration}, Aghanim, Akrami, Ashdown, Aumont, Baccigalupi, Ballardini,
  Banday, Barreiro, Bartolo, Basak, Battye, Benabed, Bernard, Bersanelli,
  Bielewicz, Bock, Bond, Borrill, Bouchet, Boulanger, Bucher, Burigana, Butler,
  Calabrese, Cardoso, Carron, Challinor, Chiang, Chluba, Colombo, Combet,
  Contreras, Crill, Cuttaia, {de Bernardis}, {de Zotti}, Delabrouille, Delouis,
  Di~Valentino, Diego, Dor{\'e}, Douspis, Ducout, Dupac, Dusini, Efstathiou,
  Elsner, En{\ss}lin, Eriksen, Fantaye, Farhang, Fergusson, {Fernandez-Cobos},
  Finelli, Forastieri, Frailis, Fraisse, Franceschi, Frolov, Galeotta, Galli,
  Ganga, {G{\'e}nova-Santos}, Gerbino, Ghosh, {Gonz{\'a}lez-Nuevo}, G{\'o}rski,
  Gratton, Gruppuso, Gudmundsson, Hamann, Handley, Hansen, Herranz,
  Hildebrandt, Hivon, Huang, Jaffe, Jones, Karakci, Keih{\"a}nen, Keskitalo,
  Kiiveri, Kim, Kisner, Knox, Krachmalnicoff, Kunz, {Kurki-Suonio}, Lagache,
  Lamarre, Lasenby, Lattanzi, Lawrence, Le~Jeune, Lemos, Lesgourgues, Levrier,
  Lewis, Liguori, Lilje, Lilley, Lindholm, {L{\'o}pez-Caniego}, Lubin, Ma,
  {Mac{\'i}as-P{\'e}rez}, Maggio, Maino, Mandolesi, Mangilli,
  {Marcos-Caballero}, Maris, Martin, Martinelli, {Mart{\'i}nez-Gonz{\'a}lez},
  Matarrese, Mauri, McEwen, Meinhold, Melchiorri, Mennella, Migliaccio, Millea,
  Mitra, {Miville-Desch{\^e}nes}, Molinari, Montier, Morgante, Moss, Natoli,
  {N{\o}rgaard-Nielsen}, Pagano, Paoletti, Partridge, Patanchon, Peiris,
  Perrotta, Pettorino, Piacentini, Polastri, Polenta, Puget, Rachen, Reinecke,
  Remazeilles, Renzi, Rocha, Rosset, Roudier, {Rubi{\~n}o-Mart{\'i}n},
  {Ruiz-Granados}, Salvati, Sandri, Savelainen, Scott, Shellard, Sirignano,
  Sirri, Spencer, Sunyaev, {Suur-Uski}, Tauber, Tavagnacco, Tenti, Toffolatti,
  Tomasi, Trombetti, Valenziano, Valiviita, Van~Tent, Vibert, Vielva, Villa,
  Vittorio, Wandelt, Wehus, White, White, Zacchei, \&
  Zonca}]{planckcollaborationPlanck2018Results2020}
{Planck Collaboration}, Aghanim, N., Akrami, Y., {et~al.} 2020{\natexlab{b}},
  Astronomy and Astrophysics, 641, A6

\bibitem[{{Prideaux-Ghee} {et~al.}(2023){Prideaux-Ghee}, Leclercq, Lavaux,
  Heavens, \& Jasche}]{prideaux-gheeFieldbasedPhysicalInference2023}
{Prideaux-Ghee}, J., Leclercq, F., Lavaux, G., Heavens, A., \& Jasche, J. 2023,
  Monthly Notices of the Royal Astronomical Society, 518, 4191

\bibitem[{Qin {et~al.}(2019)Qin, Howlett, \&
  {Staveley-Smith}}]{qinRedshiftspaceMomentumPower2019}
Qin, F., Howlett, C., \& {Staveley-Smith}, L. 2019, Monthly Notices of the
  Royal Astronomical Society, 487, 5235

\bibitem[{Rossi {et~al.}(2021)Rossi, Choi, Moon, Bautista, {Gil-Marin}, Paviot,
  {Vargas-Magana}, {de la Torre}, Fromenteau, Ross, Avila, Burtin, Dawson,
  Escoffier, Habib, Heitmann, Hou, Mueller, Percival, Smith, Zhao, \&
  Zhao}]{rossiCompletedSDSSIVExtended2021}
Rossi, G., Choi, P.~D., Moon, J., {et~al.} 2021, Monthly Notices of the Royal
  Astronomical Society, 505, 377

\bibitem[{Said {et~al.}(2020)Said, Colless, Magoulas, Lucey, \&
  Hudson}]{saidJointAnalysis6dFGS2020}
Said, K., Colless, M., Magoulas, C., Lucey, J.~R., \& Hudson, M.~J. 2020,
  Monthly Notices of the Royal Astronomical Society, 497, 1275

\bibitem[{Samushia {et~al.}(2012)Samushia, Percival, \&
  Raccanelli}]{samushiaInterpretingLargescaleRedshiftspace2012}
Samushia, L., Percival, W.~J., \& Raccanelli, A. 2012, Monthly Notices of the
  Royal Astronomical Society, 420, 2102

\bibitem[{S{\'a}nchez {et~al.}(2017)S{\'a}nchez, Scoccimarro, Crocce, Grieb,
  {Salazar-Albornoz}, Dalla~Vecchia, Lippich, Beutler, Brownstein, Chuang,
  Eisenstein, Kitaura, Olmstead, Percival, Prada, {Rodr{\'i}guez-Torres}, Ross,
  Samushia, Seo, Tinker, Tojeiro, {Vargas-Maga{\~n}a}, Wang, \&
  Zhao}]{sanchezClusteringGalaxiesCompleted2017}
S{\'a}nchez, A.~G., Scoccimarro, R., Crocce, M., {et~al.} 2017, Monthly Notices
  of the Royal Astronomical Society, 464, 1640

\bibitem[{Satpathy {et~al.}(2017)Satpathy, Alam, Ho, White, Bahcall, Beutler,
  Brownstein, Chuang, Eisenstein, Grieb, Kitaura, Olmstead, Percival,
  {Salazar-Albornoz}, S{\'a}nchez, Seo, Thomas, Tinker, \&
  Tojeiro}]{satpathyClusteringGalaxiesCompleted2017}
Satpathy, S., Alam, S., Ho, S., {et~al.} 2017, Monthly Notices of the Royal
  Astronomical Society, 469, 1369

\bibitem[{Schlegel {et~al.}(1998)Schlegel, Finkbeiner, \&
  Davis}]{schlegelMapsDustInfrared1998}
Schlegel, D.~J., Finkbeiner, D.~P., \& Davis, M. 1998, The Astrophysical
  Journal, 500, 525

\bibitem[{Scolnic {et~al.}(2022)Scolnic, Brout, Carr, Riess, Davis, Dwomoh,
  Jones, Ali, Charvu, Chen, Peterson, Popovic, Rose, Wood, Brown, Chambers,
  Coulter, Dettman, Dimitriadis, Filippenko, Foley, Jha, Kilpatrick, Kirshner,
  Pan, Rest, {Rojas-Bravo}, Siebert, Stahl, \&
  Zheng}]{scolnicPantheonAnalysisFull2022}
Scolnic, D., Brout, D., Carr, A., {et~al.} 2022, The Astrophysical Journal,
  938, 113

\bibitem[{Scolnic \& Kessler(2016)}]{scolnicMEASURINGTYPEIA2016}
Scolnic, D. \& Kessler, R. 2016, The Astrophysical Journal, 822, L35

\bibitem[{Smith {et~al.}(2020)Smith, Burtin, Hou, Neveux, Ross, Alam,
  Brinkmann, Dawson, Habib, Heitmann, Kneib, Lyke, {du Mas des Bourboux},
  Mueller, Myers, Percival, Rossi, Schneider, Zarrouk, \&
  Zhao}]{smithCompletedSDSSIVExtended2020}
Smith, A., Burtin, E., Hou, J., {et~al.} 2020, Monthly Notices of the Royal
  Astronomical Society, 499, 269

\bibitem[{Song \& Percival(2009)}]{songReconstructingHistoryStructure2009}
Song, Y.-S. \& Percival, W.~J. 2009, Journal of Cosmology and Astroparticle
  Physics, 2009, 004

\bibitem[{Strauss \& Willick(1995)}]{straussDensityPeculiarVelocity1995}
Strauss, M.~A. \& Willick, J.~A. 1995, Physics Reports, 261, 271

\bibitem[{Tamone {et~al.}(2020)Tamone, Raichoor, Zhao, {de Mattia}, Gorgoni,
  Burtin, {Ruhlmann-Kleider}, Ross, Alam, Percival, Avila, Chapman, Chuang,
  Comparat, Dawson, {de la Torre}, des Bourboux, Escoffier, {Gonzalez-Perez},
  Hou, Kneib, Mohammad, Mueller, Paviot, Rossi, Schneider, Wang, \&
  Zhao}]{tamoneCompletedSDSSIVExtended2020}
Tamone, A., Raichoor, A., Zhao, C., {et~al.} 2020, Monthly Notices of the Royal
  Astronomical Society

\bibitem[{Taruya {et~al.}(2012)Taruya, Bernardeau, Nishimichi, \&
  Codis}]{taruya_regpt_2012}
Taruya, A., Bernardeau, F., Nishimichi, T., \& Codis, S. 2012, Physical Review
  D, 86, 103528, arXiv: 1208.1191

\bibitem[{{Tripp}(1998)}]{tripptwopar1998}
{Tripp}, R. 1998, \aap, 331, 815

\bibitem[{Tully \& Fisher(1977)}]{tullyNewMethodDetermining1977}
Tully, R.~B. \& Fisher, J.~R. 1977, Astronomy and Astrophysics, 54, 661

\bibitem[{{Tully} {et~al.}(2022){Tully}, {Kourkchi}, {Courtois}, {Anand},
  {Blakeslee}, {Brout}, {de Jaeger}, {Dupuy}, {Guinet}, {Howlett}, {Jensen},
  {Pomar{\`e}de}, {Rizzi}, {Rubin}, {Said}, {Scolnic}, \&
  {Stahl}}]{tully_cosmicflows4_2022}
{Tully}, R.~B., {Kourkchi}, E., {Courtois}, H.~M., {et~al.} 2022, arXiv
  e-prints, arXiv:2209.11238

\bibitem[{Turner {et~al.}(2022)Turner, Blake, \&
  Ruggeri}]{turnerLocalMeasurementGrowth2022}
Turner, R.~J., Blake, C., \& Ruggeri, R. 2022, A Local Measurement of the
  Growth Rate from Peculiar Velocities and Galaxy Clustering Correlations in
  the {{6dF Galaxy Survey}}

\bibitem[{Watkins \& Feldman(2015)}]{watkins_unbiased_2015}
Watkins, R. \& Feldman, H.~A. 2015, Monthly Notices of the Royal Astronomical
  Society, 450, 1868

\bibitem[{Zarrouk {et~al.}(2018)Zarrouk, Burtin, {Gil-Mar{\'i}n}, Ross,
  Tojeiro, P{\^a}ris, Dawson, Myers, Percival, Chuang, Zhao, Bautista,
  Comparat, {Gonz{\'a}lez-P{\'e}rez}, Habib, Heitmann, Hou, Laurent, Le~Goff,
  Prada, {Rodr{\'i}guez-Torres}, Rossi, Ruggeri, S{\'a}nchez, Schneider,
  Tinker, Wang, Y{\`e}che, Baumgarten, Brownstein, {de la Torre}, {du Mas des
  Bourboux}, Kneib, Mariappan, {Palanque-Delabrouille}, Peacock, Petitjean,
  Seo, \& Zhao}]{zarroukClusteringSDSSIVExtended2018}
Zarrouk, P., Burtin, E., {Gil-Mar{\'i}n}, H., {et~al.} 2018, Monthly Notices of
  the Royal Astronomical Society, 477, 1639

\bibitem[{Zhai {et~al.}(2017)Zhai, Blanton, Slosar, \&
  Tinker}]{zhaiEvaluationCosmologicalModels2017}
Zhai, Z., Blanton, M., Slosar, A., \& Tinker, J. 2017, The Astrophysical
  Journal, 850, 183

\bibitem[{Zonca {et~al.}(2019)Zonca, Singer, Lenz, Reinecke, Rosset, Hivon, \&
  Gorski}]{zoncaHealpyEqualArea2019}
Zonca, A., Singer, L., Lenz, D., {et~al.} 2019, The Journal of Open Source
  Software, 4, 1298

\end{thebibliography}

\begin{appendix} 

\section{Peculiar velocity estimators}
\label{ap:vestimator}
    \subsection{Derivation of the peculiar velocity estimator}
    \label{ap:vestimator:derivation}
   Hubble residuals are given as
    \begin{equation}
        \Delta \mu = \mu_{\rm obs} - \mu_{\rm model}(z_\mathrm{obs}),
    \end{equation}
    where $\mu_\mathrm{model}$ expression is
    \begin{equation}
        \mu_\mathrm{model}(z) = 5\log\left(d_{L, \mathrm{model}}(z)\right) = 5 \log\left((1 + z) r(z)\right),
    \end{equation}
    with $r(z)$ the comoving distance. 
    In the residuals, $\mu_\mathrm{model}(z)$ is evaluated at $z = z_\mathrm{obs} = (1 + z_p) (1 + z_\mathrm{cos}) - 1$.
    
  The first-order Taylor's expansion of $d_{L, \mathrm{model}}(z_\mathrm{obs}) $ with respect to $z_p$ is
    \begin{equation}
        d_{L, \mathrm{model}}(z_\mathrm{obs}) 
        \simeq d_{L, \mathrm{model}}(z_\mathrm{cos}) + \left.\frac{\partial z}{\partial z_p}\frac{\partial d_{L, \mathrm{th}}}{\partial z}\right|_{z_p = 0} z_p.
    \label{eq:ap1:taylor1}
    \end{equation}
    We can develop the second term of 
    \eqref{eq:ap1:taylor1}
    \begin{align}
         \left.\frac{\partial z}{\partial z_p} \frac{\partial d_{L, \mathrm{th}}}{\partial z}\right|_{z_p = 0} &\simeq (1 + z_\mathrm{cos}) \notag\\
         &\phantom{{}\simeq} \times\frac{\partial}{\partial z}\left.\left[(1 + z)\frac{c}{H_0}\int_0^{z}\frac{dz'}{E(z')}\right]\right|_{z_p=0}\notag\\
        &\simeq (1 + z_\mathrm{cos}) \notag\\
        &\phantom{{}\simeq}\times \left(\frac{c}{H_0}\int_0^{z_\mathrm{cos}}\frac{dz'}{E(z')} + \frac{c(1 + z_\mathrm{cos})}{H_0E(z_\mathrm{cos})}\right)\notag\\
        &\simeq (1 + z_\mathrm{cos})\left(r(z_\mathrm{cos}) + \frac{c(1 + z_\mathrm{cos})^2}{H(z_\mathrm{cos})}\right)\notag\\
        &\simeq d_{L, \mathrm{model}}(z_\mathrm{cos}) \left( 1  + \frac{c(1 + z_\mathrm{cos})}{r(z_\mathrm{cos})H(z_\mathrm{cos})}\right).
        \label{eq:ap1:taylor2}
    \end{align}
    Injecting \eqref{eq:ap1:taylor2} in \eqref{eq:ap1:taylor1} we obtain
    \begin{align*}
        d_{L, \mathrm{model}}(z_\mathrm{obs})&\simeq  d_{L, \mathrm{model}}(z_\mathrm{cos}) \\ 
        &\phantom{{}\simeq}\times\left[1 + \left( 1  + \frac{c(1 + z_\mathrm{cos})}{r(z_\mathrm{cos})H(z_\mathrm{cos})}\right)z_p \right],
    \end{align*}
      where $d_{L, \mathrm{model}}(z_\mathrm{cos})$ can be replaced by $\frac{d_{L, \mathrm{obs}}}{(1 + z_p)^2}$ to give
    \begin{align}
         d_{L, \mathrm{model}}(z_\mathrm{obs})&\simeq \frac{d_{L, \mathrm{obs}}}{(1 + z_p)^2}\left[1 + \left( 1  + \frac{c(1 + z_\mathrm{cos})}{r(z_\mathrm{cos})H(z_\mathrm{cos})}\right)z_p \right]\notag\\
        &\simeq d_{L, \mathrm{obs}}\left[1 + \left(\frac{c(1 + z_\mathrm{cos})}{r(z_\mathrm{cos})H(z_\mathrm{cos})} - 1\right)z_p \right].
        \label{eq:ap1:taylor4}
    \end{align}
    With \eqref{eq:ap1:taylor4} we can write the relative variation of luminosity distances
    \begin{equation}
        \delta{d_L} = \frac{d_{L,\mathrm{obs}} - d_{L, \mathrm{model}}(z_\mathrm{obs})}{d_{L, \mathrm{model}}(z_\mathrm{obs})} \simeq \left(1- \frac{c(1 + z_\mathrm{cos})}{r(z_\mathrm{cos})H(z_\mathrm{cos})}\right)z_p.
        \label{eq:ap1:taylor6} 
    \end{equation} 
    This last equation is equivalent to the Eq. 15 from \cite{huiCorrelatedFluctuationsLuminosity2006}.
    Using $\Delta\mu\simeq \frac{5}{\ln(10)} \delta d_L$ and $v_p \simeq c z_p$ we finally get 
    \begin{equation}
        v_p \simeq -\frac{\ln(10)c}{5} \left(\frac{c(1 + z_\mathrm{cos})}{r(z_\mathrm{cos})H(z_\mathrm{cos})} - 1\right)^{-1}\Delta\mu .
    \end{equation}
    As a first order development, this derivation is valid for velocities such that $z_p$ is small compared to $z_\mathrm{cos}$. In Appendix~\ref{ap:vestimator:diffest} we discuss further approximations and biases of this estimator.
    \subsection{The different estimators and their bias}
    \label{ap:vestimator:diffest}
    
        In the literature we can find variants of the peculiar velocity estimator \eqref{eq:vest}:
        \begin{equation}
            \hat{v}_1 = - \frac{\ln(10) c}{5} \left(\frac{(1 + z_{\rm cos})c}{H(z_{\rm cos}) r(z_{\rm cos})} - 1 \right)^{-1} \Delta \mu.
        \label{eq:v1}
    \end{equation}
        Here we describe which approximations are made and their consequences on peculiar velocity estimation.
       From $\hat{v}_1$ we can make two approximations. Firstly, since $\frac{(1 + z_{\rm cos}) c}{H(z_{\rm cos})r(z_{\rm cos})} > 10$ for $z < 0.11$ at low-redshift we can make the approximation:
        \begin{equation}
            \hat{v}_2 = -\frac{\ln(10)c}{5} \frac{H(z_{\rm cos})r(z_{\rm cos})}{1 + z_{\rm cos}} \Delta \mu.
            \label{eq:Hest}
        \end{equation}
    The form \eqref{eq:Hest} is used in \cite{howlett_measuring_2017}, \cite{huterer_testing_2017} and \cite{laiUsingPeculiarVelocity2023}. 
        Secondly, at the low-redshift regime we can also approximate the Hubble law 
        \begin{equation}
            H(z_{\rm cos})r(z_{\rm cos}) \simeq cz_\mathrm{mod},
        \end{equation}
        where $cz_\mathrm{mod}$ is a development of the Hubble law  at a given order. Here, we use the first order development $cz_\mathrm{mod} = cz_\mathrm{cos}$\footnote{In the literature we can also find higher order developments of the Hubble law : $H(z_{\rm cos})r(z_{\rm cos}) \simeq cz_{\rm mod}= cz_{\rm cos}\left[1 + \frac{1}{2}(1-q_0)z_{\rm cos}-\frac{1}{6}(1-q_0-3q_0^2+j_0)z_{\rm cos}^2\right]$ where $q_0$ and $j_0$ are respectively the deceleration and jerk parameters.}, then \eqref{eq:vest} becomes:
         \begin{equation}
            \hat{v}_3 = -\frac{\ln(10)c}{5} \left(\frac{1 + z_{\rm cos}}{z_{\rm mod}} - 1 \right)^{-1} \Delta \mu .
        \end{equation}
        Using those approximations together, we obtain:
        \begin{equation}
            \hat{v}_4 = -\frac{\ln(10)c}{5} \frac{z_{\rm mod}}{1 + z_{\rm cos}} \Delta \mu .
        \end{equation}
        This last estimator corresponds to the Watkins estimator \cite{watkins_unbiased_2015}.
        
         \begin{figure*},
            \centering
            \includegraphics[width=\textwidth]{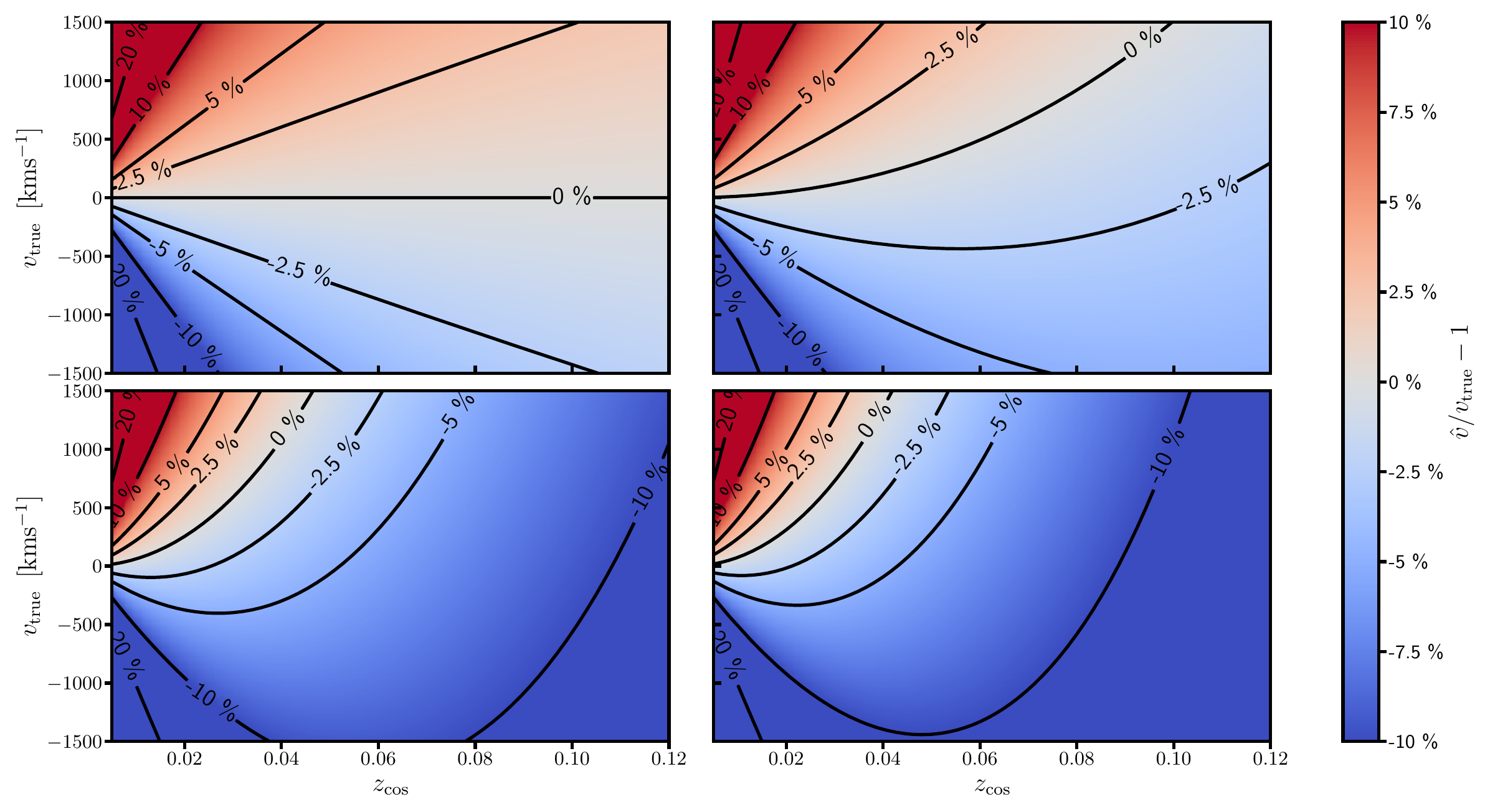}
            \caption{Bias on velocity estimator as a function of cosmological redshift ($z_\mathrm{cos} \in [0.005, 0.12]$) and true velocity. \textit{Upper left panel}: $\hat{v}_1$ estimator. \textit{Bottom left panel}: $\hat{v}_2$ estimator. \textit{Upper right panel}: $\hat{v}_3$ estimator. \textit{Bottom right panel}: $\hat{v}_4$ estimator. See discussion in Appendix~\ref{ap:vestimator:diffest}.}
            \label{fig:estimator}
        \end{figure*}
        
        As we have seen in Appendix~\ref{ap:vestimator:derivation}, the derivation of these estimators make the assumption that $z_p$ is small enough compared to $z_\mathrm{cos}$. This statement tends to be less valid at very low redshift when $z_\mathrm{cos}$ is of the same order of magnitude than $z_p$. Moreover we stated in Sect.~\ref{sec:methodology:data_vector:pecvel} that we can not have access to the cosmological redshift $z_{\rm cos}$, hence we need to evaluate our estimator using the observed redshift $z_{\rm obs}$. On Fig.~\ref{fig:estimator}, we see in the \textit{upper left panel}, that these approximations leads to a biased velocity estimation especially at very low redshift $z < 0.02$ where the bias magnitude for a velocity of $v \sim 300$ km/s is above $\sim 5\%$ for $\hat{v}_1$.
        \begin{figure}
            \includegraphics[width=0.5\textwidth]{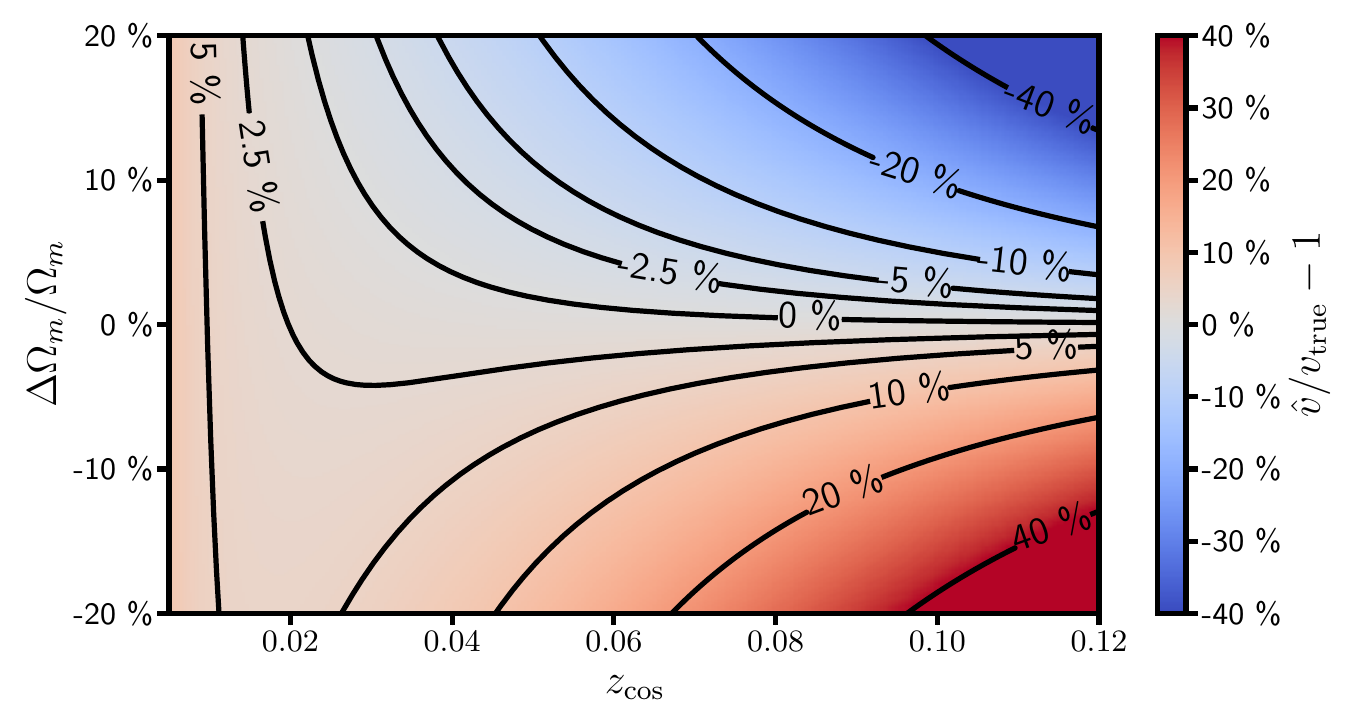}
            \caption{Bias on velotity estimator $\hat{v}_1$ as a function of cosmological redshift and $\Omega_m$ used in the estimator. The true velocity is fixed to $v_{\rm true} = 300\ {\rm km}.{\rm s}^{-1}$. See discussion in \ref{ap:vestimator:estcosmo}.}
            \label{fig:vbiascosmo}
        \end{figure}
        Comparing the estimators, we see that neglecting the "-1" term as in $\hat{v}_2$ and $\hat{v}_4$, leads to a global velocity underestimation when the redshift increases. Using the linear approximation of the Hubble law, used in $\hat{v}_3$, also tends to give underestimated velocities compared to using the real Hubble law. The first approximation is the dominant effect on the bias for the combined approximation of the $\hat{v}_4$ estimator.
        \FloatBarrier
        
        \subsection{Estimator dependence on $\Omega_m$ assumptions}
        \label{ap:vestimator:estcosmo}
        Using the Hubble law in our estimator, we have to fix a cosmology, $\Omega_m$ in Flat-$\Lambda$CDM. Using a $\Omega_m$ that differs from the true cosmology leads to a biased estimation of velocities. In Fig.~\ref{fig:vbiascosmo} we show the bias for a velocity fixed to $v_{\rm true} = 300 \ {\rm km}.{\rm s}^{-1}$ as a function of the $\Omega_m$ assumption and redshift $z_{\rm cos}$. As previously stated, biases at low-redshift are dominated by the loss of accuracy of the first order development when $z_\mathrm{cos} \sim z_p$. The bias due to the mis-estimation of $\Omega_m$ appears with increasing redshift. However with the current precision of $\Omega_m$ from \cite{planckcollaborationPlanck2018Results2020} of $\sim 2\%$ the bias stays below $5 \%$ for redshifts $z < 0.06$.
        \FloatBarrier

        \subsection{Estimators Gaussianity}
        \label{ap:vestimator:estgauss}
        The likelihood we used to estimate $\fsig$ assumes that the velocity estimator has a Gaussian distribution. We have to check that the estimator preserves the Gaussian form of the errors on $\Delta\mu$. In order to check this, we used a "toy model". We drew  N peculiar velocities from a normal distribution $v_p \sim \mathcal{N}(0, 300)$ as well as N cosmological redshifts. After computing $\mu_{\rm obs}$ using \eqref{eq:muobs} we add a Gaussian scattering with $\sigma_\mu \sim \mathcal{N}(0, 0.12)$. 
        
\begin{figure}
            \centering
            \includegraphics[width=\columnwidth]{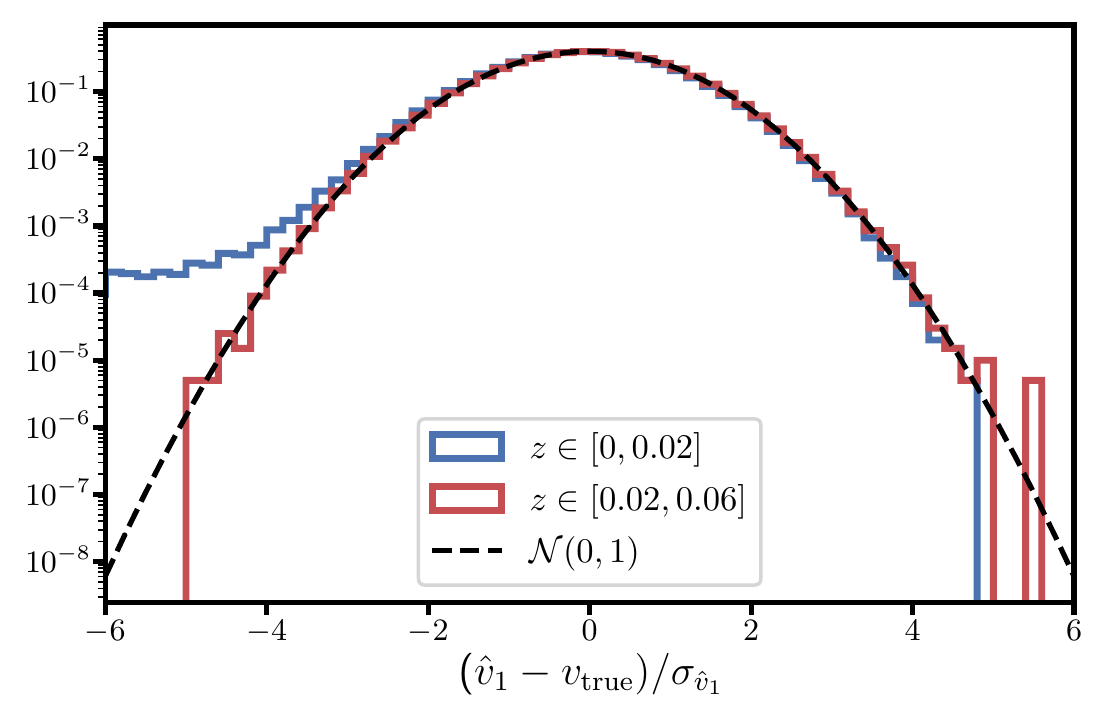} 
            \caption{Pull of peculiar velocity estimator for $\hat{v}_1$ on a "toy model" drawing $N = 10^6$ \sns. The blue line represents the pull for \sns\ in the redshift range $[0, 0.02]$ and the red line represents the pull for \sns\ in the redshift range $[0.02, 0.06]$. We see that for low redshift \sns\ the velocity estimator deviates from Gaussianity due to peculiar velocity contamination of $z_{\rm obs}$.
            }
            \label{fig:gauss_vest}
        \end{figure}

        In the \textit{top panel} of Fig.~\ref{fig:gauss_vest}, we can see that the pull of the velocity estimator $\hat{v}_1$ seems to preserve the Gaussian distribution of $\sigma_\mu$ for redshift range $z \in [0.02, 0.06]$.
        In the \textit{bottom panel} of Fig.~\ref{fig:gauss_vest}, we see that below $z < 0.02$ the velocity distribution deviates from Gaussianity due to more important effect from peculiar velocity redshift contamination.
        \FloatBarrier

\section{Power spectrum convergence}
\label{ap:pwconv}

\begin{figure}
\centering
    \includegraphics[width=\columnwidth]{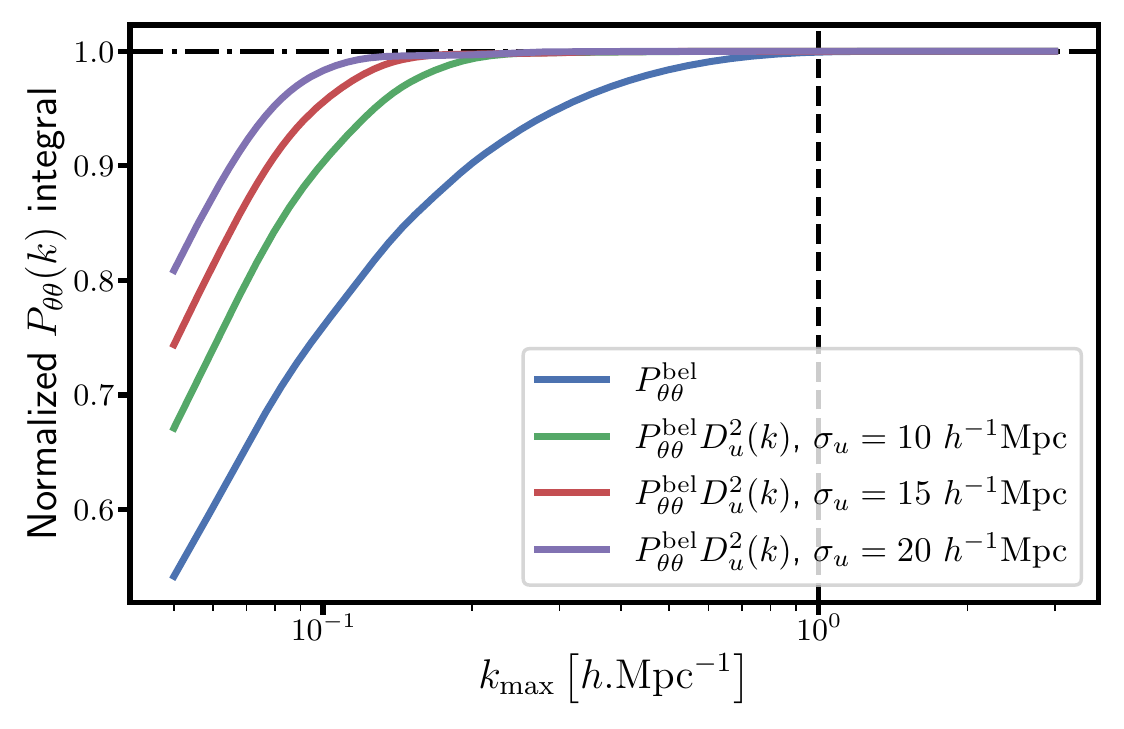}
    \caption{Normalized integral of the power spectrum  as a funtion of integration upper bound $k_{\rm max}$.}
    \label{fig:pwconv}
\end{figure}

As stated in Sect.~\ref{sec:methodology:covariance:numerical} we chose the integration limit of the power spectrum such as the integral has converged. In Fig.~\ref{fig:pwconv} we show the integral of the power spectrum as a function of the $k_{\rm max}$ upper bound of integration normalized by the integral with $k_{\rm max} = 10 h {\rm Mpc}^{-1}$. At $k_{\rm max} = 1 h {\rm Mpc}^{-1}$ we see that the integral has converged.
\FloatBarrier

\end{appendix}
\end{document}